\documentclass[aps,pra,reprint,amsmath,amssymb,superscriptaddress,onecolumn,longbibliography,nofootinbib,notitlepage]{revtex4-2}
\usepackage{mathtools}
\usepackage{siunitx}
\usepackage[hidelinks]{hyperref}
\usepackage{svg}
\usepackage{caption}
\usepackage{minitoc}
\usepackage{subcaption}
\usepackage{booktabs}
\usepackage{tabularx}
\usepackage[braket, qm]{qcircuit}
\usepackage{bbm}
\usepackage{float}
\makeatletter
\let\newfloat\newfloat@ltx    
\let\float@end\float@end@ltx   
\makeatother
\usepackage{algorithm}
\usepackage{algpseudocode}
\usepackage{amsmath}

\usepackage[justification=raggedright,singlelinecheck=false]{caption} 

\usepackage{array}
\newcommand{\eqcontrib}{These authors contributed equally to this work.}
\usepackage{xcolor}

\begin{document}
\title{Programmable nonlinearity within nanophotonic waveguides}
\author{Benjamin~A.~Ash}
\thanks{\eqcontrib}
\affiliation{School of Applied and Engineering Physics, Cornell University, Ithaca, NY 14853, USA}

\author{Ryotatsu~Yanagimoto}
\thanks{\eqcontrib}
\affiliation{School of Applied and Engineering Physics, Cornell University, Ithaca, NY 14853, USA}
\affiliation{Physics \& Informatics Laboratories, NTT Research, Inc., Sunnyvale, CA 94085, USA}

\author{Mandar~M.~Sohoni}
\thanks{\eqcontrib}
\affiliation{School of Applied and Engineering Physics, Cornell University, Ithaca, NY 14853, USA}

\author{Yang~Xu}
\affiliation{School of Applied and Engineering Physics, Cornell University, Ithaca, NY 14853, USA}

\author{Yiqi~Zhao}
\affiliation{School of Applied and Engineering Physics, Cornell University, Ithaca, NY 14853, USA}

\author{Martin~M.~Stein}
\thanks{Present address: Department of Applied Physics, Yale University, New Haven, CT 06520, USA}
\affiliation{School of Applied and Engineering Physics, Cornell University, Ithaca, NY 14853, USA}

\author{Fan~O.~Wu}
\affiliation{School of Applied and Engineering Physics, Cornell University, Ithaca, NY 14853, USA}

\author{Marc~Jankowski}
\affiliation{Physics \& Informatics Laboratories, NTT Research, Inc., Sunnyvale, CA 94085, USA}
\affiliation{E.\,L. Ginzton Laboratory, Stanford University, Stanford, CA 94305, USA}

\author{Logan~G.~Wright}
\thanks{Present address: Department of Applied Physics, Yale University, New Haven, CT 06520, USA}
\affiliation{School of Applied and Engineering Physics, Cornell University, Ithaca, NY 14853, USA}
\affiliation{Physics \& Informatics Laboratories, NTT Research, Inc., Sunnyvale, CA 94085, USA}

\author{Tatsuhiro~Onodera}
\affiliation{School of Applied and Engineering Physics, Cornell University, Ithaca, NY 14853, USA}
\affiliation{Physics \& Informatics Laboratories, NTT Research, Inc., Sunnyvale, CA 94085, USA}

\author{Peter~L.~McMahon}
\thanks{Contact: baa77@cornell.edu, ryotatsu.yanagimoto@ntt-research.com, mms477@cornell.edu, pmcmahon@cornell.edu}
\affiliation{School of Applied and Engineering Physics, Cornell University, Ithaca, NY 14853, USA}
\affiliation{Kavli Institute at Cornell for Nanoscale Science, Cornell University, Ithaca, NY 14853, USA}

\begin{abstract}
Nonlinear nanophotonics enables the engineering of optical functions using light-matter interactions in nanoscale structures with carefully designed geometries. Phase matching in nonlinear optics is typically essential for realizing high efficiency for desired processes and engineering of sophisticated functions, but it can be challenging to reliably achieve in nanophotonic waveguides. Conventionally, phase matching in nanophotonics is achieved through nanofabrication, e.g., by poling a device's nonlinear material using electrodes with fixed geometry. This approach has two drawbacks: the optical functions a device can perform are fixed at the time of fabrication, and the device's performance can be degraded---sometimes drastically---by fabrication imperfections. Here, we circumvent these limitations by engineering programmable $\chi^{(2)}$ holograms inside nanophotonic waveguides, enabling reconfigurable quasi-phase matching that can be updated after device fabrication in approximately one second. By projecting structured optical illumination onto photoconductive electrodes, we generated spatial patterns of electric-field-induced $\chi^{(2)}$ to exert spectral, modal, and polarization control over three-wave mixing processes. This programmability also allowed us to employ closed-loop in situ optimization to compensate for phase mismatch in complex waveguide structures, such as width-modulated waveguides and spiral waveguides exceeding \SI{10}{cm} in length. Furthermore, we show that programmable nonlinearity enables powerful in situ diagnostics, including direct dispersion measurements of various transverse modes and on-chip wavefront phase tomography. Our work opens up the possibility of building cascaded nonlinear nanophotonic systems where yield issues due to fabrication variations can be sidestepped while retaining the benefits of etched waveguides such as confinement and dispersion engineering.

\end{abstract}

\maketitle

\section{Introduction}
\label{sec:intro}
Nanoscale engineering of optical materials has broadened the realm of possibility for photonics, enabling dense integration of photonic components~\cite{koenderink2015nanophotonics, zhu2021integrated}, extreme enhancement of light-matter interactions~\cite{aoki2006strongcoupling,armani2007label}, and emergent functions and phenomena, such as photonic bandgaps and bound states in the continuum~\cite{joannopoulos1997photoniccrystals, hsu2016BIC}. In particular, nonlinear optics has benefited from a strong synergy with nanophotonics: carefully nanofabricated structures not only enabled significant enhancement of nonlinear interactions~\cite{zhao2022ingap, wang2019towards-single-photon} but they also unlocked unique capabilities for dispersion engineering~\cite{jankowski2021dispersion}, phase matching~\cite{fiore1998phase-matching}, and system-level integration~\cite{zhao2024frequency-division, stokowski2024integrated-comb}. These advancements span both classical and quantum applications~\cite{nehra2022few, guidry2022quantum} in narrowband as well as broadband regimes~\cite{kippenberg2011microresonator}. However, as the field of nonlinear nanophotonics has matured, inherent challenges of the conventional nanophotonics paradigm have started to surface. Nanophotonic structures are hard-etched into materials and cannot be modified post-fabrication. This inflexibility restricts potential applications and leaves devices vulnerable to fabrication errors and environmental drift, ultimately hindering large-scale integration~\cite{zhao2023unveiling}. Even state-of-the-art remedies, such as poling adapted to the measured geometry of each individual waveguide~\cite{chen2024adapted}, must be implemented device-by-device during fabrication and these structures remain fixed thereafter. A recent demonstration of multi-timescale frequency-phase matching~\cite{mehrabad2025multi} achieves robustness to fabrication imperfections using a sophisticated coupled-resonator architecture, but the approach is tied to that specific design and does not readily extend to generic device geometries. These are essential hurdles to overcome before nonlinear photonics can be deployed at a large scale with reliable, high-yield performance---a critical prerequisite for creating complex nonlinear systems for optical networking~\cite{cheng2018datacenter}, optical computing~\cite{mcmahon2023physics}, and integrated quantum information processing~\cite{wang2020integrated}.

To tackle these issues, programmable nonlinear photonics has emerged as a powerful new paradigm in which devices possess a large number of parameters that can be tuned post-fabrication. Such programmability not only enables the realization of multiple nonlinear-optical functions on a single setup~\cite{genty2021ML-photonics,tzang2018adaptive-shaping, wetzel2018customizing-SCG, qiu2024spectral-customization} but also allows nonlinear photonics to adopt powerful operating paradigms demonstrated in the preceding development of programmable physical systems in general, such as hardware error correction~\cite{miller2015perfect, bandyopadhyay2021hardware}, in situ inverse design~\cite{judson1992teaching-laser}, and physical computations~\cite{wright2022deep, tanaka2019recent}. Programmable nonlinear photonics is a young and rapidly growing field, but a remarkably diverse set of physical phenomena and platforms has already been explored, including ferroelectric liquid crystals~\cite{sultanov2024tunable}, photogalvanic effects~\cite{nitiss2022optically, lu2021photogalvanic, hickstein2019photogalvanic, li2025photogalvanic}, field-programmable nonlinearities~\cite{Ryo-paper} and amplifiers~\cite{wu2023lithography-free,wu2025fieldprogrammable}, nonlinear-optical liquids~\cite{hofmann2025nonlinearliquid}, and micro-heater arrays~\cite{hu2026efficient-QFC}. A powerful engineering tool that has recently emerged in this context is programmable quasi-phase matching (QPM)~\cite{nitiss2022optically, lu2021photogalvanic, hickstein2019photogalvanic, li2025photogalvanic, Ryo-paper}. Conventionally, QPM is realized by manufacturing periodic spatial variations of the $\chi^{(2)}$ optical nonlinearity~\cite{armstrong1962interactions,fejer1992-QPM}, e.g., by periodic poling of ferroelectric materials~\cite{boes2021efficient} or orientation patterning of semiconductors~\cite{eyres2001orientation-patterning}, and is known to enable precise control over the spatial~\cite{ellenbogen2009nonlinear}, spectral~\cite{imeshev1998engineerable, arbore1997pulse}, and even quantum features of a three-wave mixing process~\cite{weiss2025domain-engineering, serino2023realization}. The ability to render QPM programmable allows us to combine insights from this rich literature on QPM engineering with the strengths of programmable photonics. Still, many of the existing approaches constrain the structures of the QPM gratings that can be realized, and demonstrations of fully arbitrary programmable QPM have been limited to simple device geometries~\cite{Ryo-paper}, leaving the full potential of the interplay between programmable nonlinearity and nanophotonics largely unexplored.

In this work, we explore the potential of embedding programmable QPM within etched nanophotonic devices, demonstrating a rich interplay between static nanophotonic structures and real-time reconfigurable nonlinearity. To equip a given photonic device with programmable nonlinearity, we adapt the scheme developed in Refs.~\cite{Ryo-paper, Hiro-paper}, which uses optical illumination to program the device. This approach draws inspiration from preceding works on using illumination to reconfigure photonic devices, including light-induced photorefractive effects~\cite{brady1991holographic-photorefractive, psaltis1990holography-photorefractive, fleischer2003photorefractive}, field programming of optical gain~\cite{wu2023lithography-free,wu2025fieldprogrammable}, and laser writing of ferroelectric domains~\cite{xu2022femtosecond-LN}. Here, we use light to control a bias electric field, which in turn modifies the material properties of the waveguide core. Specifically, we deposit a layer of photoconductive material and a transparent electrode on the top cladding and apply a bias voltage between the top electrode and the substrate while shining patterned illumination onto the device. This renders the photoconductive layer conductive where the illumination intensity is high, creating an electric-field distribution inside the waveguide core that follows the illumination pattern. The electric field then induces an effective $\chi^{(2)}$ nonlinearity from the $\chi^{(3)}$ nonlinearity of the material via the electric-field-induced second-harmonic (EFISH) effect~\cite{terhune1962-EFISH}. It has been demonstrated that EFISH enables the implementation of a QPM grating by patterning the electric field, e.g., with lithographically defined electrodes~\cite{kashyap1989-EFISH-fiber,timurdogan2017electric, heydari2023efish}. Our approach extends these ideas to a realization of free-form programmable electrodes, allowing us to realize arbitrarily programmable QPM on various nanophotonic devices.

Using programmable QPM on a straight waveguide, we first demonstrated paradigmatic examples of nonlinear optics, realizing flexible control of the spectral, polarization, and modal features. To demonstrate spectral control, we realized second-harmonic generation (SHG), sum-frequency generation (SFG), and difference-frequency generation (DFG) with tunable output wavelengths. We achieved polarization-selective SHG by selectively activating Type-0, Type-I, or Type-II phase matching via the programmable QPM~\cite{Boyd_NLO_2008}. Furthermore, tuning the QPM grating period enabled the selective excitation of distinct transverse waveguide modes. We then cascaded multiple QPM gratings to create arbitrary coherent superpositions of transverse modes, allowing us to actively generate different orbital-angular-momentum (OAM) states~\cite{yao2011OAM}. Throughout this work, we highlight that the real-time reconfigurability of our devices enabled closed-loop in situ optimization and inverse design of device functionality based on feedback from experiments. This capability is exemplified by efficient phase matching on a spiral waveguide of more than \SI{10}{cm} in length and on wavy waveguides with spatially varying widths. To our knowledge, arbitrary QPM in such complex geometries has not been demonstrated with conventional approaches. We also demonstrated arbitrary engineering of the pulse-pumped SHG spectra by inverse designing QPM gratings. Here, the interplay with the native $\chi^{(3)}$ nonlinearity of the waveguide caused extra spectral broadening, leading to \SI{80}{nm} of bandwidth over which the spectrum can be engineered. Furthermore, we highlight that programmable QPM could be a powerful tool for in situ device characterization, demonstrating direct measurements of the waveguide dispersion and phase-sensitive tomography of spatial modes on chip. These results establish programmable QPM as a versatile engineering tool for future nonlinear nanophotonic systems.

\section{Operating principles and device designs}
\begin{figure}[h]
    \centering
    \includegraphics[width=\linewidth]{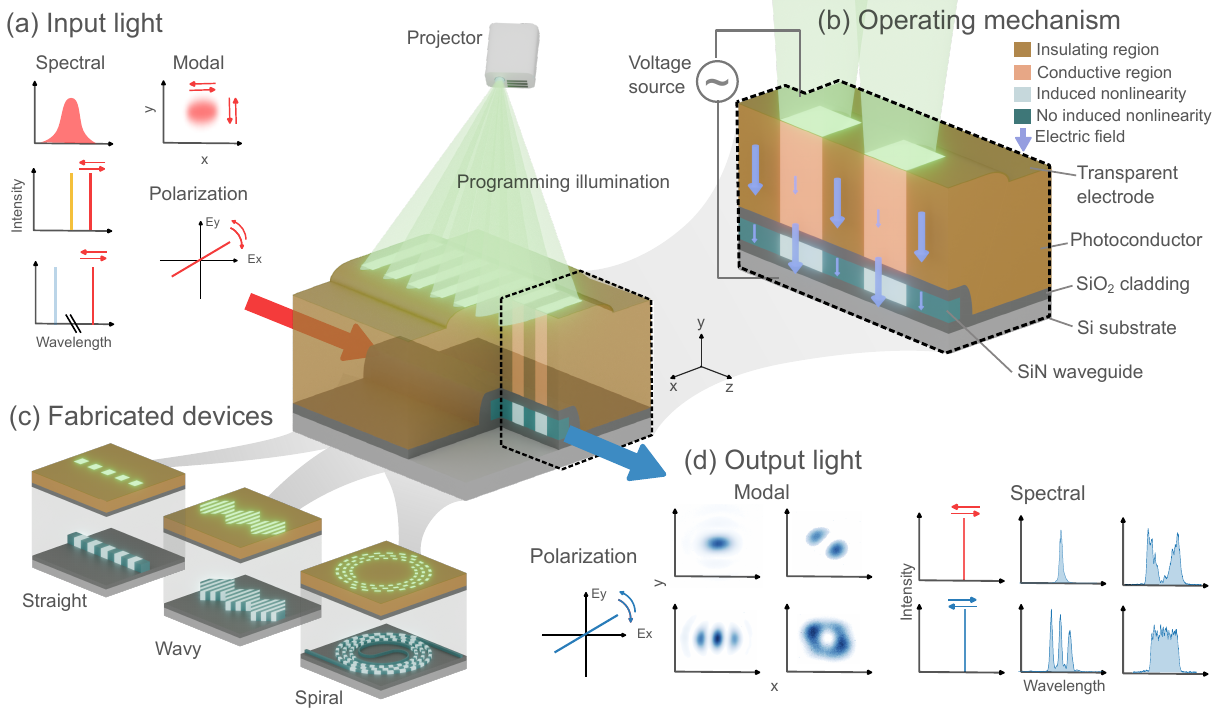}
    \caption{\textbf{Programmable nonlinear photonics with diverse device geometries and applications.} (a) Input light to the programmable nonlinear waveguide can have various features in the spectral, modal, and polarization domains. (b) A schematic of the structure and operating mechanism of a programmable nonlinear waveguide. The waveguide consists of a SiN waveguide, a photoconductor, and a transparent electrode. The illumination projected onto the device renders the photoconductor locally conductive wherever the illumination intensity is high, letting the electric field through. The electric field impinging on the SiN waveguide core then creates an electric-field-induced $\chi^{(2)}$ nonlinearity and a periodic spatial modulation of the $\chi^{(2)}$ nonlinearity, which allows us to realize programmable nonlinear dynamics with changing illumination. (c) Programmable nonlinear photonic geometries fabricated in this work. (d) The programmable QPM allows the device to selectively access the input features, process them, and generate output with engineered features. }
    \label{fig:Figure 1}
\end{figure}

In Fig.~\ref{fig:Figure 1}, we show the operating mechanism and design of the programmable nonlinear waveguide. The fabrication process for our programmable waveguides is as follows. On a conductive silicon substrate, we deposited a $\SI{1}{\micro m}$-thick thermal oxide layer as the bottom cladding and a $\SI{2}{\micro m}$-thick silicon nitride (SiN) layer as the waveguide core\footnote{In this work, we used two different kinds of PECVD-deposited SiN, one provided by Silicon-Valley Microelectronics (SVM) and the other made in-house.}. We then performed photolithography and etched the SiN to form waveguides with various geometries, including straight, spiral, and wavy shapes with varying widths. We capped the SiN with $\SI{1}{\micro m}$ of silicon dioxide ($\mathrm{SiO}_2$) as the top cladding and then performed high-temperature annealing to reduce propagation losses.\footnote{For the in-house SiN, the resulting propagation losses were \SI{0.39}{dB/cm} (\SI{1580}{nm}) and \SI{0.67}{dB/cm} (\SI{802}{nm}). The SVM-provided SiN exhibited losses of \SI{0.35}{dB/cm} and \SI{1.35}{dB/cm}, respectively.} Finally, to impart programmable nonlinearity in this device, we deposited $\SI{9}{\micro m}$ of photoconductor (silicon-rich silicon nitride (SRN)) and sputtered a transparent electrode (indium tin oxide) on top. Appendix~\ref{sec:Fabrication} goes into more detail about device fabrication.

During device operation, we connected the substrate and the top electrode to an AC voltage source, applying a vertical electric field through the entire stack. We then illuminated the device with a structured pattern. This pattern, which we refer to as programming illumination, features a two-dimensional intensity distribution $I(x,z)$ that is generated by a spatial light modulator (SLM) and a \SI{532}{nm} laser source. The wavelength of the programming illumination was chosen to be above the bandgap of the photoconductor, exciting charge carriers and rendering the material locally conductive in regions of high optical intensity. This allowed the electric field to vertically penetrate through the illuminated region~\cite{chiou2005tweezer}, creating a programmable distribution of the bias electric field $E_\text{bias}(x,z)$ inside the core. By breaking the inversion symmetry of the core material, this bias electric field induced an effective second-order nonlinearity, $\chi^{(2)}(x,z)=3\chi^{(3)}E_\text{bias}(x,z)$, from the native $\chi^{(3)}$ response. Consequently, the spatial intensity profile of the programming illumination, $I(x,z)$, directly defined the distribution of the bias field, $E_\text{bias}(x,z)$, thereby transferring the optical pattern into a spatially resolved $\chi^{(2)}(x,z)$ profile. As a result, we could program (i.e., set and change) two-dimensional distributions of $\chi^{(2)}(x,z)$ on our device. This two-dimensional spatial control allows us to dynamically program the nonlinearity along arbitrary waveguide paths, a capability crucial for phase-matching complex planar layouts such as spiral waveguides. The maximum dynamic range of our programmable nonlinearity is \SI{0.29}{pm/V} with a spatial resolution of \SI{3.8}{\micro m}, which is currently set by the resolution limit of the optics for the programming illumination. The current update speed of $\sim$\SI{1}{Hz} is limited by the response time of the SLM. We have not observed any signature of memory effects or fatigue over the course of a few weeks of continuous operation.

Generally, the $\chi^{(2)}$ optical nonlinearity mediates three-wave mixing (TWM) interactions among light waves with frequencies $\omega_j~(j\in\{1,2,3\})$ that fulfill energy conservation $\omega_3=\omega_1+\omega_2$~\cite{Boyd_NLO_2008}. In practice, however, the propagation constants (i.e., wavenumbers) of the corresponding modes, $k_j$, have a nonzero phase mismatch due to dispersion
\begin{align}
\label{eq:delta-k}
    \Delta k = k_3-k_1-k_2
\end{align}
which hinders interaction of the waves. Quasi-phase matching (QPM) is a technique that periodically modulates the distribution of $\chi^{(2)}$ nonlinearity to compensate for the phase mismatch and enable efficient TWM. In standard QPM, a spatial modulation period $\Lambda$ supplies a momentum kick that compensates for a phase mismatch of $\Delta k=2\pi/\Lambda$. More broadly, with more complex spatial profiles of the $\chi^{(2)}$ nonlinearity, one can flexibly activate desired nonlinear processes, which we leverage with the programmable nonlinearity on our platform.

In the following sections, we present the experimental results demonstrated on the programmable nonlinear waveguides.

\section{Programmable periodic poling for quasi-phase matching}
\label{sec:cw-nlo}
\begin{figure}[H]
    \centering
    \includegraphics[width=\linewidth]{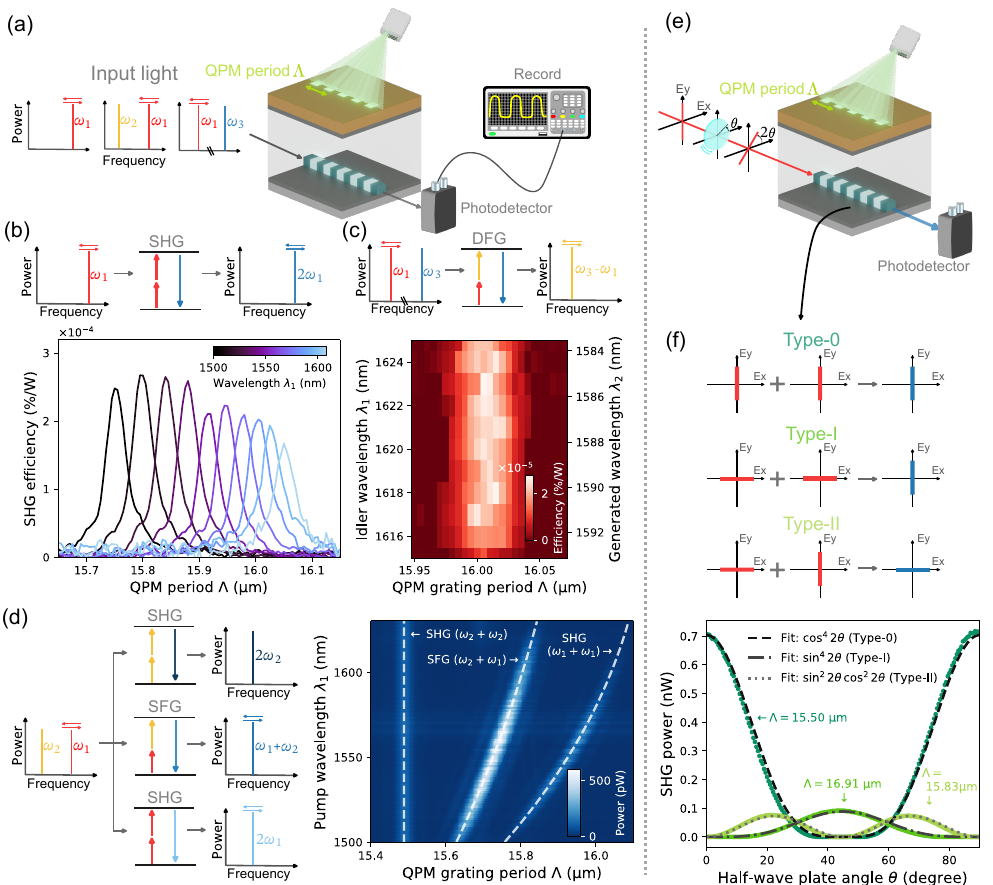}
    \caption{\textbf{Programmable continuous-wave wave-mixing and polarization-dependent phase-matching of straight channel waveguides.} (a) A schematic of the experiment. We send CW input light with a tunable wavelength or polarization into a straight programmable nonlinear waveguide. The programming illumination implements a monotonic QPM grating with period $\Lambda$. (b) The SHG efficiency for various pump wavelengths $\lambda_1$, where the QPM grating period $\Lambda$ was tuned to realize efficient phase matching. The efficiency was normalized with respect to the pump power. (c) Efficiency of the DFG. We used two CW pump lasers, one with a constant wavelength $\lambda_3\approx\SI{802}{nm}$ and power \SI{1}{mW}, and the other with a tunable wavelength $\lambda_1$ between \SI{1615}{nm} and \SI{1625}{nm} with nominal power between \SI{0.5}{mW} and \SI{0.7}{mW} on chip. (d) The power of the SFG signal pumped by two CW lasers, one with a constant wavelength $\lambda_2=\SI{1442.5}{nm}$ and the other with a tunable wavelength $\lambda_1$ between \SI{1500}{nm} and \SI{1630}{nm}, with nominal powers of \SI{5.1}{mW} and \SI{1.0}{mW} to \SI{9.0}{mW} on chip, respectively. (e) A schematic of the experiment used to demonstrate polarization-dependent phase-matching. (f) The output power of the polarization-sensitive SHG, shown as a function of the half-wave plate angle $\theta$ used to rotate the polarization of CW pump light with a wavelength of \SI{1580}{nm} and a power of \SI{24.7}{mW} measured after the waveguide. Based on prior characterizations, we estimated the QPM grating periods that are required to realize Type-0, Type-I, and Type-II phase matching. By setting the grating period $\Lambda$ to each of the respective values, we selectively activated each phase matching process. The efficiencies as functions of $\theta$ are fit to theoretical predictions. For all the measurements, we used straight programmable waveguides made of in-house SiN with a bias voltage of \SI{800}{V} at \SI{10}{Hz}. See Appendix~\ref{sec:appendix CW-NLO} for more details. }
    \label{fig:CW-NLO}
\end{figure}

In this section, we present results demonstrating paradigmatic nonlinear-optical processes in straight programmable nonlinear waveguides, highlighting the unique functionalities enabled by the programmable QPM grating. Figure~\ref{fig:CW-NLO}(a) provides an overview of the experiments performed. We pump the waveguide with continuous-wave (CW) lasers while scanning the wavelength, with the grating period of the programmable QPM selecting which nonlinear process occurs. To test the basic functionality of programmable QPM, we first implemented SHG with a tunable phase-matching condition (see Fig.~\ref{fig:CW-NLO}(b)). We coupled a CW laser into the waveguide and projected a monotonic grating pattern with period $\Lambda$ on top to create a QPM grating (i.e., $\chi^{(2)}\propto \operatorname{sgn}[\cos(2\pi z/\Lambda)]$). For any pump wavelength $\lambda_1$ between \SI{1500}{nm} and \SI{1630}{nm}, we found a $\Lambda$ that optimally phase-matched the corresponding SHG process, generating light at $\lambda_3=\lambda_1/2$ (more details can be found in Appendix~\ref{subsec:vanilla SFG}). Because of the sinusoidal oscillation of the bias electric field, the power of the SHG also oscillates in time, and we report the peak power or peak conversion efficiency. The tunability of the SHG could, in principle, extend well beyond what we demonstrate here since the bandwidth of our experimental demonstration was set by the available wavelength range of the pump laser. 

Expanding this dynamic phase-matching capability, we also demonstrated DFG (see Fig.~\ref{fig:CW-NLO}(c)). For this experiment, we employed a tunable CW source spanning $\lambda_1 = \SI{1615}{nm}$ to $\SI{1625}{nm}$, co-propagating with a fixed-wavelength pump laser at $\lambda_3\approx\SI{802}{nm}$. By configuring the modulation period $\Lambda$ to satisfy the requisite phase-matching condition, these incident fields undergo DFG, yielding an output wave at $\lambda_2=(\lambda_3^{-1}-\lambda_1^{-1})^{-1}$. Appendix~\ref{subsec:DFG} describes the DFG experiments in greater detail. Complementing the SHG and DFG experiments, we also demonstrated tunable SFG. This involved co-propagating two CW inputs: a tunable signal spanning $\SI{1500}{nm}$ to $\SI{1630}{nm}$ ($\lambda_1$), alongside a fixed-wavelength source at $\lambda_2=\SI{1442.5}{nm}$. Figure~\ref{fig:CW-NLO}(d) shows a two-dimensional heatmap of the detected SFG signal over $\lambda_1$ and $\Lambda$, in which three distinct curves are visible. The curves correspond to the SHG of $\lambda_1$, the SHG of $\lambda_2$, and the SFG between $\lambda_1$ and $\lambda_2$. The origin of each can be identified from its dependence on $\lambda_1$, as labeled in Figure~\ref{fig:CW-NLO}(d). These results show that any one of the three processes can be selected through the choice of $\Lambda$. We expect such selectivity to be difficult to achieve with approaches based on the photogalvanic effect for reconfigurable QPM, where the different generation processes would compete with one another~\cite{hickstein2019photogalvanic, wang2026green}. These results highlight the unique strength of our programmable QPM to simultaneously harness multiple nonlinear-optical processes. 

The optimal QPM period $\Lambda$ for a given pair of pump wavelengths is set by the dispersion of the interacting waveguide modes through Eq.~\eqref{eq:delta-k}. Conversely, measuring the optimal period $\Lambda_\text{opt}$ as a function of pump wavelength lets us retrieve the waveguide dispersion. We demonstrated this using the programmable SFG data of Fig.~\ref{fig:CW-NLO}(d): by jointly fitting quadratic curves to $\frac{2\pi}{\Lambda_\text{opt}(\lambda_1,\lambda_1)}$ and $\frac{2\pi}{\Lambda_\text{opt}(\lambda_1,\lambda_2)}$, we retrieved the dispersion of the fundamental TM modes ($\mathrm{TM_{00}}$) around \SI{775}{nm} and \SI{1550}{nm} up to second order. The estimated waveguide parameters are summarized in Table~\ref{tab:dispersion_main} with further experimental details in Appendix~\ref{sec:GVD_fitting}. At present, we expect the dominant source of error in our measurement to be the calibration of the imaging system's magnification ratio, which we could not determine to better than $\sim1\%$. Since this calibration sets the length scale of $\Lambda$ and therefore scales all of our dispersion estimates linearly, we report those values to three significant digits throughout this work. 
\noindent
We further cross-validated these results using SHG/SFG involving higher-order spatial mode $\mathrm{TM_{20}}$ for the second harmonic (see Sec.~\ref{sec:modal} for details on nonlinear optics with higher-order spatial modes). These independent measurements agreed well within the measurement uncertainty on the group-velocity dispersion of the $\mathrm{TM_{00}}$ mode at \SI{1550}{nm}, underscoring the reliability of the retrieved parameters. Taken together, these results show that the augmented programmable nonlinearity can serve as a powerful in situ metrology tool for nanophotonic waveguides.

\begin{table}[h]
    \centering
    \begin{tabular}{lcc}
         \hline
         \hline
         Modes used for measurement
         &$\mathrm{TM_{00}}$(FH)$~\leftrightarrow~$$\mathrm{TM_{00}}$(SH)& $\mathrm{TM_{00}}$(FH)$~\leftrightarrow~$$\mathrm{TM_{20}}$(SH) \\
        \hline
         Phase mismatch between $\mathrm{TM_{00}}$(FH) and $\mathrm{TM_{00}}$(SH) & \SI{394}{rad/mm} & ---\\
         Phase mismatch between $\mathrm{TM_{00}}$(FH) and $\mathrm{TM_{20}}$(SH) & --- & \SI{316}{rad/mm}\\
         GVM between $\mathrm{TM_{00}}$(FH) and $\mathrm{TM_{00}}$(SH)& \SI{44.4}{fs/mm} & ---\\
         GVM between $\mathrm{TM_{00}}$(FH) and $\mathrm{TM_{20}}$(SH)& --- & \SI{80.4}{fs/mm}\\
        GVD of $\mathrm{TM_{00}}$(SH)& \SI{135}{fs^2/mm} & ---\\
        GVD of $\mathrm{TM_{20}}$(SH)& --- & \SI{103}{fs^2/mm}\\
        GVD of $\mathrm{TM_{00}}$(FH)& \SI{-104}{fs^2/mm} & \SI{-104}{fs^2/mm}\\
        \hline
    \end{tabular}
    \caption{\textbf{Dispersion parameters extracted via programmable on-chip sum-frequency generation.} Measured phase mismatch, group velocity mismatch (GVM), and group velocity dispersion (GVD) of a straight programmable nonlinear waveguide. Parameters were derived by fitting second-order polynomials to the optimal QPM grating period for SHG and SFG across various pump wavelengths (see Appendix \ref{sec:GVD_fitting}). FH and SH stand for the fundamental harmonic at \SI{1550}{nm} and the second harmonic at \SI{775}{nm}, respectively. For the SHG and SFG processes, we used two pairs of modes, one with $\mathrm{TM_{00}}$(FH) and $\mathrm{TM_{00}}$(SH), and the other with $\mathrm{TM_{00}}$(FH) and $\mathrm{TM_{20}}$(SH), to independently estimate their dispersions.}
    \label{tab:dispersion_main}
\end{table}

In all the experiments described so far, all the modes involved have been TM modes. With programmable QPM, we can also selectively activate frequency conversion processes involving transverse electric (TE) modes, enabling flexible polarization control. As shown in Fig.~\ref{fig:CW-NLO}(e), we used a half-wave plate (HWP) to rotate the polarization axis of the input CW light with a wavelength of $\lambda=\SI{1580}{nm}$. When the polarization is aligned along the $x$ axis, the $\mathrm{TE_{00}}$ mode is excited; when the polarization is aligned along the $y$ axis, the $\mathrm{TM_{00}}$ mode is excited. Here, we varied the half-wave plate angle $\theta$ continuously to excite a superposition of the fundamental TM and TE modes. As the bias field broke the symmetry of the core material predominantly in the $y$ direction, three components of the effective $\chi^{(2)}$ nonlinear tensor were nonzero, i.e., $\chi^{(2)}_{yyy}$, $\chi^{(2)}_{yxx}$, and $\chi^{(2)}_{xyx}$, each of which predominantly contributes to Type-0, Type-I, and Type-II phase-matched processes, respectively. Because of the dispersion of the waveguide modes, each type of phase matching required a different QPM grating period. In the experiment, we identified $\Lambda=\SI{15.50}{\micro m}$, $\SI{16.91}{\micro m}$, and $\SI{15.83}{\micro m}$ for Type-0, Type-I, and Type-II phase matching, respectively. The data in Fig.~\ref{fig:CW-NLO}(f) shows the measured SHG power as we varied $\theta$, while we set $\Lambda$ to activate each phase-matching process. From the relative magnitude of the SHG conversion efficiency, we could estimate the ratios between the electric-field-induced nonlinear tensor components as $|\chi^{(2)}_{yxx}|/|\chi^{(2)}_{yyy}|=0.36$ and $|\chi^{(2)}_{xyx}|/|\chi^{(2)}_{yyy}|=0.32$, which are close to the condition $\chi^{(2)}_{yxx}=\chi^{(2)}_{xyx}$ expected under Kleinman symmetry~\cite{Boyd_NLO_2008}. Note that this estimate approximates the TE mode to be horizontally polarized and the TM mode to be vertically polarized. These results highlight that programmable QPM can be a powerful means of harnessing the polarization degrees of freedom in nonlinear optics. For all subsequent experiments, only $\mathrm{TM}$ modes were used.

\section{Programmable quasi-phase matching on complex geometries}
\label{sec:spiral}
\begin{figure}[h!]
    \centering
    \includegraphics[width=0.92\linewidth]{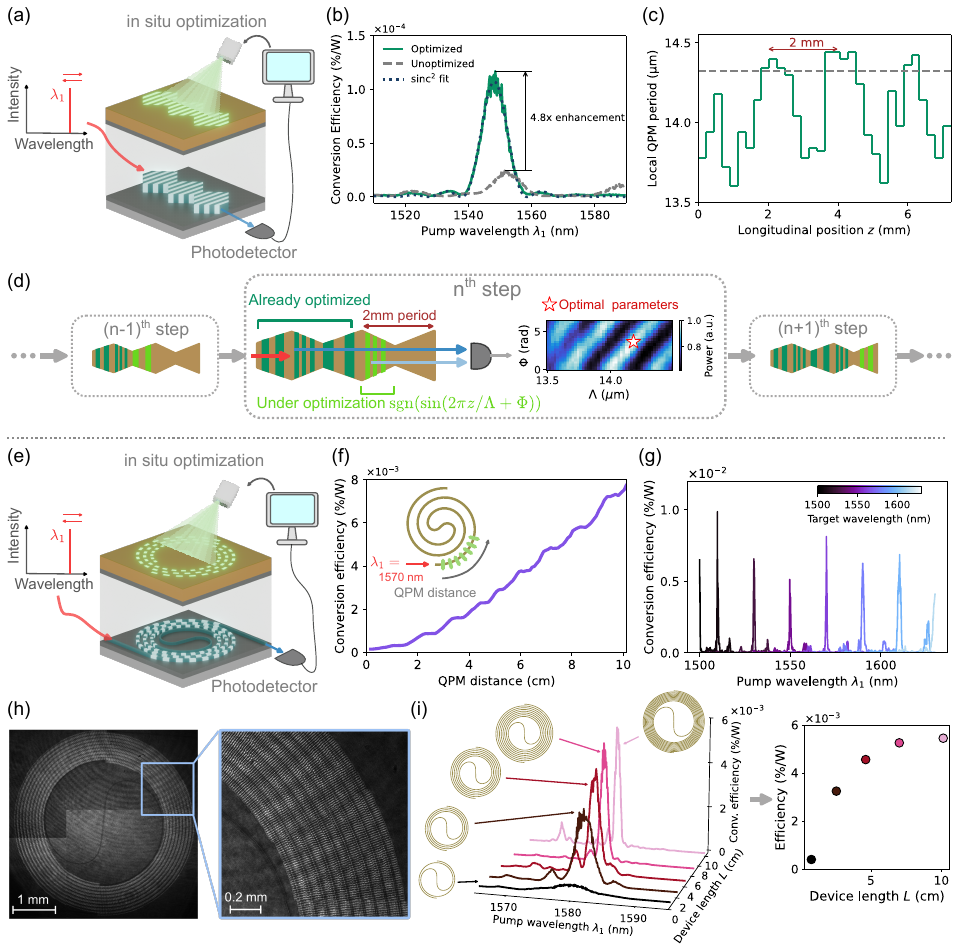}
    \caption{\textbf{Programmable quasi-phase-matching across complex waveguide geometries.} (a) Schematic of the in situ optimization of the QPM grating on a wavy waveguide geometry. (b) Measured SHG conversion efficiency as a function of the pump wavelength. The optimized QPM grating exhibited $4.8$ times more conversion efficiency compared to a QPM grating with a constant period. The dotted lines represent a fit to a $\operatorname{sinc}^2$ function. (c) Optimized local QPM period $\Lambda$ along the longitudinal axis of the waveguide. The red arrow indicates the \SI{2}{mm} period of the width modulation of the waveguide. (d) We partitioned the waveguide into 32 sections and performed iterative stepwise optimizations of the QPM grating in each section. The period $\Lambda$ and phase $\Phi$ of the QPM grating in the target section were varied while the output SHG power was measured. We chose the best combination of $\Lambda$ and $\Phi$ and stepped to the next section. (e) Schematic of the in situ optimization of the QPM grating on a spiral waveguide geometry. (f) Measured SHG conversion efficiency at \SI{1570}{nm} as a function of the length of the QPM grating on a spiral waveguide. (g) The SHG conversion efficiency measured as a function of pump wavelength for QPM gratings optimized for various target wavelengths on a \SI{10.1}{cm} long spiral. (h) Top image of a \SI{10.1}{cm} long spiral waveguide with the programming illumination projected on the top surface. (i) The SHG conversion efficiencies as functions of the pump wavelength for spiral waveguides with various device lengths. The inset shows the SHG conversion efficiency with a \SI{1580}{nm} pump (which the QPM was optimized for) as a function of the propagation length. See Appendix~\ref{sec:spirals} and Appendix~\ref{sec:Wavy Waveguides} for full details of the experiment.}
    \label{fig:spiral}
\end{figure}

In this section, we present experimental results on augmenting programmable nonlinearities in complex device geometries beyond straight waveguides, demonstrating operations of nonlinear optics that would be challenging to realize on conventional non-programmable platforms. To showcase how programmable nonlinearity enabled efficient QPM on nontrivial geometries, we fabricated a waveguide with a wavy structure. As shown in Fig.~\ref{fig:spiral}(a), the width of the waveguide adiabatically varied between \SI{5}{\micro m} and \SI{10}{\micro m} with a period of \SI{2}{mm}. Our motivation for considering this particular geometry is twofold. First, it serves as an extreme example of fabrication-induced structural variations. Under constant-period poling, such spatial variations degrade frequency conversion; consequently, specialized adaptive poling techniques are frequently employed to correct for these unintended geometric fluctuations~\cite{chen2024adapted}. Realizing efficient QPM on such a radically varying waveguide geometry would highlight the capability of our platform to correct extreme device imperfections. Second, introducing spatial variations is known to be a powerful engineering tool in nanophotonics, routinely used for adiabatic frequency conversion, dispersion engineering, and inverse design of nonlinear optics~\cite{lee2024inverse, chi2026adiabatic-taper, singh2019supercontinuum}. Augmenting such structures with programmable nonlinearity can, in principle, unlock a unique interplay of functionalities.

In this experiment illustrated in Fig.~\ref{fig:spiral}(a), we coupled a CW pump laser into the device, swept its wavelength between \SI{1510}{nm} and \SI{1590}{nm}, and measured the SHG power under various programming illumination patterns. As a baseline, a uniform, constant-period grating was chosen to maximize the SHG power when pumped with \SI{1550}{nm} (see Appendix~\ref{sec:Wavy Waveguides}). However, as depicted in Fig.~\ref{fig:spiral}(b), when the programmed QPM grating was chosen to be a uniform grating, the SHG power as a function of pump wavelength exhibited multiple fragmented peaks, deviating significantly from the ideal $\operatorname{sinc}^{2}$ shape. To recover efficient SHG, we performed in situ optimization of the QPM grating as shown in Fig.~\ref{fig:spiral}(d). For this process, we first partitioned the total length of the waveguide into 32 sections, each with a length of $\sim$\SI{0.226}{mm}, and set the wavelength of the CW pump to be \SI{1550}{nm}. A single step of the optimization involved finding the ideal period ($\Lambda$) and phase ($\Phi$) of a `local' QPM grating projected on one section of the waveguide. This was achieved by systematically sweeping $\Lambda$ and $\Phi$ while keeping all previously optimized sections fixed, and then locking in the values of $\Lambda$ and $\Phi$ that generated the maximum total SHG power. Physically, this power-maximization optimization step ensures constructive interference between the generated second-harmonic wave from the active section and the field accumulated from preceding sections. This routine was repeated until all waveguide sections were optimized. As shown in Fig.~\ref{fig:spiral}(b), projecting this optimized QPM grating yielded a pronounced $\operatorname{sinc}^{2}$ peak centered at \SI{1550}{nm} that was significantly higher than that obtained with the uniform, constant-period grating. Furthermore, as presented in Fig.~\ref{fig:spiral}(c), the local period of the optimized QPM grating oscillates with a spatial period matching the \SI{2}{mm} modulation of the waveguide’s width. These results demonstrate that in situ optimization of a programmable nonlinearity can automatically identify and compensate for the underlying topographic variations without requiring prior knowledge of the waveguide geometry. More details about the optimization process can be found in Appendix~\ref{sec:Wavy Waveguides}.

Other scenarios where conventional, non-programmable approaches to periodic poling struggle are extended-length and/or curved waveguides. Efficient frequency conversion over long distances requires unbroken constructive interference. Curved structures, which are critical for extending propagation distances in a compact spatial footprint, introduce geometry-dependent phase shifts among the interacting waves, demanding precise tailoring of the QPM grating phase. Consequently, conventional approaches to QPM have so far been limited to geometries without these challenges, such as straight waveguides or resonators with a constant bending radius~\cite{zhu2021integrated}. The ability to realize QPM on long \emph{and} curved waveguide geometries will enable nanophotonic devices that simultaneously leverage enhanced nonlinear interactions, compact spatial footprints, and extended temporal propagation that can host many temporal modes~\cite{gray2025large-time-multiplexed, lee2012-delay-line}.

To demonstrate this capability, we fabricated programmable nonlinear waveguides with circular spiral geometries spanning \SI{10.1}{cm} of propagation distance and realized efficient QPM on them. The spatial footprint of the programmable region was within a \SI{4.3}{mm} by \SI{4.3}{mm} area. We used the same in situ optimization algorithm as for the wavy waveguide, maximizing the total SHG power through iterative optimization of $\Lambda$ and $\Phi$ in each waveguide partition (see Fig.~\ref{fig:spiral}(e)). After QPM optimization for a CW pump at \SI{1570}{nm}, we observed the SHG signal monotonically increasing as a function of the distance over which the QPM grating is implemented (see Fig.~\ref{fig:spiral}(f)), suggesting that we could realize unbroken constructive interference of the SH wave over a length of \SI{10.1}{cm}. Repeating this process to optimize the SHG for different target pump wavelengths, we observed peaks of comparable conversion efficiency, as shown in Fig.~\ref{fig:spiral}(g). This demonstrates our platform's ability to phase-match SHG for any pump wavelength within a tunable bandwidth of \SI{130}{nm} (between \SI{1500}{nm} and \SI{1630}{nm}). The full width at half maximum (FWHM) of the SHG transfer functions (i.e., conversion efficiencies as functions of the pump wavelength) were as narrow as \SI{0.96}{nm} because of the length of the spiral waveguide. 

Then, to systematically characterize the SHG conversion efficiency as a function of the overall device length (as opposed to the distance of the QPM grating on a waveguide with fixed propagation distance), we evaluated a series of five spiral waveguides with varying total device lengths, where Fig.~\ref{fig:spiral}(i) shows the resultant SHG transfer functions. We observed that the FWHM of the peak continuously narrowed as the interaction length increased, while the conversion efficiency increased up to \SI{10}{cm} and then started to saturate. This deviation of the conversion efficiency from the ideal $(\mathrm{length})^2$ scaling was due to propagation loss, which diminished the return from extending the interaction length beyond a certain point (see Appendices \ref{subsec:superlinear} and \ref{subsec:transfer function} for more details). Nevertheless, our results show that in situ optimization successfully maintained constructive interference over a \SI{10.1}{cm} spiral geometry. Although the absolute conversion efficiency is much lower than that of conventional devices owing to the weak programmable nonlinearity in the devices presented here, these results nonetheless represent a step toward scaling nonlinear photonics by overcoming a critical bottleneck in conventional periodic poling. As fabrication improves and propagation loss decreases, the advantages of longer interaction lengths can be fully appreciated. Notably, ultra-low-loss SiN platforms have already demonstrated \SI{3}{dB} attenuation over lengths exceeding one meter, offering a clear pathway to exploit the quadratic ($L^2$) scaling of conversion efficiency over macroscopic distances~\cite{Lipson-complete-SiN}.  We note here that the results presented in the rest of this work are based on straight waveguides.

\section{Transverse mode engineering}
\label{sec:modal}
\begin{figure}[h]
    \centering
    \includegraphics[width=\linewidth]{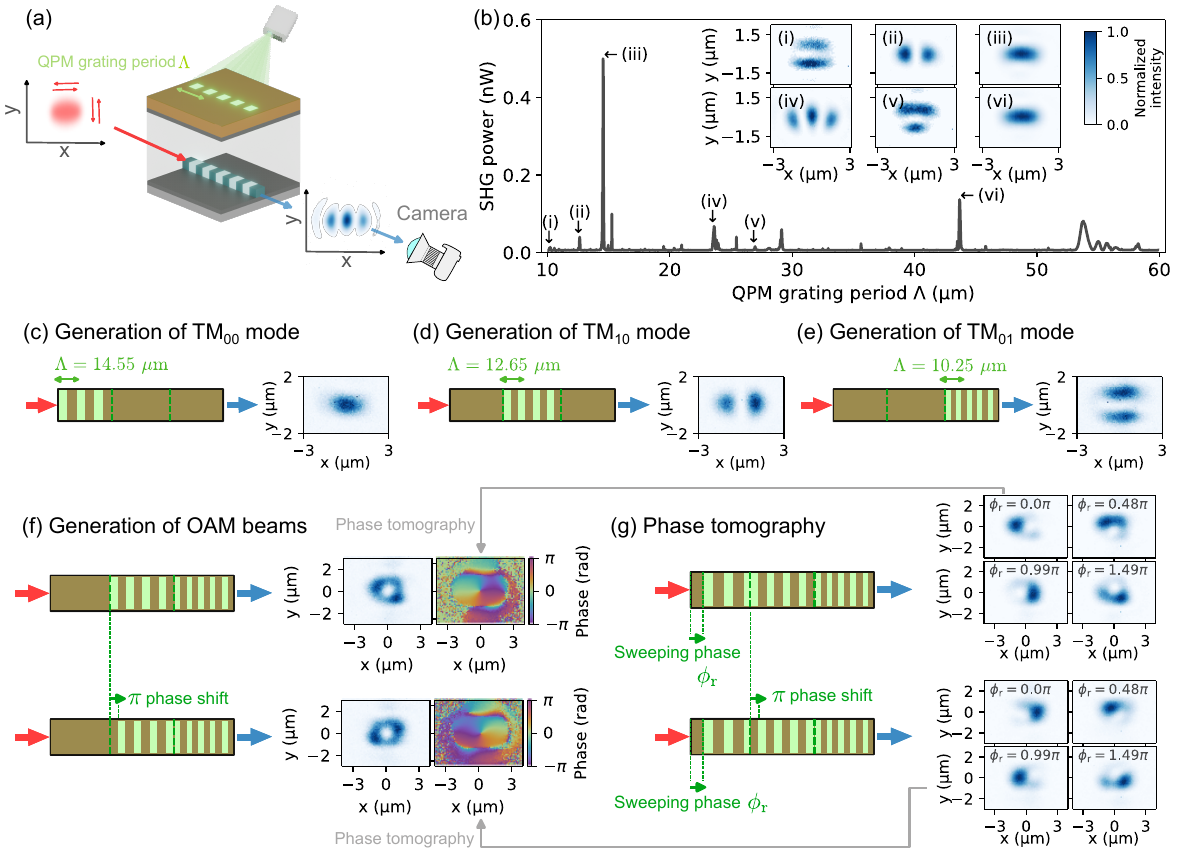}
\caption{\textbf{Programmable transverse mode engineering and on-chip phase tomography.} (a) Schematic of the experiment. We pump a straight programmable nonlinear waveguide with a CW pump laser at a wavelength of \SI{1590}{nm} with a nominal on-chip power of \SI{26.6}{mW} and measure the spatial profiles of the output SHG using a camera as we vary the QPM grating period $\Lambda$. (b) The phase-matching landscape for spatial modes. We measured the output SHG power for various $\Lambda$, showing the spatial profiles of the output at each peak. We note that peak (iii) is saturated due to its high brightness. (c)--(e) We partition the waveguide into 3 subsections and project QPM gratings with periods of $\Lambda=\SI{14.55}{\micro m}$, $\SI{12.65}{\micro m}$, and $\SI{10.25}{\micro m}$, which produce SHG in the TM$_\mathrm{00}$, TM$_\mathrm{10}$, and TM$_\mathrm{01}$ modes, respectively. (f) Cascaded generation of distinct spatial modes and their superpositions. We simultaneously project QPM grating patterns on the sections for the TM$_\mathrm{10}$ and TM$_\mathrm{01}$ modes to produce their coherent superpositions with tunable relative phases, which produces a beam with an OAM of $\pm 1$. (g) Phase tomography operation to retrieve the phase of the light generated in (f). Here, we activate the section assigned to generate the $\mathrm{TM_{00}}$ mode as a phase reference beam with a tunable phase $\phi_\text{r}$. The interference with the phase reference allows us to directly map the phase of the signal on chip. For all the measurements, we used a bias voltage of \SI{800}{V} at \SI{10}{Hz}. See Appendix~\ref{sec:spatial modes} for full details of the experiment.}
    \label{fig:modal}
\end{figure}

While previous sections exclusively used the fundamental spatial modes of the programmable waveguides, with the exception of the in situ dispersion characterization with SFG in Sec.~\ref{sec:cw-nlo}, higher-order spatial modes offer a rich set of additional degrees of freedom. The potential applications of spatially-multimode photonics have caught attention for their unique capabilities, e.g., for high-capacity OAM multiplexed communication channels~\cite{ma2023scaling}, modal phase matching of nonlinear-optical processes~\cite{moutzouris2003second}, unique multimode interferometers~\cite{soldano1995optical-multimode-interferometer}, optical computation \cite{on2024programmable}, and quantum information processing \cite{Mohanty2017quantum}. In this section, we demonstrate the ability of a programmable nonlinear waveguide to control these transverse modal degrees of freedom via SHG.

To investigate the phase-matching parameter space for SHG processes involving higher-order spatial modes, we scanned the QPM grating period $\Lambda$ under CW pumping at \SI{1590}{nm} (see Fig.~\ref{fig:modal}(a)). We intentionally coupled the pump light off from the center axis to break symmetry and excite odd-order transverse modes. As shown in Fig.~\ref{fig:modal}(b), this sweep revealed a series of distinct conversion peaks, each corresponding to an SHG process phase-matched to a specific higher-order spatial mode, with the generated transverse intensity profiles shown in the figure insets. By repeating these measurements for different pump wavelengths, we tracked the positions of each phase-matching peak in the spectrum, from which we measured the dispersion for these higher-order transverse modes (see Sec.~\ref{sec:cw-nlo} and Appendix~\ref{sec:GVD_fitting}).

Beyond exciting isolated spatial modes, the programmability of our platform enables on-chip generation of arbitrary superpositions of these modes. We demonstrated this by partitioning the waveguide into distinct segments that independently generate different SH spatial modes along the optical path, whose relative phases can be tuned by translating the spatial position of each QPM grating. As shown in Fig.~\ref{fig:modal}(f), we applied this technique to generate a superposition of the $\mathrm{TM_{10}}$ and $\mathrm{TM_{01}}$ modes. By equalizing their amplitudes (via the lengths of their respective partitions), and tuning their relative phase to $\pm \pi/2$, we synthesized OAM states with angular momentum $l = \pm 1$. We note that our current devices featured a rectangular cross-section, which resulted in slight geometric asymmetries of the $\mathrm{TM_{10}}$ and $\mathrm{TM_{01}}$ modes. This manifested as minor sidelobes, which could be fully suppressed by utilizing a square waveguide cross-section.

In verifying the helical phase structure of the generated OAM states, we demonstrated another feature of the programmable nonlinear waveguide: on-chip phase tomography. We added a third partition of the waveguide to generate the fundamental $\mathrm{TM_{00}}$ mode, which would serve as a phase reference\footnote{Since the $\mathrm{TM_{00}}$ mode possesses a spatially uniform phase profile across the waveguide core, it serves as an ideal intrinsic local oscillator.} (local oscillator, i.e., LO). By systematically sweeping the relative phase ($\phi_{\text{r}}$) between the LO and the OAM state (see Fig.~\ref{fig:modal}(g)), we captured the interference patterns. Using these interference patterns, we could successfully reconstruct the phase of the OAM states as shown in Fig.~\ref{fig:modal}(f) (see Appendix~\ref{subsec:OAM generation} for more details).  Ultimately, the ability to generate and coherently interfere higher-order spatial modes establishes the programmable nonlinear waveguide as a tool for high-dimensional state engineering and optical signal processing~\cite{singh2026progress-spatial-modes}. 

\section{In situ inverse design of spectrum}
\label{sec:spectrum}
\begin{figure}[h]
    \centering
    \includegraphics[width=\linewidth]{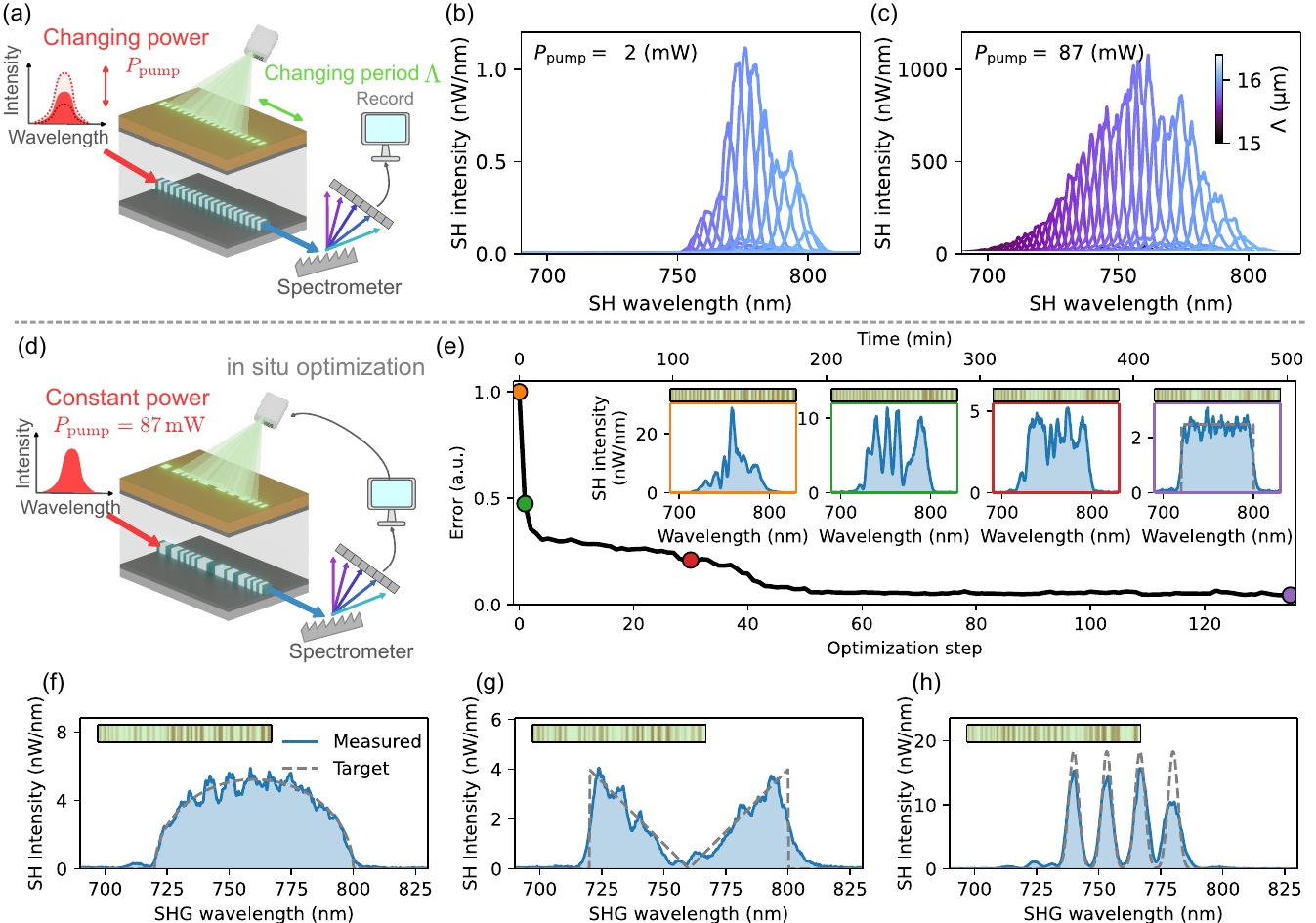}
    \caption{\textbf{In situ inverse design of arbitrary second-harmonic generation spectra.} (a) A schematic of the experiment used to measure the output SHG spectra for various monotonic QPM grating periods $\Lambda$. The pump was an ultrashort pulsed laser with different average powers of (b) \SI{2}{mW} and (c) \SI{87}{mW}. As we increased the pump power, we observed the bandwidth of SHG increasing because the native $\chi^{(3)}$ nonlinearity of the SiN waveguide broadened the spectrum of the pump. (d) A schematic of the closed-loop optimization protocol used to engineer different SHG spectra. (e) Evolution of the error between the experimentally measured and target spectra, with the bottom axis showing the corresponding optimization step count. The insets show the measured SHG spectra at several stages of the optimization, above which the stripes represent the optimized programming illumination patterns, longitudinally downsampled at a period of \SI{14.5}{\micro m}. Experimentally measured spectra for additional target shapes are shown in (f) a semicircle, (g) twin peaks, and (h) four peaks. The gray dashed lines present the target shapes.  For all the measurements, we used a bias voltage of \SI{800}{V} at \SI{10}{Hz}.}
    \label{fig:spectral}
\end{figure}

So far, our results have utilized CW lasers as pump sources. On the other hand, QPM grating engineering is known to be a powerful way to harness the broadband dynamics of nonlinear-optical interactions, enabling arbitrary pulse shaping~\cite{imeshev1998engineerable}, separable biphoton generation~\cite{xin2022spectrally}, and multi-port quantum pulse gates~\cite{serino2023realization}. However, conventional QPM gratings are fixed at fabrication, which limits each device to a single function, and the performance is vulnerable to fabrication imperfections, environmental drift, and variations in experimental conditions. To circumvent these limitations, Ref.~\cite{Ryo-paper} demonstrated a programmable QPM grating on a nonlinear slab waveguide, enabling real-time optimization and reconfigurable control of broadband SHG spectra over a bandwidth of approximately \SI{40}{nm}. The accessible bandwidth of the SHG was primarily limited by the original bandwidth of the pump laser (ELMO-HP; Menlo Systems). Here, using the same pump laser, we show that we can overcome this bandwidth limitation through the interplay between the native $\chi^{(3)}$ nonlinearity and the programmable $\chi^{(2)}$ nonlinearity on a straight programmable nonlinear waveguide.

An illustration of the spectral engineering experiment is shown in Fig.~\ref{fig:spectral}(a). A straight programmable nonlinear waveguide was pumped by an ultrashort pulse laser, and the output SHG spectrum was measured by a spectrometer. To characterize the programmable bandwidth, we first measured the SHG spectrum for various monotonic QPM grating periods $\Lambda$ at low (\SI{2}{mW}) and high (\SI{87}{mW}) pump powers. When the pump power is low, the generated SHG wavelength cannot go beyond a window of approximately \SI{40}{nm}, even with values of $\Lambda$ that could, in principle, phase match SHG outside this window. On the other hand, as we increased the pump power, we observed SHG beyond this original window. This is because the native $\chi^{(3)}$ nonlinearity broadened the pump bandwidth, which was then upconverted by the programmable $\chi^{(2)}$ nonlinearity. 

We then leveraged this spectral broadening to realize an SHG spectrum with an extended bandwidth. Here, the goal was to optimize the QPM grating pattern such that the output spectrum takes a desired shape. As shown in Fig.~\ref{fig:spectral}(d), this was achieved by feeding information about the discrepancy between the target and the experimental outcome back to the programming illumination and iteratively updating the illumination pattern. In Fig.~\ref{fig:spectral}(e), we show the evolution of the SHG spectrum and its deviation from the target shape as a function of the optimization step, over which the programming illumination patterns are iteratively updated. The same method was applied to other target shapes, such as (f) a semicircle, (g) twin peaks, and (h) four peaks, spanning up to \SI{80}{nm} of bandwidth. Such extended programmable bandwidths might benefit a variety of applications, including the frequency conversion of ultrashort pulses without temporal distortion~\cite{arbore1997engineerable} and the generation of ultra-broadband entangled photon pairs for quantum networks~\cite{nasr2008ultrabroadband}.

\section{Discussion}
\label{sec:discussion}

\subsection{Summary of the results}
We have presented a platform for real-time, reconfigurable $\chi^{(2)}$ optical nonlinearity within nanophotonic waveguides. Through spatial engineering of the QPM gratings, we demonstrated comprehensive control over paradigmatic nonlinear-optical processes across spectral (i.e., SHG, SFG, and DFG), polarization (i.e., Type-0, Type-I, and Type-II phase matching), and transverse-mode degrees of freedom. Ultimately, this platform enables arbitrary, real-time switching between nonlinear processes and functions on a single integrated device. A central theme of this work is realizing nonlinear optical operations that are largely inaccessible to conventional, static nonlinear devices. For example, utilizing SFG in tandem with a software-tunable QPM grating provided a metrology tool to extract the dispersion parameters of specific waveguide modes. Beyond exciting fundamental modes, programmable QPM allowed us to selectively excite coherent superpositions of desired transverse modes, enabling the generation and on-chip characterization of an OAM beam. Equally important to the spatial and metrological capabilities is the platform's capacity to optimize the device performance via real-time experimental feedback. We leveraged this in situ optimization to account for geometric variations and realized efficient phase matching on nontrivial device structures, such as spatially varying wavy waveguides and spiral waveguides with lengths exceeding \SI{10}{cm}, for a desired pump wavelength anywhere within a \SI{130}{nm} window (constrained solely by the limits of our tunable pump laser). The in situ optimization could handle even more complicated targets, such as arbitrary spectral profiles for broadband pulse-pumped SHG, where we demonstrated \SI{80}{nm} of programmable bandwidth for the output. Notably, the yield of the nonlinear functions in this work was essentially 100\%: regardless of nanofabrication variations, the programmable nonlinearity could always be reconfigured to realize efficient QPM.

\subsection{Limitations and potential improvements}
Here, we discuss limitations of the devices used in our proof-of-concept experiment and how they could be mitigated in the future. We expect these are the challenges that have to be addressed in order for the technology developed in this work to be widely applied to real-world applications.

One of the most important limitations is the weak optical nonlinearity. In this work, we estimated a maximum value of  $\sim$\SI{0.29}{pm/V} (see Appendix~\ref{sec:chi2 estimation}) of programmable $\chi^{(2)}$ nonlinearity from the SiN core, which falls within the variations of the previously reported values of electric-field-induced $\chi^{(2)}$ nonlinearity in SiN~\cite{sultanov2024tunable,nitiss2022optically, lu2021photogalvanic, hickstein2019photogalvanic,li2025photogalvanic,Ryo-paper}. Specifically, a programmable $\chi^{(2)}$ nonlinearity of \SI{0.47}{pm/V} was demonstrated previously using the same architecture as this work~\cite{Ryo-paper}, where the difference is primarily attributed to the lower bias voltage used in this work. We expect there are mainly two possible routes to mitigate this challenge. The first approach is to engineer the device geometry. Both tighter transverse field confinement and a longer waveguide length improve the efficiency of the device. Indeed, compared to the previous demonstration of a programmable nonlinear waveguide on a slab geometry~\cite{Ryo-paper}, our spiral waveguide with a smaller mode area and a length of \SI{10.1}{cm} exhibited a more than 100-fold improvement in efficiency. Currently, the main factor limiting further progress on this front is the propagation loss of the waveguide, which diminishes the returns from an extended interaction length. Notably, higher-quality SiN is known to offer much lower propagation loss than what we observed~\cite{Lipson-complete-SiN}. One meter long spiral waveguides with \SI{2.4}{dB/m} loss are reported in the literature~\cite{liu2021highyield}, and our results show that realizing a QPM over such an extended length is possible with programmability. Fabricating a resonant structure could further extend the interaction length. Combining the state-of-the-art loss values for SiN with the nonlinearities available on programmable nonlinear photonics, we estimate that an SHG conversion efficiency as high as \SI{1e7}{\% W^{-1}} could, in principle, be realized on a micro-ring resonator, which would exceed the best value in integrated photonics~\cite{Ryo-paper}. These numbers suggest fertile potential for device improvement. The second route is to explore materials with larger nonlinearities. The literature on electric-field-induced $\chi^{(2)}$ nonlinearity is still scarce, and the values for many materials remain unknown. Still, $\chi^{(2)}$ nonlinearities of \SI{41}{pm/V}~\cite{timurdogan2017electric} and \SI{22.7}{pm/V}~\cite{lin2019observation-EFISH} have been reported for silicon and silicon-rich silicon nitride, respectively. Although their smaller bandgaps limit their use in the visible range~\cite{yuan2025green, wang2026green}, their superior transmission at longer wavelengths makes them promising for mid-infrared applications~\cite{lafforgue2022supercontinuum}. Large-bandgap semiconductors like SiC, GaN, and AlN are also promising candidates due to their large transparency window, high breakdown field, and relatively strong nonlinearities~\cite{modal-phase-matching2, zheng2022integratedGaN, guo2016AlN}. Finally, we note that we could increase the effective nonlinearity by a factor of $2$ by dynamically switching the programming illumination in sync with the bias voltage, which also enables continuous device operation (see Appendix~\ref{sec:dynamic-illumination}).

Aside from the magnitude of the nonlinearity, it is worth summarizing three other key metrics of our device performance. First, as the QPM grating period decreases, the fringing effects of the electric fields inside the waveguide would reduce the effective nonlinearity. This limit could be relaxed by employing a photoconductive waveguide core, which would allow a higher resolution of the QPM grating~\cite{Ryo-paper}. Alternatively, the programmable QPM could be implemented on top of conventional phase-matching techniques, such as birefringent or modal phase matching, to significantly relax the required grating period (See Appendix~\ref{sec:hybrid-phase-matching}). It is also possible to use lithographically defined electrodes to apply the bias voltage~\cite{timurdogan2017electric, heydari2023efish}, where photoconductors can be used to selectively activate the electrodes to enable on-chip control of the bias field. Second, the current update speed of the device is rather slow, $\sim\SI{1}{s}$, a timescale set by the speed of the spatial light modulator (SLM) used to generate the programming illumination, whereas state-of-the-art SLMs have demonstrated speeds on the order of tens of megahertz~\cite{panuski2022SLM, wei2026SLM}. Once the projector speed is no longer a limitation, the next speed limit is imposed by the RC time constant of the device, $\sim\SI{0.1}{s}$. We expect that an update speed exceeding $\SI{1}{MHz}$ could be possible with the use of more conductive photoconductors, e.g., amorphous Si~\cite{kumata1986a-Si} (see Appendix~\ref{sec:photoconductor-improvement}). Third, the projector setup requires bulky free-space optics, increasing the footprint of the entire system. Bonding a micro light-emitting diode (LED) display to the device surface could alleviate this problem while providing sufficient resolution and illumination intensity to maintain device performance~\cite{Hiro-paper}.

\subsection{Prospective applications}
\label{sec:applications}
Even with the level of nonlinearity demonstrated in this work, various applications are expected to become viable with improved low-loss device fabrication. More concretely, Ref.~\cite{Ryo-paper} showed that the following four applications may be feasible: (i) on-chip arbitrary pulse shapers; (ii) reconfigurable quantum frequency converters; (iii) widely wavelength-tunable integrated light sources; and (iv) quantum light sources with programmable entanglement structures~\cite{zhao2025scalable-network, weiss2025domain-engineering, imeshev1998engineerable, harris1969tunable}. By demonstrating the effective operation of programmable nonlinearity on nanophotonic devices, our current work makes an important step toward and beyond this goal. For instance, our demonstration of tunable Type-II phase matching in SHG shows the potential to realize programmable polarization-entangled photons via the reverse process, i.e., Type-II spontaneous parametric down-conversion~\cite{couteau2018SPDC}. Similarly, frequency-tunable DFG and SFG, for which we demonstrated up to \SI{130}{nm} of tunable bandwidth, constitute central building blocks for scalable quantum networking~\cite{zhao2025scalable-network}. Furthermore, the ability to dynamically synthesize and switch between distinct spatial mode superpositions equips this architecture to serve as an active, on-demand encoding or routing node for spatial-mode multiplexed quantum communication networks~\cite{cozzolino2019high, wang2018multidimensional}.

More broadly, programmable nonlinearity could be augmented onto existing photonic platforms to exploit its interplay with their native functions. Our result on $\chi^{(2)}$-$\chi^{(3)}$-hybrid arbitrary spectrum generation in Sec.~\ref{sec:spectrum} demonstrates one form of such synergy, and even more sophisticated control becomes possible when the underlying system is more complex. For instance, Ref.~\cite{lee2024inverse} demonstrated that an inversely designed wavy $\chi^{(3)}$ waveguide can shape the spectrum of supercontinuum generation, and programmable QPM could be overlaid on such a structure to add reconfigurable $\chi^{(2)}$ functionality, as we demonstrated in Sec.~\ref{sec:spiral}. In Ref.~\cite{hu2022photo-comb}, they showed that the operation of a SiN microcomb can be initiated and switched using a $\chi^{(2)}$ nonlinearity induced by the photogalvanic effect. We expect that fully programmable QPM gratings could offer even more flexible control of such dynamical systems, with potential applications to programmable frequency combs, computation, and light sources~\cite{li2026segment}.

Another unique strength of electric-field-induced $\chi^{(2)}$ nonlinearity is that it can realize QPM in essentially any material. Among the materials with otherwise desirable optical properties, such as a large transparency window, low optical loss, and industrial scalability, only a few are compatible with second-order nonlinear optics, owing to the absence of a native $\chi^{(2)}$ nonlinearity or of a means to perform periodic poling. The ability to realize QPM on demand in such materials would therefore have significant implications. For instance, Refs.~\cite{yuan2025green, wang2026green} realized SiN-integrated sources that bridge the ``green gap,'' the wavelength range that is difficult to access with semiconductor lasers. This was enabled by the large bandgap of SiN together with the electric-field-induced $\chi^{(2)}$ nonlinearity arising from the photogalvanic effect. Our approach could extend programmable QPM to an even wider range of materials, including those without a photogalvanic response, opening up unique pathways to nonlinear optics in spectral ranges that are otherwise hard to access. For example, large-bandgap fluorides~\cite{kraemer2026toward} could enable vacuum-ultraviolet nonlinear optics, while operation in the mid-infrared could be enabled by silicon, germanium, or certain III-V semiconductors~\cite{lafforgue2022supercontinuum}. 

Beyond applications as light sources, nonlinear optics has long been known as a powerful tool for characterization and metrology, e.g., for the temporal waveforms of ultrashort pulses~\cite{trebino1997measuring}, losses in nanophotonic waveguides~\cite{chen2026universalloss}, and intra-cavity quantum dynamics~\cite{choi2025observingcavity}. We expect that the added programmability of the nonlinearity will further enhance the capability of nonlinear optics for in situ characterization. To provide a concrete example, we demonstrated that the dispersion of a desired transverse mode of the waveguide can be measured directly by leveraging SFG with a programmable QPM (see Sec.~\ref{sec:cw-nlo}). We note that estimating such dispersion parameters from first-principles simulations is sometimes challenging; in our work, the high-temperature annealing caused cracking in the SiN slab regions, which made it difficult to measure the refractive index with standard techniques such as ellipsometry. In Sec.~\ref{sec:spiral}, we showed that the local variation of the phase-matching condition could be resolved with a programmable QPM, underscoring the potential for spatially resolved characterization of the waveguide parameters. Finally, the phase-sensitive tomography of an OAM mode in Sec.~\ref{sec:modal} leveraged the ability of the programmable nonlinearity to generate light at a desired location with an arbitrary phase and transverse mode. Such light could serve as an on-demand, programmable local oscillator, enabling on-chip heterodyne measurement of signal light, e.g., for the characterizations of the other components in the system.

\subsection{Conclusions}
In this work, our primary focus has been to demonstrate the potential of augmenting nanophotonic devices with programmable nonlinearity. We believe one of the most important contributions of our work is to point toward a new development and manufacturing paradigm beyond the conventional one-device-one-function approach, which could eventually enable truly scalable integrated nonlinear photonics. A programmable device can be deployed as a ``blank slate'' that then inversely designs its own function, in situ, based on real-time feedback from the environment in which it operates. This would not only enable the robust engineering of complex functions with 100\% yield but also allow the device to remain fully adaptive to changing environments and demands, potentially transcending nonlinear nanophotonics beyond its present regime of small circuit depths toward systems of much greater scale.

As the technology matures, the exact means of realizing programmable nonlinearity may well evolve. Our approach based on a photoconductor and free-space programming illumination was convenient for a proof of concept, but to maximize integrability, it may turn out to be advantageous to employ lithographically defined electrodes and integrated circuitry to control the bias electric fields. Unlike conventional nonlinear nanophotonics, which relies on specialized materials, all of these approaches could in principle be implemented within complementary metal-oxide-semiconductor (CMOS)-compatible fabrication, further underscoring the potential for scalability~\cite{heydari2023efish, timurdogan2017electric}.

Although much development is needed before reaching that stage, we emphasize that this would inherently be a paradigm with implications beyond any single application. The one-device-many-functions paradigm means that a wide range of nonlinear nanophotonic systems, spanning different applications and contexts, can be built on the same general-purpose device. In analogy with the transistor, this large market could justify the extensive manufacturing and design optimization that might be prohibitive for more specialized nonlinear nanophotonic devices. The proof-of-concept results presented here offer a glimpse of what this new paradigm for nonlinear nanophotonics could look like.

\section*{Data and code availability}
All simulation data, experimental data, and analysis code needed to reproduce the main results presented in this manuscript are available at \url{https://doi.org/10.5281/zenodo.23072388}

\section*{Acknowledgements}
We thank NTT Research for their financial and technical support. We gratefully acknowledge the Air Force Office of Scientific Research for funding under Award Number FA9550-22-1-0378, and the National Science Foundation for funding under Award Number CCF-1918549. This work was performed in part at the Cornell NanoScale Facility, a member of the National Nanotechnology Coordinated Infrastructure (NNCI), which is supported by the National Science Foundation (Grant NNCI-2025233). P.L.M. gratefully acknowledges financial support from a David and Lucile Packard Foundation Fellowship. The authors thank Christopher Alpha, Maxwell Anderson, Garry Bordonaro, Thibault Chervy, Jeremy Clark, Phil Infante, Oscar Jaramillo, Saeed Khan, Vladimir Kremenetski, Guilherme Marega, Jeff Moses, Mathieu Ouellet, Sridhar Prabhu, Giovanni Sartorello, Purnendu Sen, Michael Skvarla, Chirag Vaswani, Tianyu Wang, Wayne Wang, Aaron Windsor, Yiyang Liu, Yongqi Zhang, and Ruomin Zhu for helpful comments and discussions.

\section*{Author contributions}
B.A.A., R.Y., M.M.Sohoni, T.O., L.G.W., and P.L.M. conceived the project. B.A.A., R.Y., M.M.Stein, T.O., and L.G.W. designed the devices. B.A.A., R.Y., and M.M.Sohoni designed the experiments with advice from M.J., Y.X., and F.O.W. B.A.A. fabricated the devices with aid and recipe development from R.Y., M.M.Stein, T.O., and Y.Z. R.Y., M.M.Sohoni, and B.A.A. built the imaging setup to program the $\chi^{(2)}$ nonlinearity pattern. B.A.A., M.M.Sohoni, and R.Y. performed all the experiments. B.A.A., R.Y., and M.M.Sohoni analyzed the experimental data. R.Y., B.A.A., and M.M.Sohoni wrote the manuscript with input from all authors. P.L.M. supervised the project.

\section*{Competing interests}
M.M.Stein, L.G.W., T.O. and P.L.M. are listed as inventors on a patent application (US20250314915A1) on 2D-programmable waveguides.

\bibliographystyle{mcmahonlab}
\bibliography{references}

\newpage

\appendix

\makeatletter

\section*{Appendix Contents}
\@starttoc{aptoc}

\let\oldsection\section
\let\oldsubsection\subsection

\renewcommand{\section}[1]{%
  \oldsection{#1}%
  \addcontentsline{aptoc}{section}{%
    \protect\numberline{\thesection}\protect\textbf{#1}%
  }%
}

\renewcommand{\subsection}[1]{%
  \oldsubsection{#1}%
  \addcontentsline{aptoc}{subsection}{\protect\numberline{\thesubsection}#1}%
}

\renewcommand{\Call}[2]{\textproc{#1}\if\relax\detokenize{#2}\relax\else(#2)\fi}

\makeatother

\section{Appendix Introduction}
This document provides appendix information related to the results presented in the main text. The sections in this appendix document are organized into the following five categories:

\begin{itemize}
    \item \textbf{Device fabrication and specifications}: Sec.~\ref{sec:fab+device specs} provides details on how to fabricate the nonlinear programmable waveguide device presented in this work and details about the various device specifications (waveguide geometry, material) used in this work.

    \item \textbf{Theoretical model and performance characterizations}: Sec.~\ref{sec:Device Design and Characterization} discusses the operating principle, device design considerations, and performance characterizations that provide intuition for how the nonlinear programmable waveguide devices are designed.

    \item \textbf{Experimental apparatus}: Sec.~\ref{sec:optical setup} describes the optical and electrical equipment and setup required to operate and characterize the nonlinear programmable waveguide.

    \item \textbf{Experimental results}: The experimental results in the main text are presented in greater detail. Sec.~\ref{sec:chi2 estimation} details our procedure for estimating the induced $\chi^{(2)}$ in our devices. Sec.~\ref{sec:appendix CW-NLO} describes SFG, DFG, dispersion characterization, and polarization-mixing experiments using CW light. Sec.~\ref{sec:spirals} and Sec.~\ref{sec:Wavy Waveguides} describe how active feedback is used to optimize the $\chi^{(2)}$ pattern for non-straight waveguide geometries (specifically the spiral and wavy geometries in this work).  Sec.~\ref{sec:spatial modes} describes the generation of higher-order spatial modes, using modal interference to generate OAM modes, and performing on-chip phase tomography.  Lastly, Sec.~\ref{sec:spectral shaping} shows how in situ feedback and $\chi^{(3)}$ spectral broadening are used to shape an SH pulse over 80 nm of bandwidth.

    \item \textbf{Future device improvements}: Sec.~\ref{sec:Device Improvements} details future improvements to the electrical or optical characteristics of programmable nonlinear waveguide devices to enhance performance.
    
\end{itemize}

\section{Device Fabrication} \label{sec:fab+device specs}
\subsection{Fabrication Procedure}
\label{sec:Fabrication}

\begin{figure}
    \centering
    \includegraphics[width=1\linewidth]{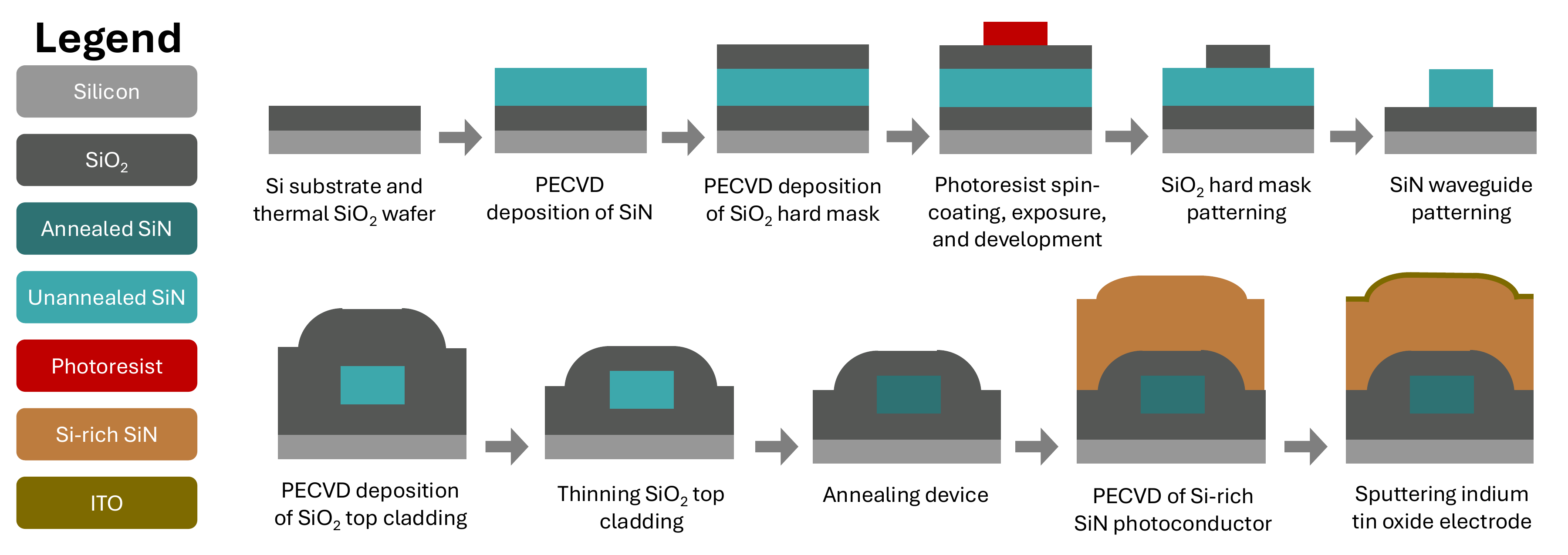}
    \caption{Process flow to fabricate the programmable channel waveguide device.  The thickness of the Si-rich \(\mathrm{SiN}\) photoconductor is depicted as thinner in this cartoon process flow than is realized for physical devices.  The text underneath each cartoon cross-section denotes the change from the previous fabrication step.}
    \label{fig:fabrication}
\end{figure}

In the following paragraphs, we describe the fabrication procedure and mention any modifications required for \(\mathrm{SiN}\) deposited in-house or by Silicon-Valley Microelectronics (SVM).  This process is similar to the fabrication procedure used in previous work\cite{Ryo-paper}. Fig.~\ref{fig:fabrication} shows the fabrication procedure for the programmable nonlinear waveguide device. 

We begin our fabrication process using a p-type silicon wafer with 1 \si{\micro\meter} of thermal oxide.  For \(\mathrm{SiN}\) films deposited in-house, we deposit 2 \si{\micro\meter} of plasma-enhanced chemical vapor deposition (PECVD) \(\mathrm{SiN}\) using 3 sccm of \(\mathrm{SiH_4}\), 40 sccm of \(\mathrm{H_2}\), and 2000 sccm of \(\mathrm{N_2}\) in an Oxford PlasmaPro 100 PECVD.  We adjusted the \(\mathrm{SiH_4}\) flow to control the film’s stoichiometry and loss.  Other devices used 2 \si{\micro\meter} thick PECVD \(\mathrm{SiN}\) deposited by SVM on a positively doped silicon wafer with 1 \si{\micro\meter} of thermal oxide.  We used PECVD \(\mathrm{SiN}\) instead of low-pressure chemical vapor deposition (LPCVD) \(\mathrm{SiN}\) due to its lower mechanical stress.  This allowed us to deposit thicker \(\mathrm{SiN}\) films using PECVD than would be easily possible using LPCVD.  The drawback of our approach is that PECVD \(\mathrm{SiN}\) has higher intrinsic optical absorption, as the lower deposition temperature leaves more defects in the film \cite{Lipson-PECVD, Lipson-complete-SiN}.  

Before patterning the \(\mathrm{SiN}\) film, we deposited 1.1 \si{\micro\meter} of PECVD \(\mathrm{SiO_2}\) for use as a hard mask.  We then spin-coated a 60 nm thick deep ultraviolet (DUV) anti-reflection coating and an 800 nm thick DUV resist using a Suss MicroTec Gamma Resist Cluster.  To expose our resist, we used a DUV stepper (PAS 5500; ASML) and used MIF 726 to develop the resist. We patterned our stepper masks in-house using a Heidelberg DWL 2000 system to expose resist on a six-inch \(\mathrm{Cr}\) mask.  We then developed the resist, wet-etched the exposed \(\mathrm{Cr}\) layer, and stripped the resist before using the mask in the ASML stepper system.  

After developing the resist on our wafers, we performed an \(\mathrm{O_2}\) descum and used a \(\mathrm{CH_2F_2}\) and \(\mathrm{He}\) dry etching chemistry in an inductively coupled plasma (ICP) etcher (PlasmaPro 100 RIE; Oxford Instruments) to pattern the \(\mathrm{SiO_2}\) hard mask.  We used this \(\mathrm{SiO_2}\) etching chemistry because of its high oxide-to-resist selectivity and straight sidewalls.  We then stripped the remaining resist using an ICP \(\mathrm{O_2}\) descum in a YES Ecoclean system.  Next, we ICP dry etched the \(\mathrm{SiN}\) layer using a \(\mathrm{CHF_3}\), \(\mathrm{O_2}\), and \(\mathrm{N_2}\) chemistry, which enhanced nitride-to-oxide selectivity while minimizing the polymer residues from the etch \cite{Lipson-complete-SiN}.  To build our top cladding layer, we deposited 3 \si{\micro\meter} of PECVD \(\mathrm{SiO_2}\) and used a \(\mathrm{CHF_3}\) and \(\mathrm{O_2}\) ICP dry etching recipe to remove 2 \si{\micro\meter} from the top of that oxide layer.  Because the PECVD deposition is more conformal than the oxide etch, we end with 1 \si{\micro\meter} of \(\mathrm{SiO_2}\) on top of our waveguides and $\sim 1.4$ \si{\micro\meter} of \(\mathrm{SiO_2}\) on the waveguide sidewalls.  

For \(\mathrm{SiN}\) provided by SVM, we then performed a rapid thermal annealing process with a ramp of \ang{5}C/s, a temperature of \ang{800}C, and an annealing time of 5 minutes.  For in-house \(\mathrm{SiN}\), we performed a furnace anneal.  For furnace processing, we loaded our wafers at \ang{300}C and ramped to \ang{1200}C over three hours.  We annealed the wafers at \ang{1200}C for 6 hours and ramped the temperature down to \ang{300}C over 6 hours.  Both annealing processes were done in a \(\mathrm{N_2}\) ambient environment.  These annealing steps reduce intrinsic material absorption losses caused by voids and defects in the PECVD \(\mathrm{SiN}\) films.  Additionally, high-temperature annealing at \ang{1200}C removes \(\mathrm{N-H}\) bonds, which cause a known absorption peak at 1520 nm \cite{Annealing1, Annealing2}.  While slab waveguides with a 2 \si{\micro\meter} thickness would crack from tensile stress during these annealing processes, we annealed after the \(\mathrm{SiN}\) was etched into channel waveguides.  This prevents stress from accumulating and allows our 2 \si{\micro\meter} thick channel waveguides to survive this high-temperature annealing process \cite{Annealing1}.  We did not anneal \(\mathrm{SiN}\) provided by SVM  at \ang{1200}C because we empirically found it increased baseline loss compared to rapid thermal processing. 

\begin{figure}
    \centering
    \includegraphics[width=0.6\linewidth]{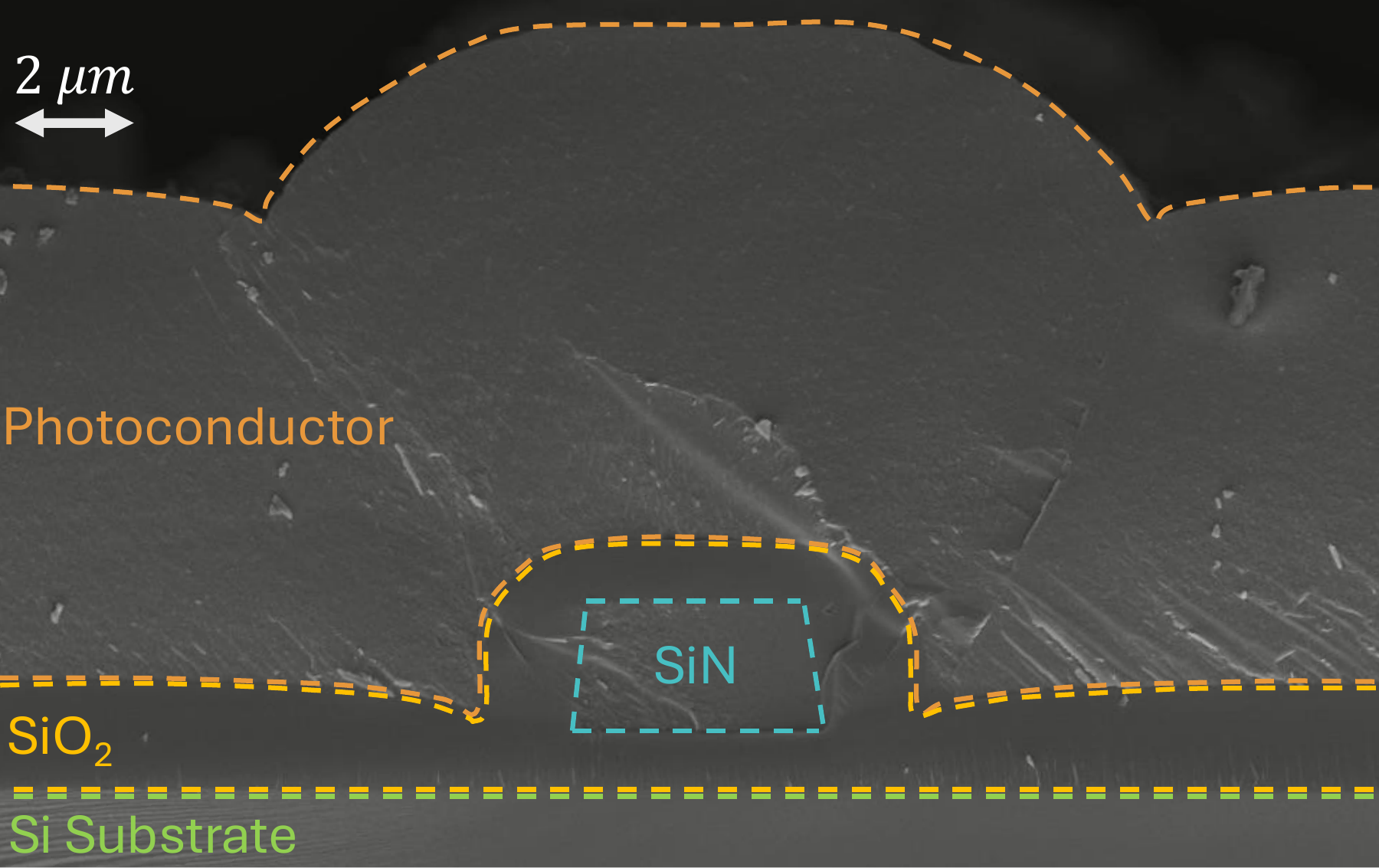}
    \caption{Labeled SEM image of a fully fabricated programmable nonlinear channel waveguide device.}
    \label{fig:SEM Cross Section}
\end{figure}

After annealing, we deposited 9 \si{\micro\meter} of PECVD \(\mathrm{Si}\)-rich \(\mathrm{SiN}\) as our photoconductive layer.  We then used a Kurt Lesker PVD system to sputter 40 nm of indium tin oxide (ITO) as an electrode on top of our photoconductor.  We cleaved the wafer to expose smooth facets for butt-coupling.  The facets were $\sim$ 2 mm away from the edge of the ITO electrode to prevent dielectric breakdown through the air when operating the devices at high voltage. Fig.~\ref{fig:SEM Cross Section} shows a labeled SEM image of a fully fabricated device cross-section.


\subsection{Implemented Device Specifications}
\label{sec:possible devices}
As Sec.~\ref{sec:Fabrication} details, there are two possible \(\mathrm{SiN}\) materials implemented in this work: those deposited in-house and furnace-annealed and those deposited by SVM and subjected to rapid thermal annealing.  While the \(\mathrm{SiN}\) deposited by SVM displays a higher baseline conversion efficiency, the \(\mathrm{SiN}\) deposited in-house maintains a lower loss over a broader bandwidth due to its higher-temperature thermal treatment.  Details on loss characterizations can be found in Sec.~\ref{sec:loss characterization}.  Furthermore, previous demonstrations \cite{Ryo-paper} using programmable $\chi^{(2)}$ were done on one device to show the full versatility of this approach to nonlinear optics.  While the devices demonstrated in this work still maintain the ability to perform multiple tasks on one waveguide, this work uses waveguides of differing geometry to show how programmable $\chi^{(2)}$ nonlinear optics can adapt to different geometries and augment the linear functionality of the device.  Tab.~\ref{tab:possible devices} shows the devices utilized in this work and characterizations of their nominal conversion efficiency.

The nominal SHG conversion efficiency reported in Tab.~\ref{tab:possible devices} averages over multiple devices and wavelength characterizations, so there is some spectral and device-specific variance that is not captured.  Because this is the conversion efficiency over several millimeters of propagation, the nominal conversion efficiency also takes the spatial average of induced $\chi^{(2)}$ over that length. Furthermore, the reported values of nominal conversion efficiency do not include wavelengths below 1550 nm to avoid incorporating data affected by the loss peak at 1520 nm. The length of the straight waveguide is extracted by fitting SHG power versus $\Delta k$ with a $\operatorname{sinc}^2$ function, where $\Delta k$ is the momentum mismatch from the point with greatest SHG conversion efficiency.  Because of inhomogeneities in the programming illumination, the real poling length is longer but includes sections with lower programming illumination power or imaging aberrations.  This same effective poling distance was used for adiabatic waveguides, but the poling distance of spiral waveguides was calculated using the underlying physical structure. As will be discussed in Sec.~\ref{subsec:spiral phase-mathcing}, inhomogeneities in the programming illumination optics and different optimization procedures make absolute comparisons between different geometries challenging.  Lastly, as will be discussed in Sec.~\ref{sec:loss characterization} and Sec.~\ref{sec:spatial modes}, the large geometry of our waveguides allows us to couple into multiple FH spatial modes.  This means that, when calculating the normalized conversion efficiency from the output FH power, we normalize over the power in all the spatial modes instead of just the fundamental spatial mode.  Therefore, it is best to interpret all values reported in this work as a lower-bound measurement of the conversion efficiency, which could be improved with better optical alignment and spatial filtering of the FH pump.

\begin{table}[h!]
\centering
\begin{tabular}{|c|c|c|c|c|}
\hline

\(\mathrm{SiN}\) Material &
Waveguide Geometry &
Poling Distance (cm) &
Nominal SHG Conv. Eff. (\%/W) &
Waveguide Cartoon \\

\hline

In-house &
Straight &
0.55 &
$1.8 \times10^{-4}$ &
\raisebox{-0.5\height}{\includegraphics[width=2cm]{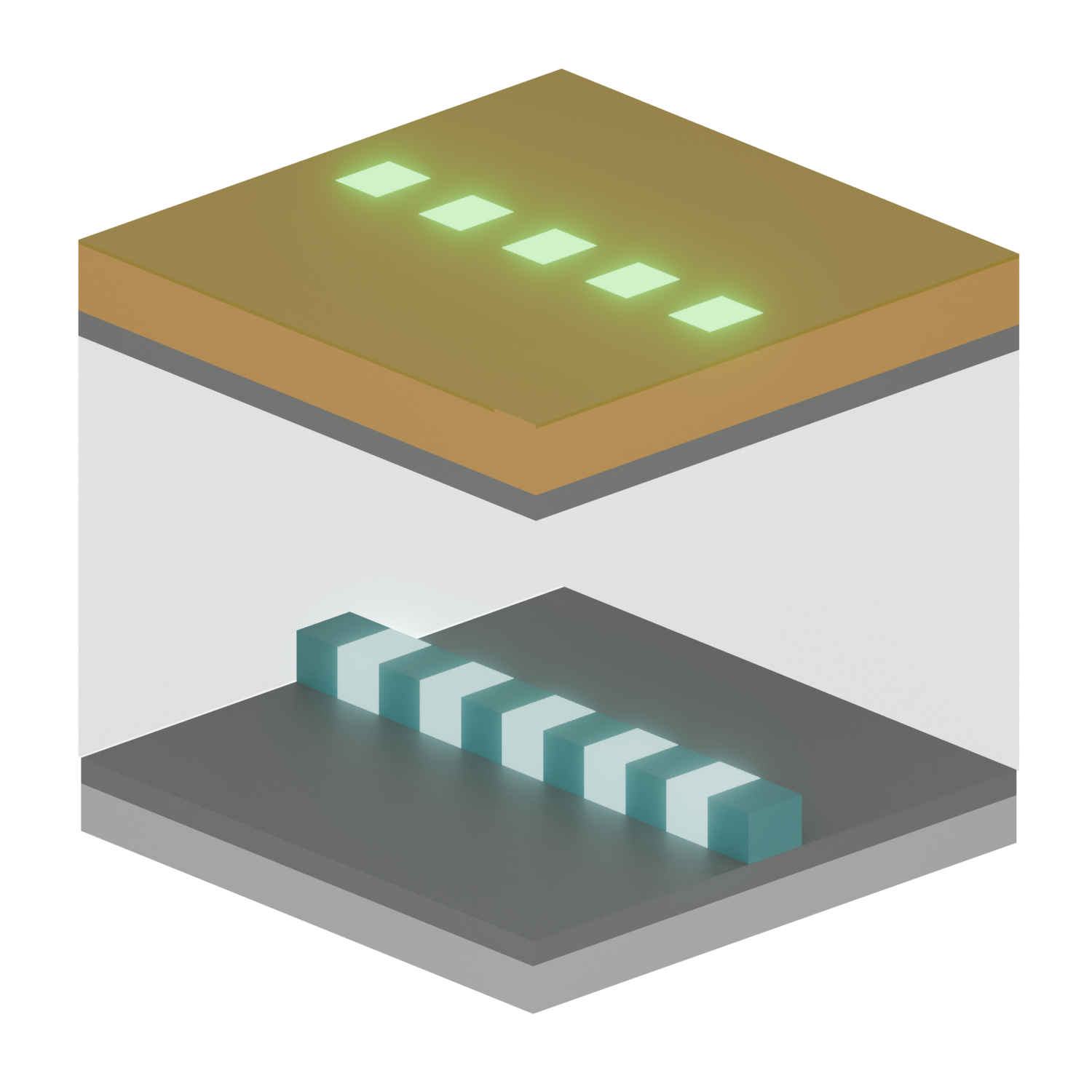}}
\\
\hline

SVM &
Straight &
0.55 &
$4.7 \times10^{-4}$ &
\raisebox{-0.5\height}{\includegraphics[width=2cm]{Appendix_Figures/Straight2.png}}
\\
\hline

In-house &
Wavy &
0.55 &
$1.3 \times10^{-4}$ &
\raisebox{-0.5\height}{\includegraphics[width=2cm]{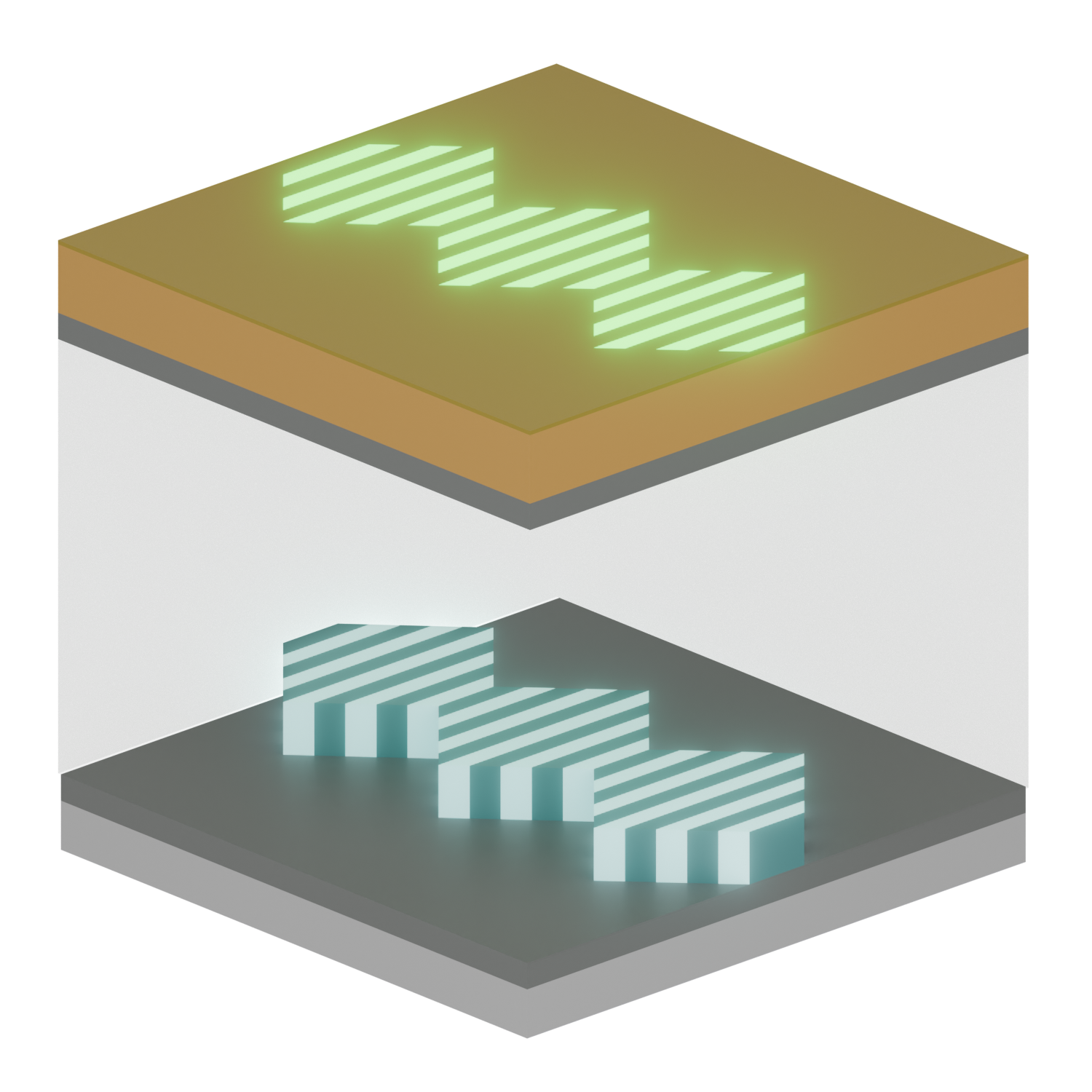}}
\\
\hline

In-house &
Spiral &
0.7, 2.6, 4.6, 7.0, 10.1, 12.1, 14.7 &
$6.7 \times10^{-3}$ &
\raisebox{-0.5\height}{\includegraphics[width=2cm]{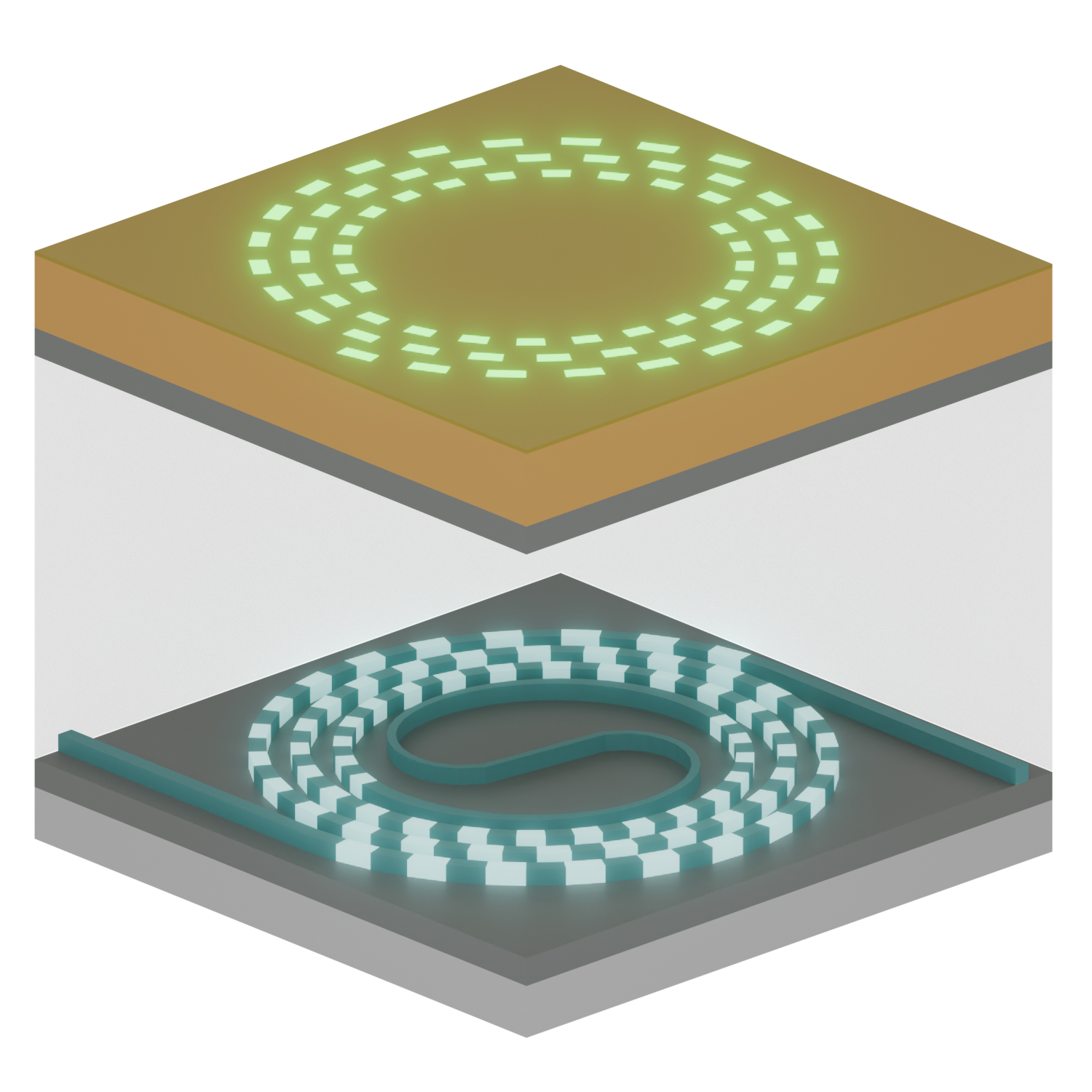}}
\\
\hline

SVM &
Spiral &
10.1 &
$1.4 \times10^{-2}$ &
\raisebox{-0.5\height}{\includegraphics[width=2cm]{Appendix_Figures/Spiral2.png}}
\\
\hline

\end{tabular}
\caption{Material, waveguide geometry, effective poling distance, nominal SHG conversion efficiency, and a cartoon of the underlying waveguide geometry for each device-type implemented in this work.  Spiral waveguides using in-house \(\mathrm{SiN}\) were tested with structures of different propagation lengths. The nominal conversion efficiency of the spiral devices is quoted for a poling distance of 10.1 cm for both \(\mathrm{SiN}\) deposited in-house and by SVM.}
\label{tab:possible devices}
\end{table}

\section{Device Design and Characterization}
\label{sec:Device Design and Characterization}
\subsection{Operating Principle}

Unlike fixed-on-fabrication approaches, our device relies on programmability enabled by the electric-field induced second harmonic (E-FISH) effect, which allows us to induce an effective $\chi^{(2)}$ by changing the electric field bias in the waveguiding layer.  Physically, this effect can be understood as using the native $\chi^{(3)}$ to create an effective $\chi^{(2)}$. E-FISH can be described using the perturbative expansion of the nonlinear polarization in terms of the applied electric field, given as
\begin{equation}
\label{eq:expansion}
    P_i = \epsilon_0(\sum_j \chi^{(1)}_{i,j}E_j + \sum_{j,k}\chi^{(2)}_{i,j,k}E_jE_k+ \sum_{j,k,l}\chi^{(3)}_{i,j,k,l}E_jE_kE_l \cdots)
\end{equation}
where $i$, $j$, $k$, and $l$ are different tensor dimensions each with three polarization dimensions \cite{Boyd_NLO_2008}. From Eq.\eqref{eq:expansion}, the generalized third-order term in the perturbative expansion of a material's polarization is
\begin{equation}
    P^{(3)}_i(\omega_1+\omega_2+\omega_3) = \epsilon_0\sum_{j,k,l} \chi^{(3)}_{ijkl}(\omega_1+\omega_2+\omega_3, \omega_1, \omega_2, \omega_3)E_j(\omega_1)E_k(\omega_2)E_l(\omega_3)
\end{equation}
where $E_j(\omega_1)$, $E_k(\omega_2)$, and $E_l(\omega_3)$ are three interacting fields with different polarization directions and frequencies.  In the programmable devices presented in this work, two of the electric fields are optical (meaning $\omega_1, \omega_2 \gg 0$) while one field is near DC (meaning $\omega_3 \approx 0$). Supposing this $E_{\text{bias}}$ field is polarized along the $y$ direction, the third order polarization becomes
\begin{equation}
    P^{(3)}_i(\omega_1+\omega_2) = \epsilon_0\sum_{j,k}(\chi^{(3)}_{ijky}(\omega_1+\omega_2, \omega_1, \omega_2, 0)+\chi^{(3)}_{ijyk}(\omega_1+\omega_2, \omega_1, 0,\omega_2)+\chi^{(3)}_{iyjk}(\omega_1+\omega_2, 0,\omega_1, \omega_2))E_j(\omega_1)E_k(\omega_2)E_{\text{bias}}
\end{equation}
as each possible permutation of the interacting fields must be accounted for. Because we use wavelengths far from the absorption bands of \(\mathrm{SiN}\), we can apply Kleinman symmetry as
\begin{equation}
    \chi^{(3)}_{ijky}(\omega_1+\omega_2, \omega_1, \omega_2, 0)=\chi^{(3)}_{ijyk}(\omega_1+\omega_2, \omega_1, 0,\omega_2)=\chi^{(3)}_{iyjk}(\omega_1+\omega_2, 0,\omega_1, \omega_2)
\end{equation}
This simplifies the expression of $P^{(3)}$ to 
\begin{equation}
\label{eq:final chi3}
        P^{(3)}_i(\omega_1+\omega_2) = 3\epsilon_0 \sum_{j,k}\chi^{(3)}_{ijky}(\omega_1+\omega_2, \omega_1, \omega_2, 0)E_{\text{bias}}E_j(\omega_1)E_k(\omega_2)
\end{equation}
Referring back to Eq.~\eqref{eq:expansion}, the second-order polarization, $P^{(2)}$, is expressed as 
\begin{equation}
\label{eq:final chi2}
    P^{(2)}_i = \epsilon_0\sum_{j,k}\chi^{(2)}_{ijk}(\omega_1+\omega_2, \omega_1, \omega_2)E_j(\omega_1)E_k(\omega_2)
\end{equation}
When comparing the form of Eq.~\eqref{eq:final chi2} and Eq.~\eqref{eq:final chi3}, the expression for $P^{(3)}$ can be interpreted as $E_{\text{bias}}$ inducing an effective $\chi^{(2)}$ via the $\chi^{(3)}$ as
\begin{equation}
\label{eq:effective chi2}
    \chi^{(2)}_{ijk}(\omega_1+\omega_2, \omega_1, \omega_2) = 3\chi^{(3)}_{ijky}(\omega_1+\omega_2, \omega_1, \omega_2, 0)E_{\text{bias}}
\end{equation}
Unlike most $\chi^{(3)}$ processes, inducing a strong $\chi^{(2)}$ requires $\chi^{(3)}$ to be large when one of the fields has a near-DC frequency. This means a material with a strong optical $\chi^{(3)}$ response may not have a strong induced $\chi^{(2)}$ response.  Furthermore, even though most of this work focuses on the $\chi^{(2)}_{yyy}$ component, Sec.~\ref{sec:Polarization Experiments} demonstrates tasks where different polarization components are mixed.

While Eq.~\eqref{eq:effective chi2} shows that an effective $\chi^{(2)}$ can be induced using a strong $E_{\text{bias}}$, a spatially uniform $\chi^{(2)}_{\text{eff}}$ is not sufficient to show most second-order nonlinear optical behavior in an efficient manner.  Energy conservation for a  $\chi^{(2)}$ sum frequency process is satisfied by 
\begin{equation}
    \omega_1 + \omega_2 = \omega_3
\end{equation}
where $\omega_1$ and $\omega_2$ are the frequencies of the input fields and $\omega_3$ is the frequency of the generated field.  To enhance the efficiency of a nonlinear process, the phase slip between the newly generated SH field and the previously generated SH field must be compensated so all of the SH fields constructively interfere through the entire propagation distance.  This phase slip is caused by a momentum mismatch between the SH and FH fields, which affects the output power as
\begin{equation}
    P_{\omega_3} \propto \chi^{(2)^2} L^2 \mathrm{sinc}^{2}(\Delta kL/2)
\end{equation}
where $L$ is the distance over which the nonlinear interaction occurs and $\Delta k$ describes the momentum mismatch as $\Delta k = k_3 -k_2-k_1$. Due to underlying material dispersion, $\Delta k$ is nonzero in most systems.  While dispersion engineering of waveguides or angle-dependent phase-matching in bulk crystals are methods to minimize $\Delta k$ (known as phase-matching), these approaches cannot be adapted to phase-match any arbitrary nonlinear process \cite{modal-phase-matching1, modal-phase-matching2, modal-phase-matching3, modal-phase-matching4}.  Quasi-phase-matching (QPM) is one of the most flexible methods to enhance the nonlinear conversion efficiency of a broad selection of nonlinear processes.  Instead of tuning the underlying dispersion, QPM relies on spatially patterning the $\chi^{(2)}$ of the propagation medium so the spatial momentum kick provided by the $\chi^{(2)}$ hologram compensates for the $\Delta k$ of the interacting waves \cite{PPLN1, PPLN2, PPLN3}.  

We create spatially reprogrammable $\chi^{(2)}$ patterns by integrating a photoconductor on top of a \(\mathrm{SiN}\) waveguide \cite{Ryo-paper, Hiro-paper}. By applying high voltage to the device stack and shining light on different parts of the photoconductor, we can change the conductivity distribution of the photoconductor in real-time.  Because a lower resistance in the photoconductor means a larger voltage (and therefore $E_{\text{bias}}$) in the waveguiding layer, we can use this programming illumination to control the spatial distribution of the induced $\chi^{(2)}$ in the waveguiding layer.  Moreover, because the photoconductive programmability employed here does not have a memory of the previously programmed states, we can easily alter our $\chi^{(2)}$ pattern by changing the programming illumination.  This allows us to achieve arbitrarily reprogrammable nonlinear photonic devices that apply different QPM gratings for a variety of nonlinear optical processes \cite{Ryo-paper}.  We use a \(\mathrm{SiN}\) waveguide because it displays a comparatively large induced $\chi^{(2)}$ \cite{heydari2023efish} while maintaining a broad transparency window \cite{Lipson-complete-SiN}.

\subsection{Circuit Model and Electrical Testing}
\label{sec:Circuit Model}

Because we use a photoconductor to electrically modify the $E_\text{bias}$ in the \(\mathrm{SiN}\) waveguiding layer, a circuit model of the device must be employed to fully understand how best to electrically engineer our film stack to maximize the programmable $\chi^{(2)}$.  This section discusses the theoretical development of the circuit model and then describes experiments that test nonlinear performance as the electrical bias conditions are changed.  By fitting the device's response with the theoretical circuit model, we can extract performance metrics that inform our design. 

\begin{figure}
    \centering
    \includegraphics[width=1\linewidth]{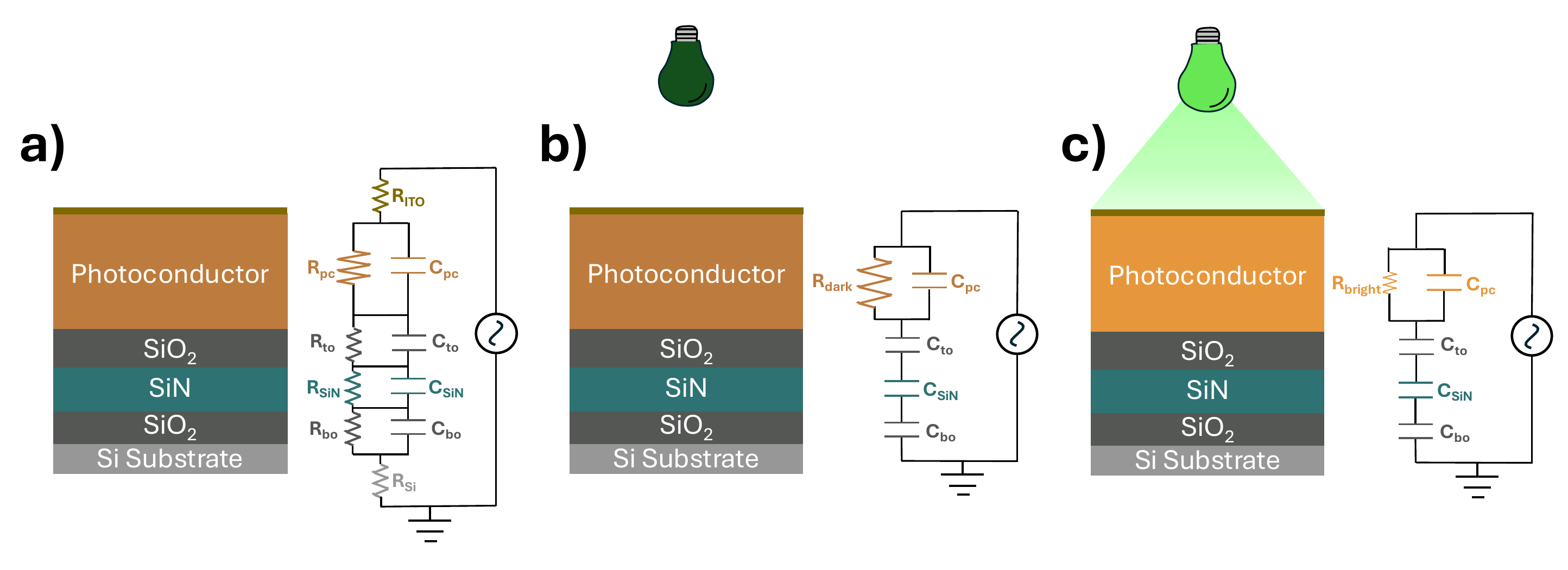}
    \caption{a) Device cross-section diagram and full circuit model treating each film separately. b) Diagram of the device stack in the dark state with the approximate circuit model. c) Diagram of the device stack in the bright state with the approximate circuit model.  The resistance of the photoconductor is lower in the bright state than the dark state.}
    \label{fig:circuit_model}
\end{figure}

To effectively model each film in the device stack, we treat each layer as a capacitor and resistor in parallel.  However, because the conductivities of the waveguiding layers (meaning the \(\mathrm{SiO_2}\) claddings and \(\mathrm{SiN}\) core) are so small by comparison to the photoconductor, these layers are treated as purely capacitive.  While the substrate, \(\mathrm{ITO}\) top electrode, and high-voltage circuitry have finite resistance, these are much smaller than the resistance of the photoconductor.  This allows us to neglect their contribution to the device's electrical behavior. Fig.~\ref{fig:circuit_model} shows the complete circuit model of our device and the approximate model we employ.  

To achieve maximal switching of the induced $\chi^{(2)}$, the impedance of the device layers should satisfy
\begin{equation}
    \lvert Z_\text{bright} \rvert < \lvert Z_\text{SiN} \rvert< \lvert Z_\text{dark}\rvert
    \label{eq:impedance}
\end{equation}
where $Z_\text{bright}$ and $Z_\text{dark}$ correspond to the impedance of the photoconductor with and without illumination, respectively.  Physically, this is because the \(\mathrm{SiN}\) waveguide is electrically in series with the rest of the device stack, which means that the voltage in the waveguide is 
\begin{equation}
\label{eq:bias voltage}
    V_{\text{wg}} = \frac{V_{\text{bias}}Z_{\text{SiN}}}{Z_{\text{SiN}}+Z_{\text{to}}+Z_{\text{bo}} + Z_{\text{pc}}}
\end{equation}
where $V_{\text{bias}}$ is the voltage amplitude applied to the device, $Z_{\text{to}}$ is the top oxide impedance, $Z_{\text{bo}}$ is the bottom oxide impedance, and $Z_{\text{pc}}$ is the photoconductor impedance (this could either be for the bright or dark state).  Since the nonlinear conversion efficiency is proportional to 
\begin{equation}
\label{eq:max nonlinear voltage}
    \eta \propto \lvert V_{\text{wg, bright}} - V_{\text{wg, dark}} \rvert^2
\end{equation}
a large contrast between the waveguide voltage in the bright state ($V_{\text{wg, bright}}$) and the waveguide voltage in the dark state ($V_{\text{wg, dark}}$) maximizes the nonlinearity $\eta$.  The voltage divider relation between the \(\mathrm{SiN}\) waveguide and photoconductor means $V_{\text{wg, bright}}$ is maximized when $Z_{\text{bright}} < Z_{\text{SiN}}$ and $V_{\text{wg, dark}}$ is minimized when $Z_{\text{SiN}} < Z_{\text{dark}}$. 

Because the photoconductor is orders of magnitude more conductive in both the bright and dark state than the waveguide and cladding layers, an AC bias voltage must be applied to the device stack to bring the impedance of the \(\mathrm{SiN}\) waveguide within the programmable range of the photoconductor. This means the SHG output of the device is not constant in time.  As we discuss in Sec.~\ref{sec:dynamic-illumination}, future devices could be operated with an output that is constant in time by temporally syncing the programming illumination with the AC electrical bias.

\begin{figure}
    \centering
    \includegraphics[width=1\linewidth]{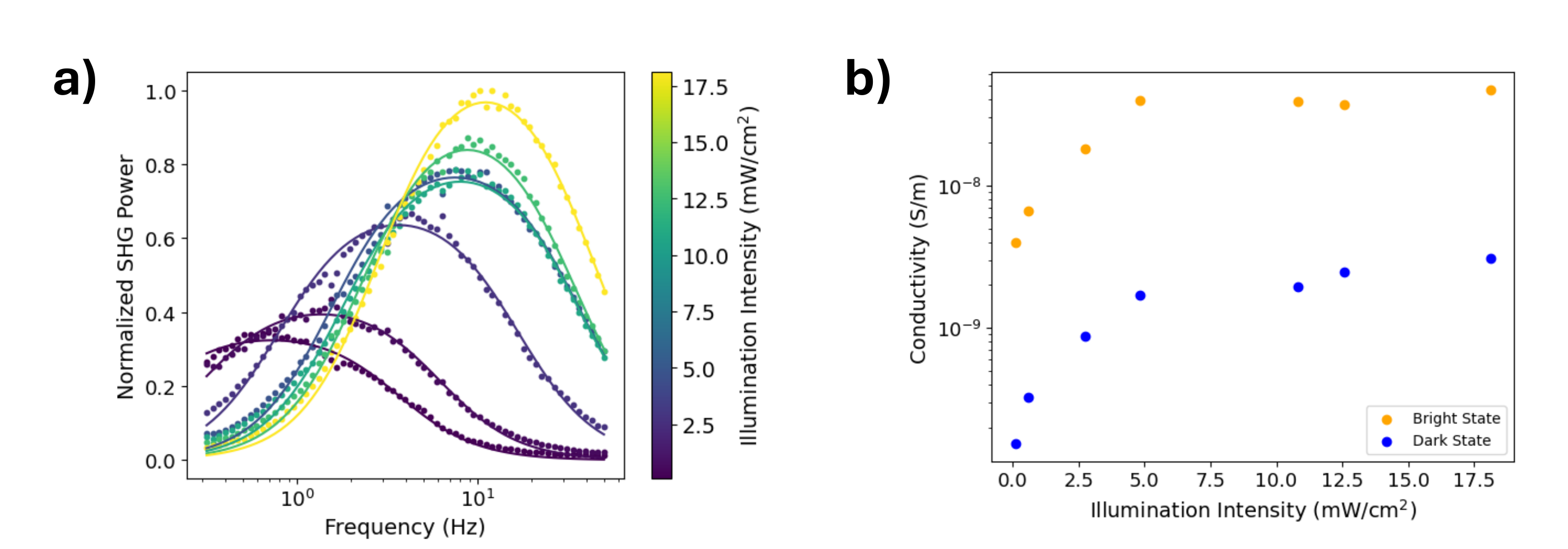}
    \caption{a) Experimentally measured SHG power (normalized to peak observed SHG power) for different bright state illumination conditions and electrical frequencies. Fits from Eq.~\eqref{eq:full cond fit} are overlaid. b) Values of the bright and dark state photoconductor conductivities obtained by fitting the curves in panel a).  Constant applied voltage was 800 V.}
    \label{fig:Illumination Contrast}
\end{figure}
\begin{figure}
    \centering
    \includegraphics[width=1\linewidth]{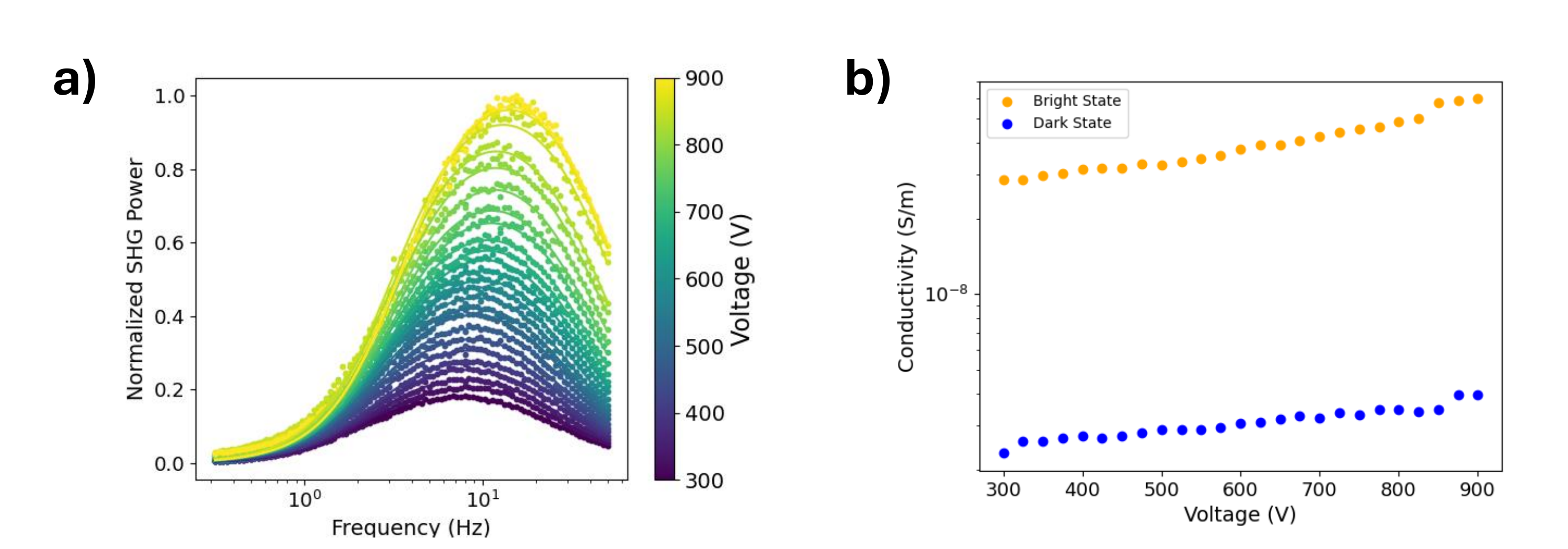}
    \caption{a) Experimentally measured SHG power (normalized to peak observed SHG power) for different applied voltage amplitudes and electrical frequencies. Fits from Eq.~\eqref{eq:full cond fit} are overlaid. b) Values of the bright and dark state photoconductor conductivities obtained by fitting the curves in panel a). Constant applied illumination intensity was 18.1 mW/(cm$^2$)}
    \label{fig:Voltage Contrast}
\end{figure}

Fig.~\ref{fig:Illumination Contrast} shows the measured SHG for different illumination intensities and Fig.~\ref{fig:Voltage Contrast} shows the measured SHG for different voltage amplitudes. The AC electrical frequency, $\omega$, was swept in Fig.~\ref{fig:Voltage Contrast} and Fig.~\ref{fig:Illumination Contrast} because the conductivity conditions of the photoconductor change with different voltage amplitudes and illumination intensities (as discussed in Sec.~\ref{subsec:photoconductor cond}).  By fitting the electrical transfer function, we can extract the bright and dark state conductivity of our photoconductor.  In order to make our fit function, we use the principle in Eq.~\eqref{eq:max nonlinear voltage} that the nonlinearity is proportional to the square of the programmable voltage difference in the \(\mathrm{SiN}\) layer.  Plugging in the values for \(\mathrm{SiN}\) waveguide bias voltage described in Eq.~\eqref{eq:bias voltage}, the SH power ($P_{\text{SHG}}$) can be fit as 
\begin{equation}
\label{eq:full cond fit}
    P_{\text{SHG}} = \kappa \left|
    \frac{Z_{\text{SiN}}}{Z_{\text{SiN}} + Z_{\text{to}} + Z_{\text{bo}} + Z_{\text{bright}}}
    -
    \frac{Z_{\text{SiN}}}{Z_{\text{SiN}} + Z_{\text{to}} + Z_{\text{bo}} + Z_{\text{dark}}}
    \right|^2
\end{equation}
where $\kappa$ is a scaling constant to assist with fitting. For different illumination conditions, $Z_{\text{pc}}$ becomes
\begin{equation}
   Z_{\text{pc}} =  Z_{\text{bright}} = \frac{1}{\frac{A\sigma_{\text{bright}}}{d_{\text{pc}}} + j\omega C_{\text{pc}}}
    \label{eq:bright PC impedance}
\end{equation}
in the bright state and
\begin{equation}
   Z_{\text{pc}} =  Z_{\text{dark}} = \frac{1}{\frac{A\sigma_{\text{dark}}}{d_{\text{pc}}} + j\omega C_{\text{pc}}}
   \label{eq:dark PC impedance}
\end{equation}
in the dark state.  The impedances of the \(\mathrm{SiN}\) and \(\mathrm{SiO_2}\) layers are just capacitive, as Fig.~\ref{fig:circuit_model} shows.  We then use the expression for $P_{\text{SHG}}$ to fit $\kappa$, $\sigma_{\text{bright}}$, and $\sigma_{\text{dark}}$ as a function of $\omega$. Fig.~\ref{fig:Voltage Contrast}(b) and Fig.~\ref{fig:Illumination Contrast}(b) plot the fitted values of $\sigma_{\text{bright}}$ and $\sigma_{\text{dark}}$ for different applied voltage and illumination conditions.

The fitted conductivity values in Fig.~\ref{fig:Illumination Contrast} show both the bright and dark state photoconductivity values increasing as the illumination power increases.  In principle, the dark state conductivity of the photoconductor should remain constant, as the increased illumination power should only be applied in the bright state. In reality, there is leakage illumination power through the SLM even when the photoconductor should be in the dark state.  This gives rise to the behavior seen in Fig.~\ref{fig:Illumination Contrast}(b), where the conductivity of the photoconductor in the dark state increases as the projected illumination power increases.  The average contrast between the bright and dark state was measured as $\approx 20 \times$.  The contrast between the fitted value of $\sigma_{\text{bright}}$ and $\sigma_{\text{dark}}$ at the optimal operating condition of 10 \(\mathrm{Hz}\), 800 \(\mathrm{V}\), and 18.1 \(\mathrm{mW/cm^2}\) was $\approx 15 \times$.

An important design implication of the programmability relation in Eq.~\eqref{eq:impedance} and Eq.~\eqref{eq:bias voltage} is that a thicker photoconductor results in a better $\Delta E_{\text{bias}}$ contrast.  This is because, unlike the cladding and core layers that act as capacitors, the photoconductor's resistive behavior should dominate the impedance over the capacitive behavior. Consequently, a larger $d_{\text{pc}}$ means a greater switching in the photoconductor's resistance as the internal conductivity is changed by illumination.

Although the \(\mathrm{SiO_2}\) layers play a critical role in optical performance (as discussed in Sec.~\ref{subsec:leakage}), they cause inefficient electrical performance.  Unlike the photoconductor, whose resistance is dramatically reduced in bright-state, the resistance of the \(\mathrm{SiO_2}\) layer is constant and large.  The low \(\mathrm{SiO_2}\) dielectric permittivity means that, when the claddings behave as capacitors, they draw large bias electric fields.  Consequently, the applied voltage does not all drop across the programmable, nonlinear \(\mathrm{SiN}\) core, leading to operational inefficiency.  Furthermore, if the internal voltage in the \(\mathrm{SiO_2}\) claddings is high enough, these are the first layers to undergo dielectric breakdown, meaning the device's nonlinearity is not fully activated.  As a result, thinner cladding layers improve the device's electrical behavior.

\subsection{Photoconductor Conductivity}
\label{subsec:photoconductor cond}
This section describes measurements that show nonlinear conductivity behavior in the photoconductors employed in this work.  As Fig.~\ref{fig:Voltage Contrast} shows, the fitted value for the bright and dark state photoconductor conductivity increases as the applied voltage on the device increases.  To further probe this nonlinear relationship, we measure the conductivity response of different \(\mathrm{Si}\)-rich \(\mathrm{SiN}\) films under a range of applied electric field and illumination conditions.  

\begin{figure}
    \centering
    \includegraphics[width=1\linewidth]{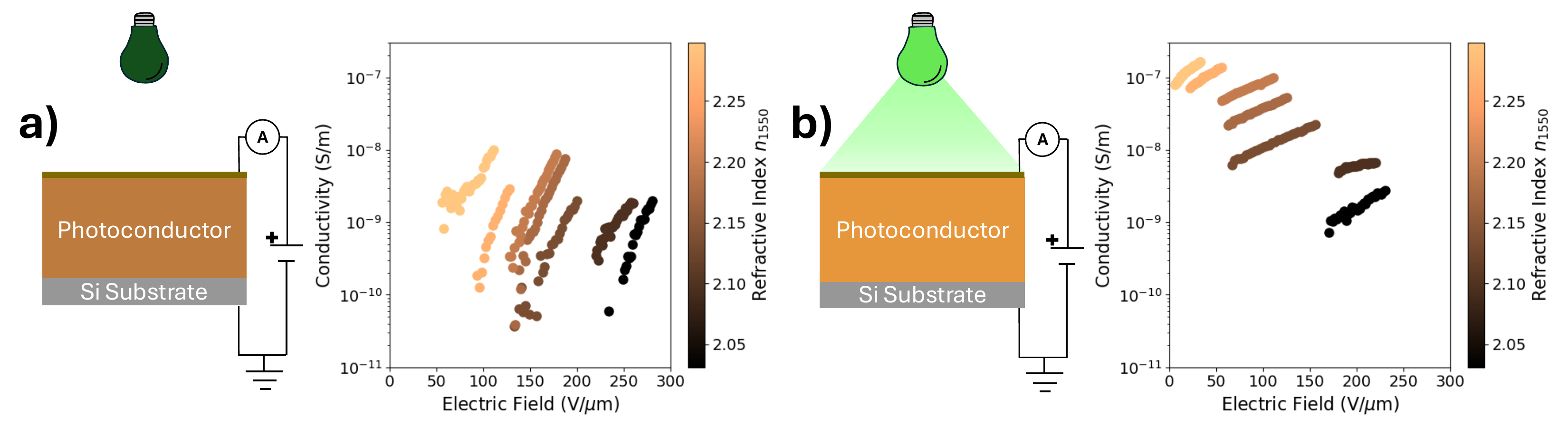}
    \caption{a) Cartoon image of the electrical setup and characterization of the electrical conductivity of different photoconductive films in the dark state.  b) Cartoon image of the electrical setup and characterization of the electrical conductivity of different photoconductive films in the bright state.  The photoconductive films are varied by adjusting the silicon content during fabrication, which increases the refractive index with greater silicon content.  The labeled refractive index is the one measured at 1550 nm.}
    \label{fig:Single Film Photoconductor}
\end{figure}

Fig.~\ref{fig:Single Film Photoconductor} shows conductivity measurements taken with different photoconductors where the photoconductive film is the only layer in the stack. The measurement in  Fig.~\ref{fig:Single Film Photoconductor}(a) shows the film in the dark state, while the measurement in Fig.~\ref{fig:Single Film Photoconductor}(b) shows the film in the bright state under the same illumination conditions as the final device.  For the final device, we used the film with the largest refractive index to leverage its heightened conductivity to operate at a higher electrical frequency. As the figure shows, there is a clear nonlinear relationship between the applied electric field and the film's measured conductivity.  One possible explanation for this phenomenon is the Poole-Frenkel effect, which is commonly observed in PECVD \(\mathrm{SiN}\) films of different stoichiometry \cite{PF-SiN1}.  Under this effect, the applied electric field excites additional carriers from trap states, which leads to an increased electrical conductivity \cite{PF-SiN2, Tao01041996}.

From these measurements, it would naively appear that we can achieve a $\approx 100 \times$ switching ratio for the same electric field.  Because of the nonlinear relationship between conductivity and internal electric field, the film possesses an internal feedback mechanism to reduce the electric field contrast between the bright and dark state.  This is why we only observe a conductivity switching ratio of $\approx15 \times$, as shown in Fig.~\ref{fig:Voltage Contrast} and Fig.~\ref{fig:Illumination Contrast}.

\subsection{Phase-Matching Properties}
\label{subsec:dispersion properties}
\begin{figure}
    \centering
    \includegraphics[width=1\linewidth]{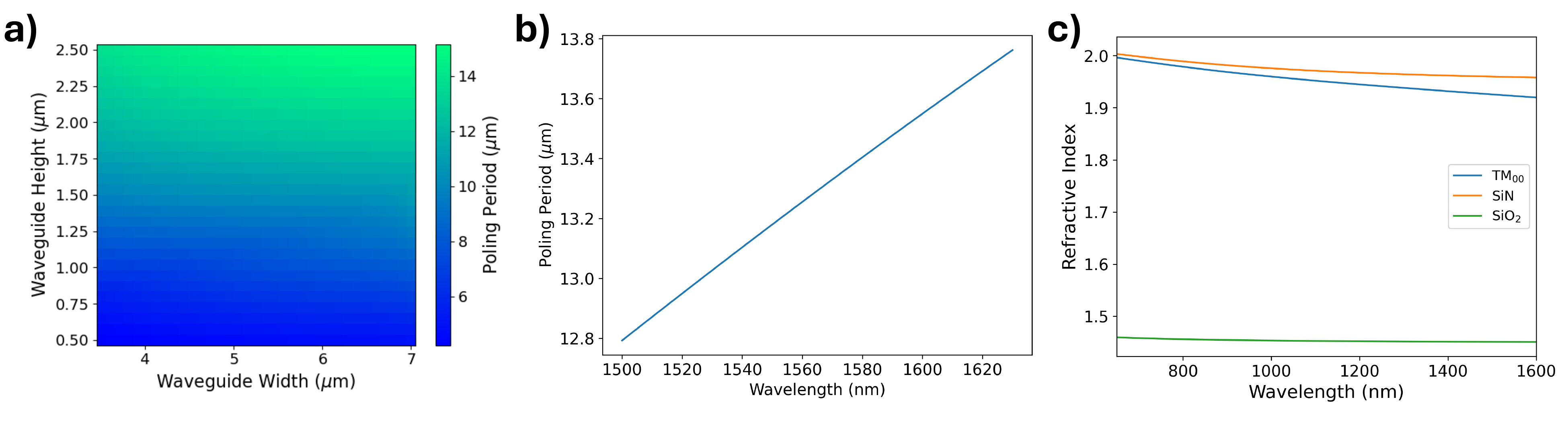}
    \caption{a) Simulated poling period of an unannealed SVM \(\mathrm{SiN}\) waveguide with different core widths and heights. b) Simulated poling period as a function of wavelength for the bandwidth of our tunable laser. c) Dispersion of the unannealed SVM \(\mathrm{SiN}\), \(\mathrm{SiO_2}\), and simulated \(\mathrm{TM_{00}}\) mode as a function of wavelength.}
    \label{fig:Dispersion Properties}
\end{figure}

This section discusses the considerations for engineering the phase-matching properties to enhance nonlinear performance.  Because of the practical limitations arising from the SLM pixel size, photoconductor activation, and electric field fringing, we designed our devices to have programmable poling periods in the range of 13 \si{\micro\meter} to 16 \si{\micro\meter} for the \(\mathrm{TM_{00}}\) + \(\mathrm{TM_{00}}\) $\xrightarrow{}$ \(\mathrm{TM_{00}}\) process.  These practical limitations are discussed more in Sec.~\ref{subsec:field fringing}, Sec.~\ref{sec:optical setup}, and Sec.~\ref{sec:Circuit Model}.  The poling period needed to phase-match an SHG process is
\begin{equation}
    \Lambda = \frac{2 \pi}{k_{\text{SH}} - 2k_{\text{FH}}}
\end{equation}
where $k_{\text{SH}}$ is the wave vector of the SH field and $k_{\text{FH}}$ is the wave vector of the FH field. Although the dispersion of the \(\mathrm{SiN}\) material can be tuned by changing the \(\mathrm{Si}\)-content, non-stoichiometric \(\mathrm{SiN}\) displays higher propagation losses \cite{Lipson-complete-SiN}.  Instead of modifying the material dispersion, we use the geometric dispersion of our channel waveguide to engineer $\Lambda$.  Using the \(\mathrm{SiN}\) material dispersion shown in Fig.~\ref{fig:Dispersion Properties}(c), we calculate how $\Lambda$ varies with the waveguide thickness and width.  This simulation assumes the etch completely clears the \(\mathrm{SiN}\) on the sidewall and the sidewall angle is $90^\circ$.  Simulations were conducted for unannealed \(\mathrm{SiN}\) films purchased from SVM.

\begin{figure}
    \centering
    \includegraphics[width=1\linewidth]{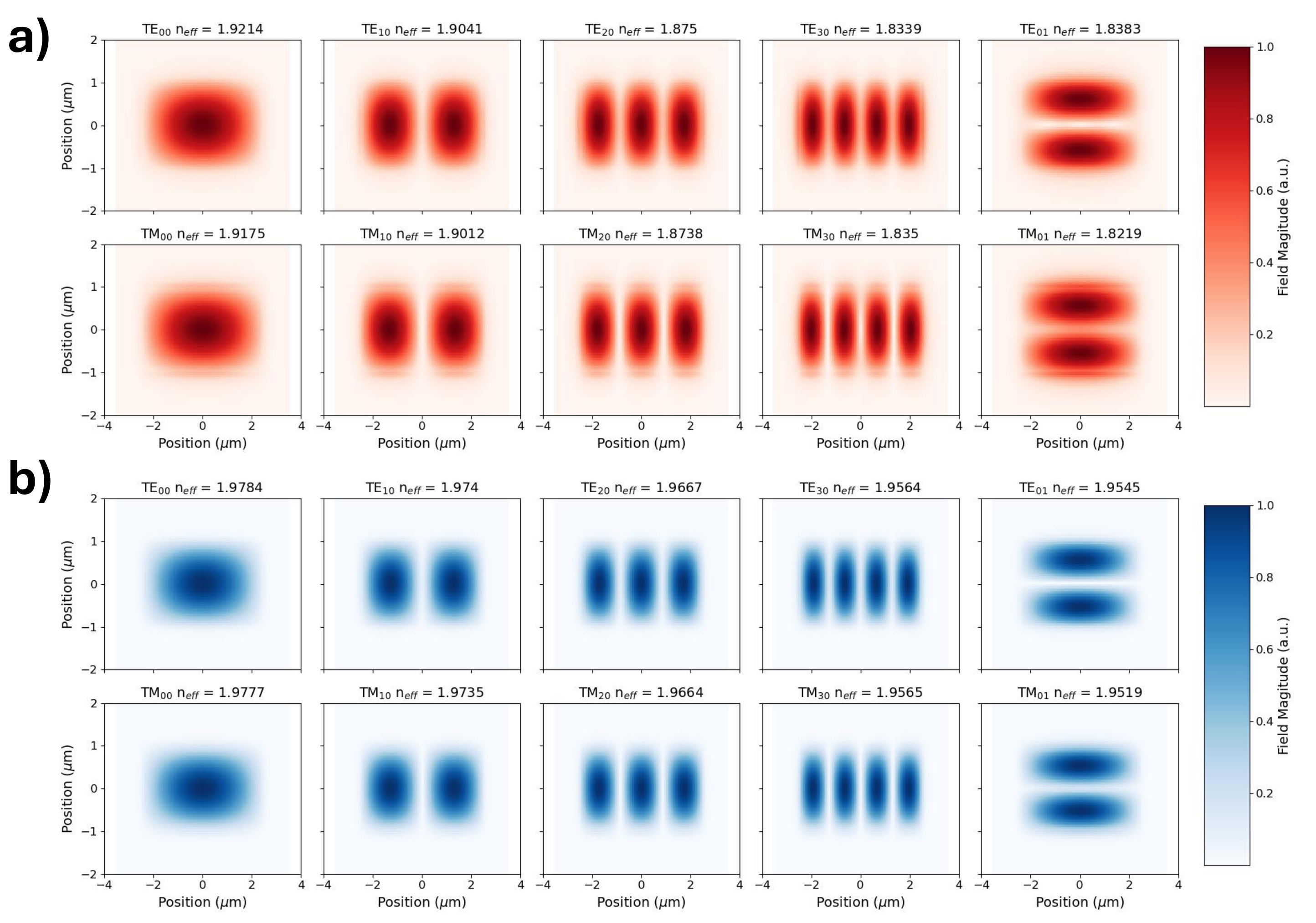}
    \caption{a) Calculated \(\mathrm{TE}\) and \(\mathrm{TM}\) field magnitudes for the FH modes. b) Calculated \(\mathrm{TE}\) and \(\mathrm{TM}\) field magnitudes for the SH modes.  Both simulations were conducted using the material dispersion model for unannealed SVM \(\mathrm{SiN}\) with waveguide dimensions of 2 \(\mathrm{\mu m}\) by 5 \(\mathrm{\mu m}\), which are the same properties as Fig.~\ref{fig:Dispersion Properties}.  The mode number and calculated effective index are also labeled in the title of each mode plot. For TM modes we plot $E_y$, for TE modes we plot $E_x$, and the plots are normalized so the largest field magnitude is 1.}
    \label{fig:mode shape}
\end{figure}

Fig.~\ref{fig:Dispersion Properties}(a) shows the relationship between $\Lambda$ and the waveguide geometry.  From these results, we find that 2 \si{\micro\meter} by 5 \si{\micro\meter} is the waveguide geometry that most safely satisfies our resolution constraints.  The calculations for $\Lambda$ in this simulation were only done for the $\text{TM}_{\text{00}} + \text{TM}_{\text{00}} \rightarrow \text{TM}_{\text{00}}$ process, but as Sec.~\ref{sec:Polarization Experiments} and Sec.~\ref{sec:spatial modes} discuss, more processes are possible.  Fig.~\ref{fig:mode shape} shows the electric field magnitude in the dominant vector direction for different SH and FH spatial modes. Subsequent experiments measured poling periods that are longer than those shown in Fig.~\ref{fig:Dispersion Properties}.  Discrepancies could be caused by imperfect magnification calibration on the imaging setup and fabrication errors resulting in waveguide geometry variation.  Additionally, the material model of \(\mathrm{SiN}\) deposited by SVM was measured before the waveguides were annealed, as annealed waveguides are etched and do not have a large slab region that our ellipsometer can characterize.  This means the material model used in Fig.~\ref{fig:Dispersion Properties} is different from that of the annealed material and \(\mathrm{SiN}\) deposited in-house.  Still, programmable $\chi^{(2)}$ allows us to adapt our devices even if the underlying dispersion models are faulty, showing the robustness of our devices. The simulations presented in Fig.~\ref{fig:mode shape} and Fig.~\ref{fig:Dispersion Properties} were conducted using Tidy3D mode solvers.  

\subsection{Substrate Leakage Management}
\label{subsec:leakage}
\begin{figure}
    \centering
    \includegraphics[width=0.5\linewidth]{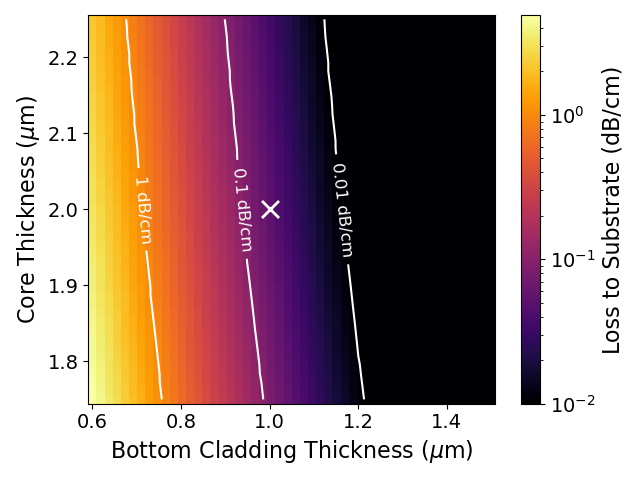}
    \caption{Theoretically predicted waveguide loss from radiative coupling to the silicon substrate as the core and bottom cladding thicknesses are adjusted.  The ``X'' marks the core and cladding thicknesses used in the final device.}
    \label{fig:Substrate_Loss}
\end{figure}

While thicker \(\mathrm{SiO_2}\) layers negatively impact electrical performance, they improve optical performance.  This is because the claddings optically isolate the waveguide mode from the \(\mathrm{Si}\)-substrate and the \(\mathrm{Si}\)-rich \(\mathrm{SiN}\) photoconductor, which both have higher refractive indices than the waveguide mode. This means lower propagation loss and higher nonlinear conversion efficiency.  Fig.~\ref{fig:Substrate_Loss} shows simulated predictions for waveguide loss into the \(\mathrm{Si}\)-substrate as the core \(\mathrm{SiN}\) and cladding \(\mathrm{SiO_2}\) thicknesses are swept.  Because the \(\mathrm{Si}\)-rich \(\mathrm{SiN}\) photoconductor has a lower refractive index than the \(\mathrm{Si}\)-substrate, the results in Fig.~\ref{fig:Substrate_Loss} can be taken as an upper bound on the radiative loss into the photoconductor.  Sec.~\ref{subsec:dispersion properties} showed that a \(\mathrm{SiN}\) core thickness of 2 \si{\micro\meter} would be ideal to manage waveguide dispersion, so we chose cladding thicknesses of 1 \si{\micro\meter} to minimize radiative losses to $\sim$ 0.1 dB/cm for this waveguide thickness.  Additionally, the results in Fig.~\ref{fig:Substrate_Loss} can be used to understand mode leakage out of the sidewalls into the photoconductor.  While the horizontal mode shape is different from a slab mode, it was experimentally found that a \(\mathrm{SiO_2}\) thickness of at least $\sim$ 1 \si{\micro\meter} minimized radiative losses through the sidewalls.  Because a thicker sidewall \(\mathrm{SiO_2}\) does not negatively impact electrical performance, we used a sidewall thickness of 1.4 \si{\micro\meter} to further suppress this power leakage.  The simulations presented here calculated radiative loss using the transfer matrix method \cite{Transfer-Matrix} and were verified using COMSOL \cite{Hiro-paper}.

\subsection{Management of Spatial Mode Out-Coupling}
\label{subsec:Mode Shapes}
\begin{figure}
    \centering
    \includegraphics[width=0.4\linewidth]{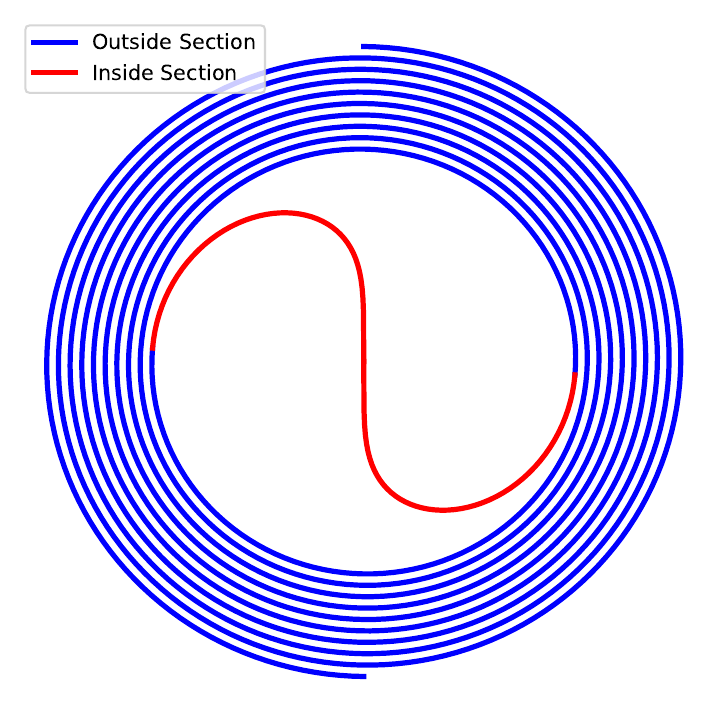}
    \caption{Cartoon rendering of the spiral geometry for poling over 10.1 cm, with each section labeled using a different color. The waveguide widths are not drawn to scale with the physical waveguide widths utilized experimentally.}
    \label{fig:Spiral Drawing}
\end{figure}
For the longer spiral waveguides we tested, the spiral was characterized by two equations in polar coordinates.  The outside section is an Archimedean spiral described as
\begin{equation}
    r(\theta) = a \theta + \theta_0
\end{equation}
and the inner section is described as
\begin{equation}
    r(\theta) = b \sin(\theta)^c
\end{equation}
To preserve the continuity of the spiral at the junction between the inner and outer sections, $r(\theta)$ and $\frac{dr(\theta)}{d\theta}$ must be continuous.  If the junction between the inner and outer sections occurs at $\theta_c$, then the continuity condition is 
\begin{align}
    a &= bc\sin(\theta_c)^{c-1} \cos(\theta_c) \\
    \theta_0 &= b\sin(\theta_c)^c - bc\sin(\theta_c)^{c-1}\cos(\theta_c)\theta_c
\end{align}
By adjusting $b$, $c$, and $\theta_c$, we can adjust the geometry of our spirals. Figure \ref{fig:Spiral Drawing} shows a drawing of the spiral geometry used for 10.1 cm of nonlinear interaction length, with the inside and outside sections separately labeled.

Because the dimensions of our channel waveguides are large, they support many spatial modes for both the FH and SH fields (as discussed in Sec.~\ref{subsec:many modes} and Sec.~\ref{subsec:dispersion properties}).  For waveguides that are straight, this presents minimal concern, as there are few ways for light in the fundamental spatial mode to couple into higher-order spatial modes. Spiral waveguides, however, change their radius of curvature gradually, meaning they can couple different spatial modes as light propagates through the structure.  This results in higher effective loss for the pump mode, reducing the device conversion efficiency.  Furthermore, the generated SHG will couple out of the \(\mathrm{TM_{00}}\) mode into higher-order spatial modes, preventing coherent interference between SHG produced throughout the waveguide.  Therefore, the waveguide geometry must minimize the higher-order spatial-mode coupling while ensuring the poling period stays within the range of our programming illumination's capabilities.

\begin{figure}
    \centering
    \includegraphics[width=1\linewidth]{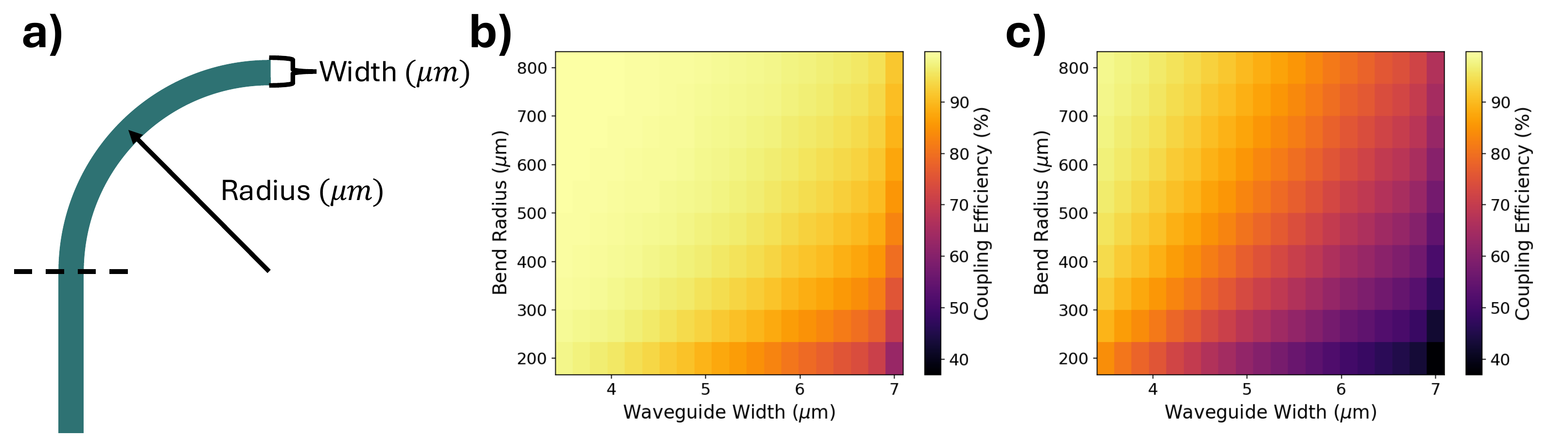}
    \caption{a) Diagram of the waveguide geometry being simulated, with labels indicating which geometric features are scanned. b) Coupling efficiency (in \%) between straight and circular waveguides for 1550 nm. c) Coupling efficiency (in \%) between straight and circular waveguides for 775 nm.  The simulations in panels b) and c) present the coupling efficiency between the \(\mathrm{TM_{00}}\) of the circular and straight waveguides at the junction. These simulations were carried out using Tidy3D mode solvers.}
    \label{fig:Curved_coupling}
\end{figure}

Fig.~\ref{fig:Curved_coupling} shows the mode overlap between the fundamental spatial mode at a straight-circular junction for different waveguide widths and bend radii.  These simulations assumed a waveguide height of 2 \si{\micro\meter}. The spirals employed in our experiment adiabatically changed their radius of curvature over a long distance, so the presented simulations should not be taken to represent the exact dynamics of our system.  Instead, they provide an approximate guiding model to understand which device geometries will likely suffer from extensive spatial mode out-coupling. 

To alleviate mode leakage out of the \(\mathrm{TM_{00}}\) mode, we employ two design techniques: keeping the radius of curvature above 500 \si{\micro\meter} when the waveguide is 5 \si{\micro\meter} wide and tapering the width of the waveguide down to 2 \si{\micro\meter} when the radius is below that threshold.  In practice, we implement this strategy by leaving the outer Archimedean section of the spiral at a width of 5 \si{\micro\meter} and tapering the inner section of the spiral to 2 \si{\micro\meter}.  This allows us to maintain long poling periods for the majority of the propagation while keeping the FH and SH fields in the fundamental spatial mode.

\subsection{Field Fringing}
\label{subsec:field fringing}
An additional consideration for the electrical behavior concerns the effects of bias fields fringing through the top cladding and the \(\mathrm{SiN}\) waveguide layer, causing the induced $\chi^{(2)}$ pattern to blur.  To better understand the effects of field fringing, we created electrostatic field simulations to model field behavior within our device. To solve for the electrostatic field, we use the following relations
\begin{equation}
    \vec{D}(\vec{r}) = \epsilon_0 \epsilon(\vec{r})\vec{E}(\vec{r})
\end{equation}
\begin{equation}
    \vec{E}(\vec{r})=- \vec{\nabla}\phi(\vec{r})
\end{equation}
where $\vec{E}$ is the electrostatic field, $\vec{D}$ is the displacement field, $\phi$ is the electrostatic potential, and $\epsilon$ is the permittivity. With no static charge, which is approximately the case for our devices, $\vec{\nabla} \cdot \vec{D}(\vec{r}) = 0$.  Combining these expressions, we find an equation that solves for the electrostatic potential of our devices as
\begin{equation}
    \vec{\nabla} \cdot(\epsilon(\vec{r})\vec{\nabla}\phi(\vec{r})) = 0
    \label{eq:fringing equation}
\end{equation}
By solving Eq.~\eqref{eq:fringing equation}, we examine how electrostatic fields fringe through the top cladding and \(\mathrm{SiN}\) waveguide for different device geometries.  When solving Eq.~\eqref{eq:fringing equation}, we assume symmetric boundary conditions in the horizontal dimension, treat the substrate and the bright state photoconductor as perfectly conducting, and treat the dark state photoconductor as perfectly insulating.  We run these simulations using a DC applied voltage even though the experiments were run using AC voltage.  This is because, in the ideal limit, the impedance of the device layers should follow the relationship in Eq.~\eqref{eq:impedance}.  While these simulations do not give exact quantitative predictions for the expected field for a given voltage, they provide insight into how fringing changes the relative contrast in the bright and dark sections of the \(\mathrm{SiN}\) waveguide. 

\begin{figure}
    \centering
    \includegraphics[width=1\linewidth]{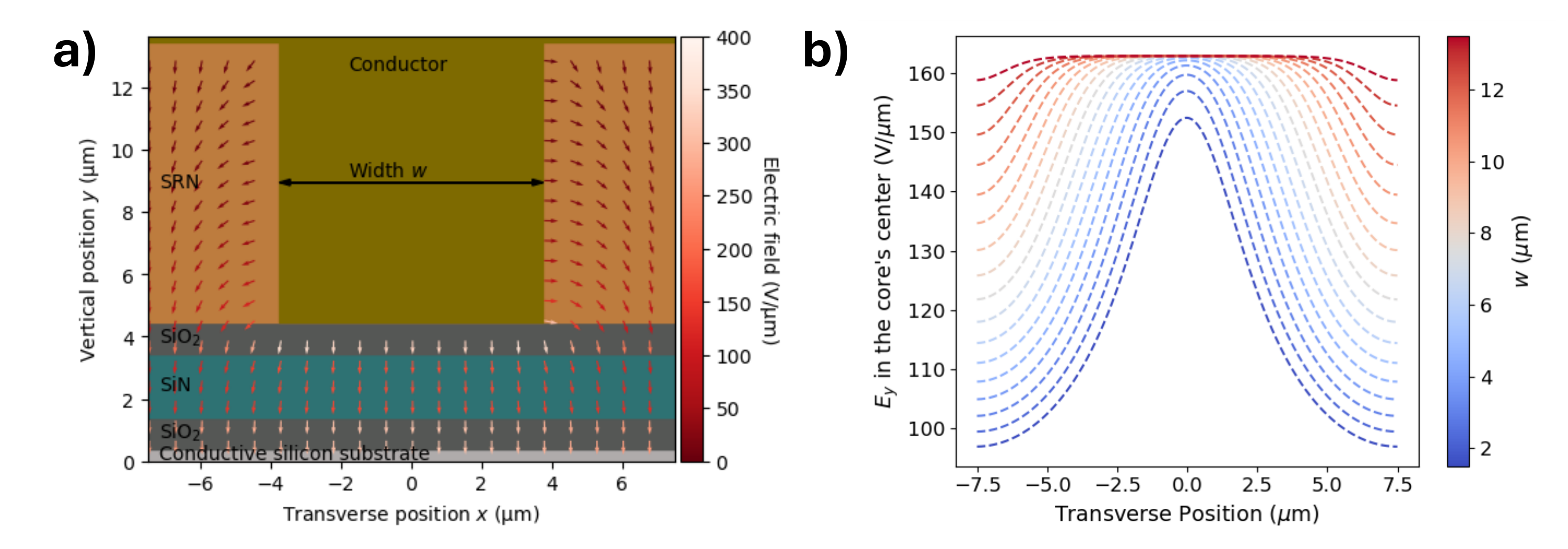}
    \caption{a) Simulated device cross-section showing electric field fringing. b) Vertical electric field \(\mathrm{E_y}\) in the center of the \(\mathrm{SiN}\) waveguide core as the width of the illumination spot changes. All the simulations in this figure were completed with \(\mathrm{V_{app}}\) = 1000 V, \(\mathrm{d_{to}}\) = 1 \(\mathrm{\mu m}\), \(\mathrm{d_{bo}}\) = 1 \(\mathrm{\mu m}\), \(\mathrm{d_{SiN}}\) = 2 \(\mathrm{\mu m}\), \(\mathrm{d_{pc}}\) = 9 \(\mathrm{\mu m}\), and a period of 15 \(\mathrm{\mu m}\).}
    \label{fig:Fringing cross section}
\end{figure}
\begin{figure}
    \centering
    \includegraphics[width=1\linewidth]{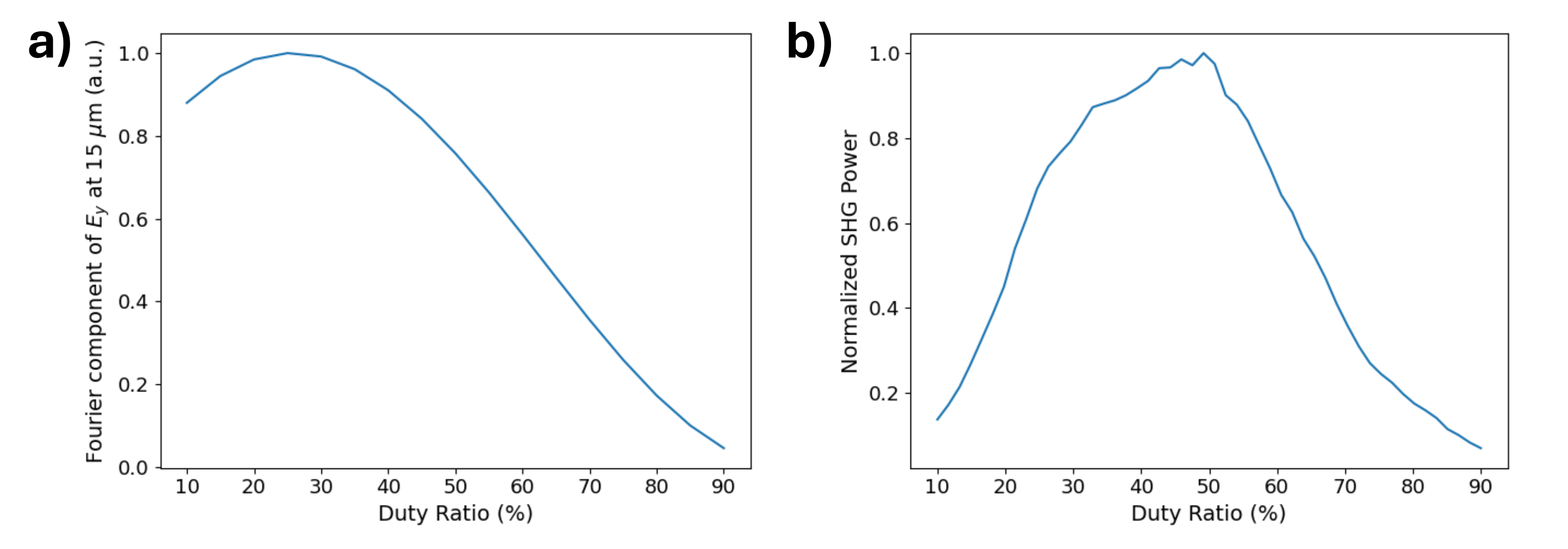}
    \caption{a) Normalized simulated Fourier component of \(\mathrm{E_y}\) in the center of the \(\mathrm{SiN}\) waveguide core at a period of 15 \(\mathrm{\mu m}\).  This plot was generated by analyzing simulation data in Fig.~\ref{fig:Fringing cross section}. b) Experimentally measured SHG at a poling period of 14.9 \(\mathrm{\mu m}\) for different duty ratios.  The y-axis is normalized to the maximum measured SHG power.}
    \label{fig:Duty Ratio}
\end{figure}

Fig.~\ref{fig:Fringing cross section} shows how \(\mathrm{E_y}\) in the center of the \(\mathrm{SiN}\) waveguide varies as the width of the illumination spot $w$ changes for a projected poling period of 15 \si{\micro \meter}. Fig.~\ref{fig:Duty Ratio} shows the simulated Fourier component at a period of 15 \si{\micro\meter} for different duty ratios, as well as the experimental measurement. While the simulated \(\mathrm{E_y}\) Fourier component at a period of 15 \si{\micro \meter} is greatest for a duty ratio of approximately 26 $\%$, we experimentally found that a duty ratio of 50 $\%$ maximizes the conversion efficiency.  The discrepancy between the simulation and experimental results presented in Fig.~\ref{fig:Duty Ratio} likely comes from the assumption that the photoconductor is completely conductive when illuminated.  The photoconductor's activation has a nonlinear dependence on illumination power and depends on the illumination spot size, as larger spot sizes suffer less from imaging resolution limits. Given the size of each spatial light modulator pixel on the waveguide as $w_{\text{pixel}} =  3.77$ \si{\micro \meter}, our spot sizes are limited by the discretization of the SLM. If the activation of our photoconductor is much higher for $w = 2w_{\text{pixel}}$ than $w =w_{\text{pixel}}$, then a duty ratio of 50 $\%$ will still provide better nonlinear performance despite the drawbacks of field fringing. 

\begin{figure}
    \centering
    \includegraphics[width=1\linewidth]{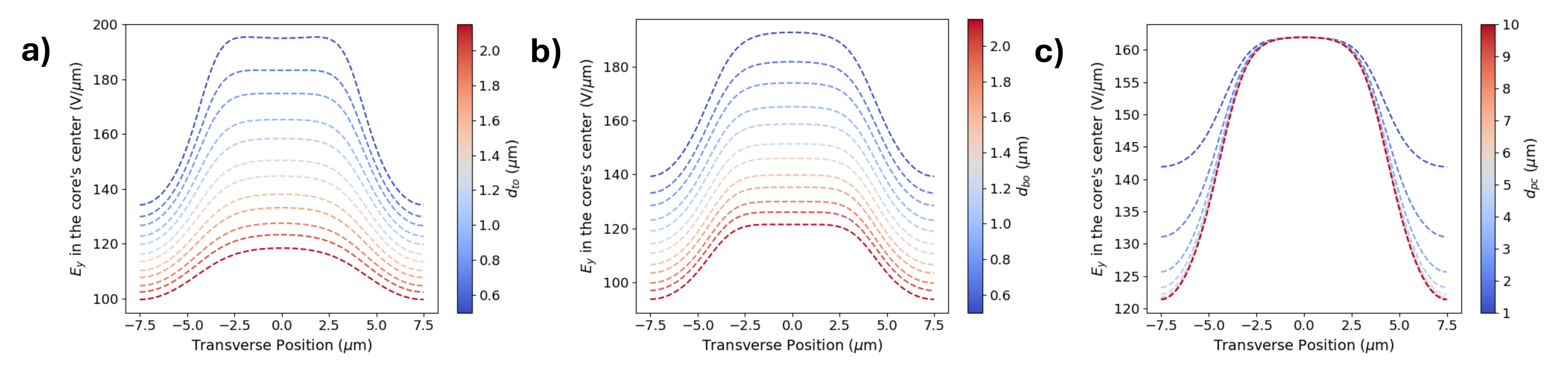}
    \caption{a) Vertical electric field \(\mathrm{E_y}\) in the center of the \(\mathrm{SiN}\) waveguide as the top oxide thickness is varied.  b) \(\mathrm{E_y}\) as the thickness of the bottom oxide is varied. c) \(\mathrm{E_y}\) as the thickness of the photoconductor is varied.  Unless otherwise stated, the simulation parameters used an illumination width of 7.5 \(\mathrm{\mu m}\) and device specifications outlined in Fig.~\ref{fig:Fringing cross section}.}
    \label{fig:Fringing thickness}
\end{figure}

Fig.~\ref{fig:Fringing thickness} shows how \(\mathrm{E_y}\) changes with different top cladding, bottom cladding, and photoconductor thicknesses.  Decreasing the thickness of the top cladding has a greater effect on the field contrast in the waveguide than decreasing the thickness of the bottom cladding.  This is because the top cladding is closer to the position of greatest field contrast at the photoconductor.  Consequently, the most fringing happens there.  Additionally, Fig.~\ref{fig:Fringing thickness} shows that increasing the photoconductor thickness has diminishing returns for increasing the electric field contrast as $d_{\text{pc}}$ approaches 8 \si{\micro\meter}.  Because the photoconductor has finite resistance in the experiment, a thicker photoconductor still accentuates switching from illumination in practice.

\begin{figure}
    \centering
    \includegraphics[width=1\linewidth]{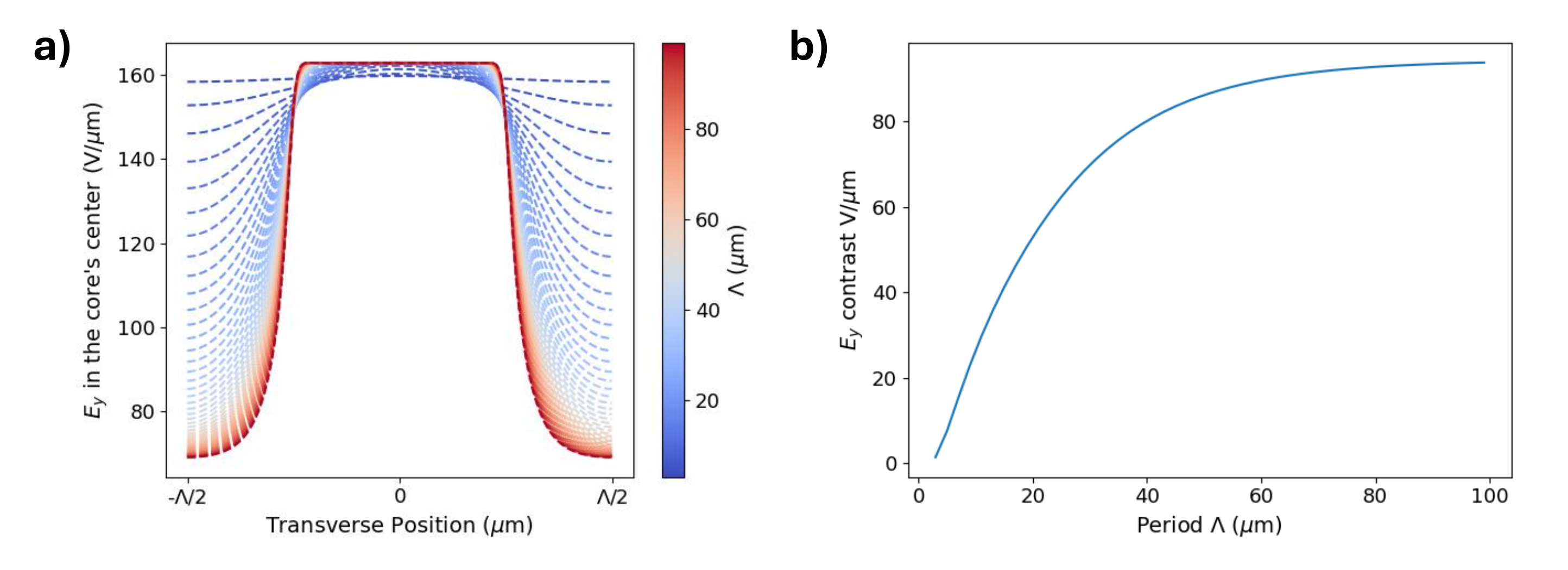}
    \caption{a) Vertical electric field \(\mathrm{E_y}\) in the center of the \(\mathrm{SiN}\) waveguide core as the period of the illumination pattern is varied.  The duty cycle is assumed to be $50 \%$ for each period. b) Maximum contrast between \(\mathrm{E_y}\) in the bright and dark state at the center of the \(\mathrm{SiN}\) waveguide as the period is swept.  Parameters used in these simulations match those used in Fig.~\ref{fig:Fringing cross section}.}
    \label{fig:Fringing Period}
\end{figure}

Fig.~\ref{fig:Fringing Period} shows the predicted \(\mathrm{E_y}\) in the center of the \(\mathrm{SiN}\) waveguide for different poling periods.  The contrast between the bright and dark state increases as the poling period increases, though the net increase in contrast diminishes with larger periods.  In addition to the limitations of the illumination setup, this simulation provides an additional reason why using a larger poling period achieves better device performance.

\subsection{Optical Loss}
\label{sec:loss characterization}
\begin{figure}
    \centering
    \includegraphics[width=0.4\linewidth]{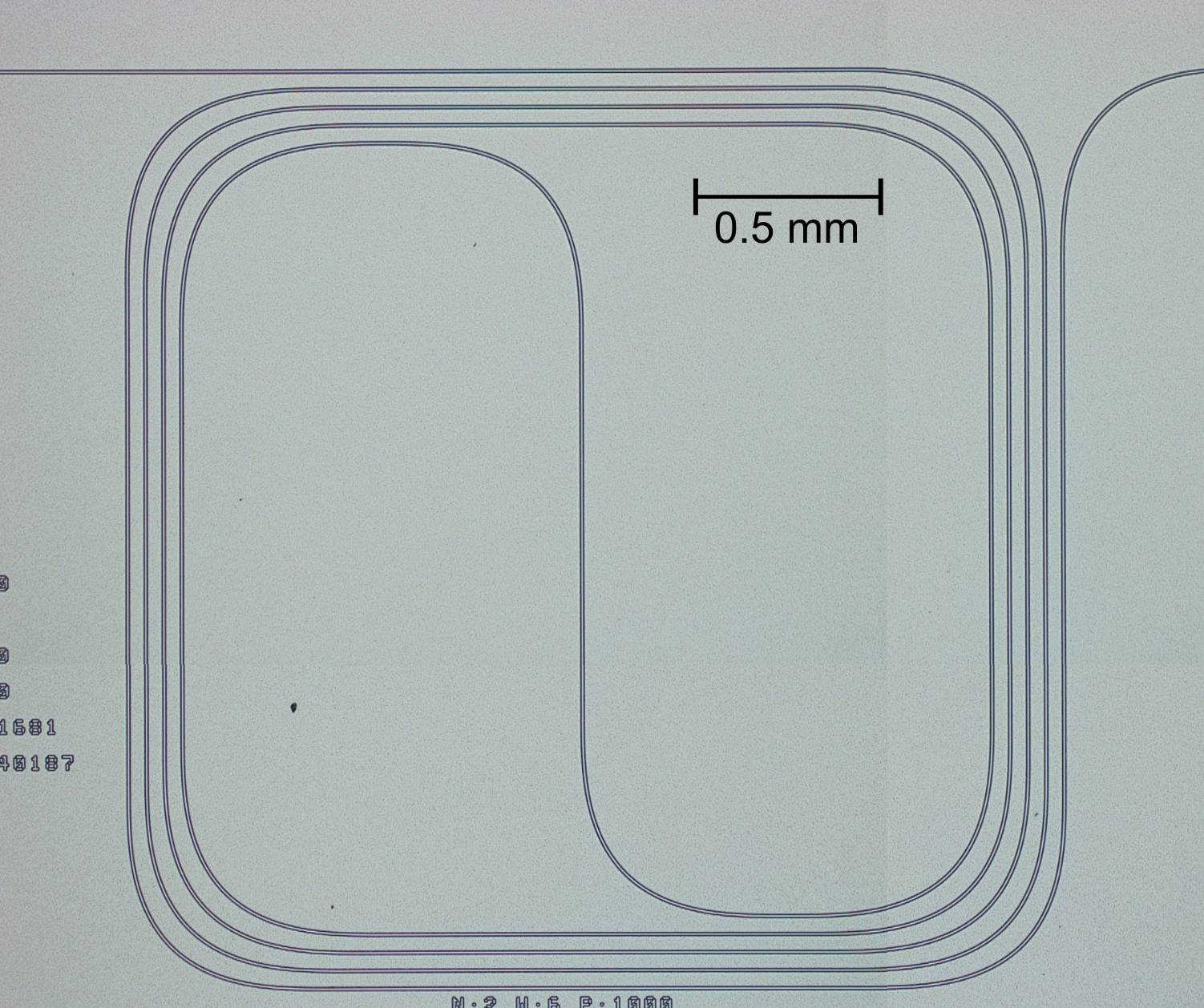}
    \caption{A microscope image of a fabricated square spiral.  This image was created by stitching together multiple images.  The same waveguide geometry and image were presented in previous work \cite{Ryo-paper}.}
    \label{fig:Euler Spiral Picture}
\end{figure}

This section discusses the characterization of optical loss in our waveguides, which is necessary to calculate conversion efficiency and nonlinear behavior over longer propagation distances. We characterize the FH and SH loss of our waveguides using the cutback method.  We implement the cutback method by measuring the transmission from waveguides with similar geometry but of different lengths.  In this case, we used a straight waveguide and a square-shaped spiral waveguide (as shown in Fig.~\ref{fig:Euler Spiral Picture}) that was 4.5 cm longer than the straight waveguide. We measure multiple waveguides of each length and take the average transmission, which allows us to minimize the negative effects of any waveguides with poor facet quality.  We then use the formula
\begin{equation}
    \mathrm{loss_{dB/cm}} = \frac{10\log(\frac{P_\text{short}}{P_\text{long}})}{L_\text{long} - L_\text{short}}
\end{equation}
to calculate the loss measured between the two waveguides. 

\begin{figure}
    \centering
    \includegraphics[width=1\linewidth]{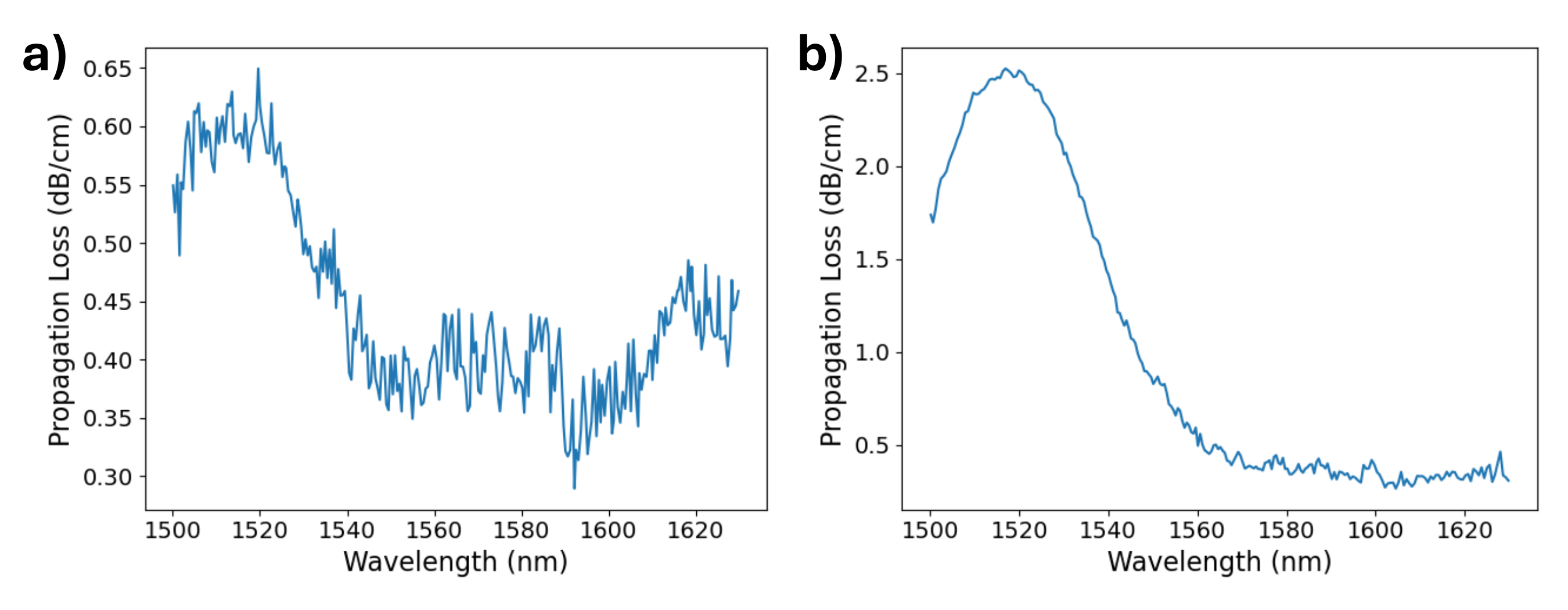}
    \caption{a) Measured loss spectrum of \(\mathrm{SiN}\) deposited in-house.  b) Measured loss spectrum of \(\mathrm{SiN}\) deposited by SVM.}
    \label{fig:Loss Spectra}
\end{figure}

Furthermore, because we measure the nonlinear conversion efficiency at many wavelengths, we record the full transmission spectrum of our waveguides from 1500 nm to 1630 nm. Fig.~\ref{fig:Loss Spectra} shows the measured loss spectrum in the infrared for \(\mathrm{SiN}\) deposited in-house and \(\mathrm{SiN}\) deposited by SVM. One source of error in these measurements is the multimode nature of our waveguides.  Higher-order spatial modes experience more loss than the fundamental spatial mode due to their increased overlap with waveguide sidewalls.  This causes higher scattering losses for the higher-order spatial modes.  Because we measure the power for all spatial modes at the waveguide output, the measured loss for the fundamental spatial mode is likely inflated by the multimode-ness of our waveguides.  This gets further exaggerated at bent structures where the modes interact more heavily with the sidewalls than on straight racetracks.  Consequently, the loss values reported here are likely upper bounds for the actual waveguide losses.

The main advantage of \(\mathrm{SiN}\) deposited in-house over \(\mathrm{SiN}\) deposited by SVM is the lower loss near 1520 nm due to high-temperature annealing.  As discussed in Sec.~\ref{sec:Fabrication}, \(\mathrm{SiN}\) deposited in-house undergoes multiple hours of furnace annealing at 1200 $^\circ\mathrm{C}$ while \(\mathrm{SiN}\) deposited by SVM only undergoes rapid thermal annealing for 5 minutes at 800 $^\circ\mathrm{C}$.  This higher annealing temperature drives out \(\mathrm{N-H}\) impurity bonds, which cause loss \cite{Lipson-complete-SiN}. In the wavelength range of 1570 nm to 1630 nm, \(\mathrm{SiN}\) deposited in-house has an average loss of 0.39 dB/cm and \(\mathrm{SiN}\) deposited by SVM has an average loss of 0.35 dB/cm.  This means, away from the \(\mathrm{SiN}\) material absorption peak, loss is dominated by sidewall roughness, common PECVD material defects, and radiative losses (as discussed in Sec.~\ref{subsec:leakage}).  We measured the SH loss solely at a wavelength of 802.5 nm due to the available light sources.  For \(\mathrm{SiN}\) deposited in-house we measured a loss of 0.67 dB/cm and for \(\mathrm{SiN}\) deposited by SVM we measured a loss of 1.35 dB/cm at 802.5 nm.

\section{Experimental Apparatus}
\label{sec:optical setup}
\subsection{Illumination Optics}
\begin{figure}
    \centering
    \includegraphics[width=0.6\linewidth]{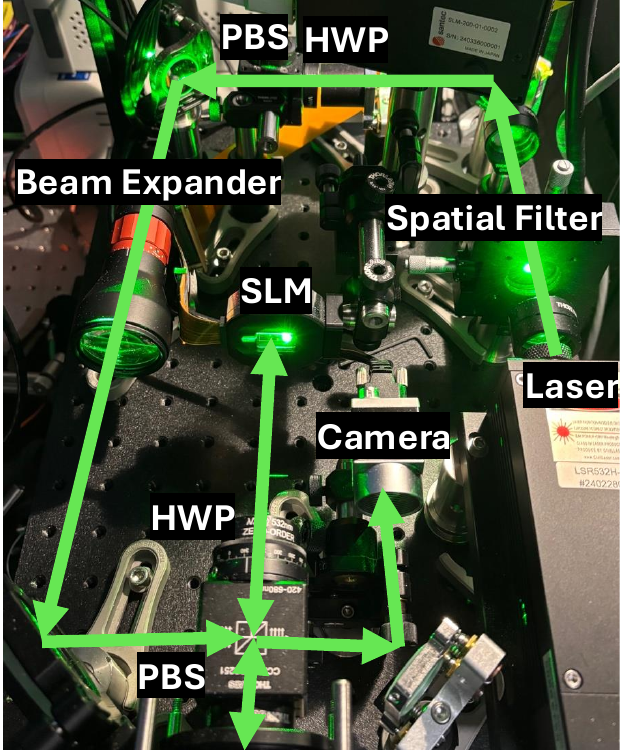}
    \caption{Labeled image of the optical setup used to project the programming illumination on the programmable waveguide.  SLM stands for spatial light modulator, HWP stands for half-wave plate, and PBS stands for polarizing beam splitter.}
    \label{fig:Imaging Setup Picture}
\end{figure}
\begin{figure}
    \centering
    \includegraphics[width=1\linewidth]{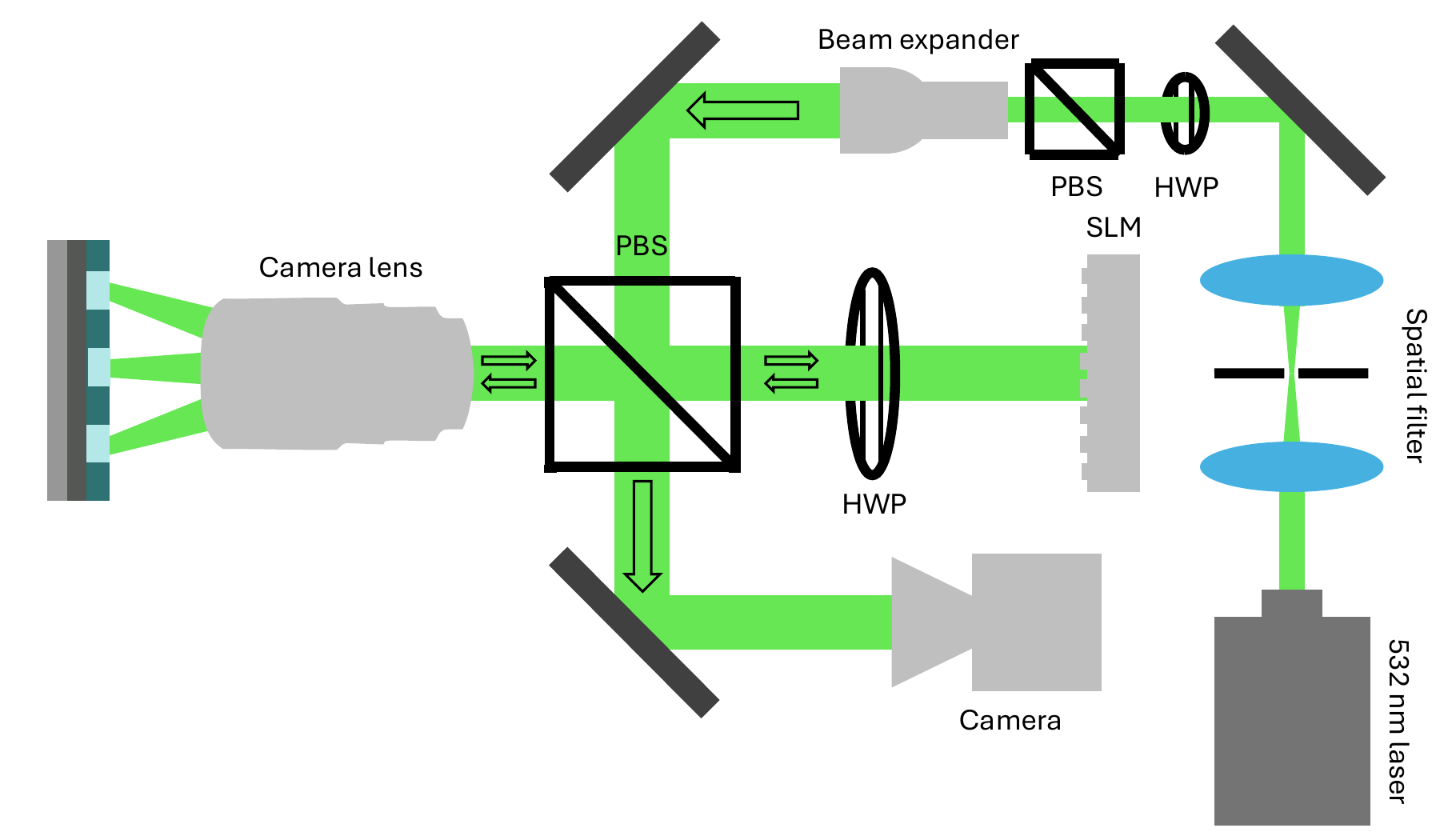}
    \caption{Labeled diagram of a simplified version of the optical setup used to project the programming illumination on the programmable waveguide.}
    \label{fig:Imaging Setup Cartoon}
\end{figure}

The following section describes the methods used to build the optics that projected the programming illumination on the photoconductor.  We used a 532 nm light source (532nm Green DPSS Laser; Civil Laser) and spatially filtered the mode using a 4F imaging setup and pinhole.  Following the spatial filter, we used a half-wave plate (HWP) and polarizing beam splitter (PBS) to tune the laser power that entered the rest of the setup and clean the laser polarization.  While this was useful for characterization, we did not adjust the power away from the maximum setting in the final experiments. For straight waveguides that propagate along one axis, we inserted an anamorphic prism into the beam path to tighten the illumination strip in the direction orthogonal to optical propagation. For spiral waveguides, which take up a larger spatial footprint in two dimensions, we did not use an anamorphic prism.  Later in the beam path, we used a beam expander to enlarge the illumination spot to cover the entire waveguide along the dimension parallel to mode propagation.

To pattern the illumination spot spatially, we used a spatial light modulator (SLM-200-01; Santec), a polarizing beam splitter, and a half-wave plate.  The spatial light modulator (SLM) has a resolution of 1920 by 1200 pixels and a pixel size of 8 \si{\micro\meter} by 8 \si{\micro\meter}.  We used a macro camera lens (Milvus 100 mm f/2M Lens; Carl Zeiss) to image the SLM onto the programmable nonlinear waveguide device.  While the imaging focus was readjusted multiple times during this study, the magnification ratio could be determined by calibrating the number of SLM pixels required to span a known physical distance lithographically defined on the device.  We measured a nominal de-magnification ratio of 2.12, meaning our average minimum resolvable spot size is $\sim$ 3.77 \si{\micro\meter}.  This limits our programmable poling periods to $\sim$ 7.5 \si{\micro\meter} at a minimum. The illumination intensity on-chip is $\sim$ 50 mW/$\text{cm}^2$.  While coherent illumination sources like the one employed in this work are known to cause speckle patterns due to small phase perturbations, we did not observe notable speckle patterns on our devices. Fig.~\ref{fig:Imaging Setup Cartoon} shows a labeled cartoon diagram of this setup, and Fig.~\ref{fig:Imaging Setup Picture} shows a labeled image of the apparatus.

\subsection{Measurement Apparatus}
\begin{figure}
    \centering
    \includegraphics[width=1\linewidth]{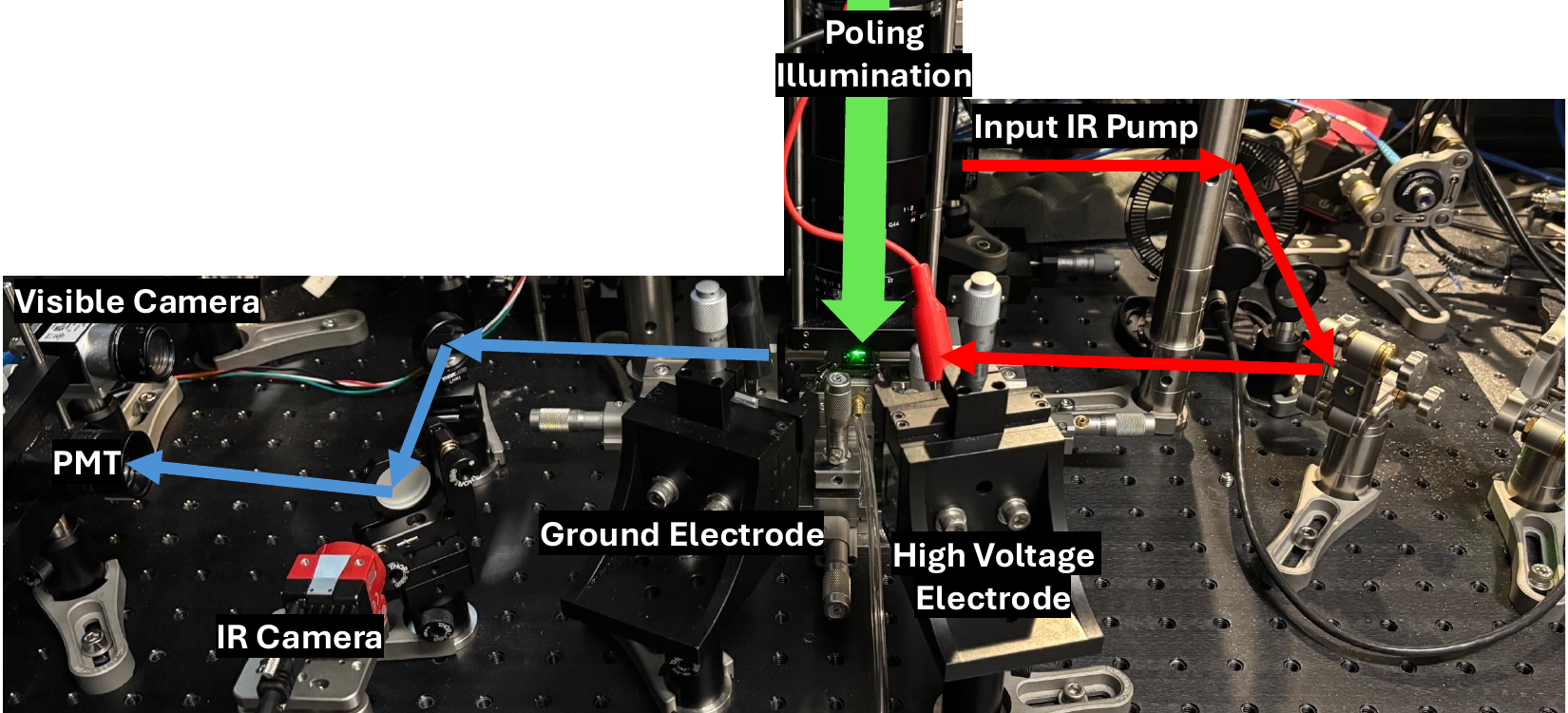}
    \caption{Labeled image of the optical components used to characterize nonlinear programmable waveguide devices.}
    \label{fig:Labeled Bottom Optics}
\end{figure}
\begin{figure}
    \centering
    \includegraphics[width=0.65\linewidth]{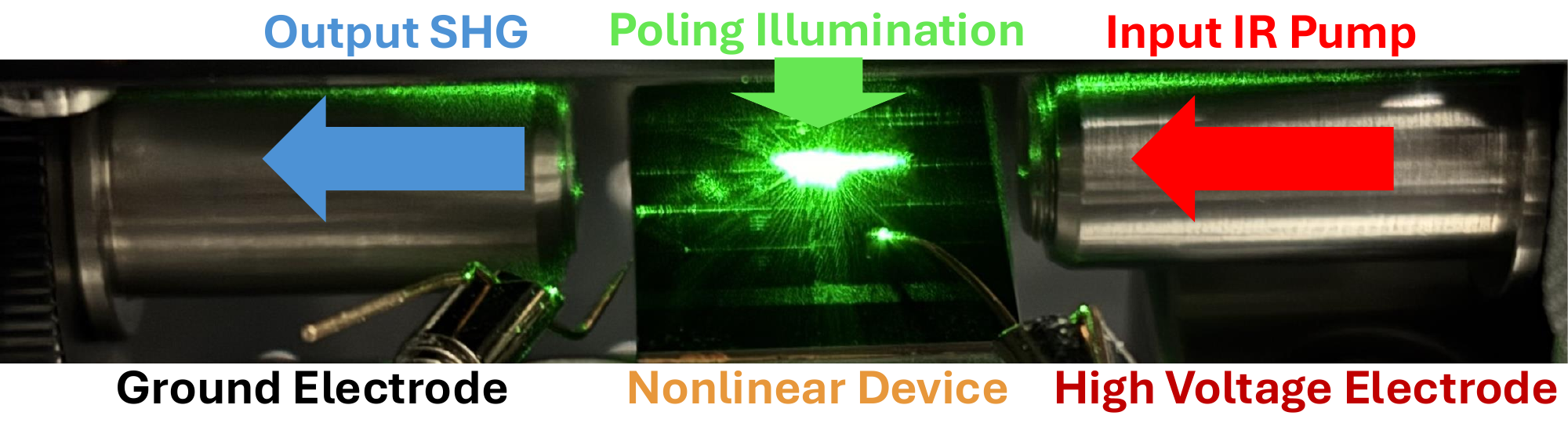}
    \caption{Labeled image of the optical inputs and electrical contacts on a nonlinear programmable waveguide device.}
    \label{fig:Coupling Image}
\end{figure}
\begin{figure}
    \centering
    \includegraphics[width=1\linewidth]{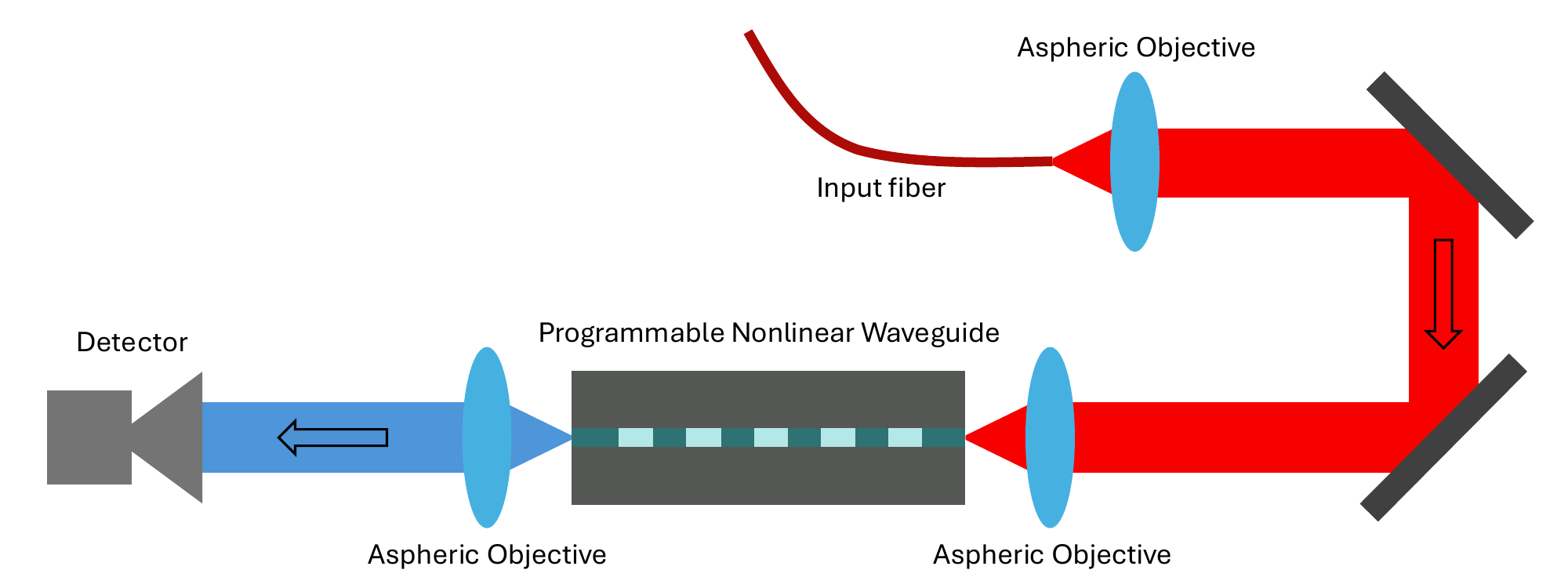}
    \caption{Labeled cartoon diagram of the optical setup used to couple light into a programmable nonlinear waveguide and measure the output generated SHG power.}
    \label{fig:Coupling Setup}
\end{figure}

Fig.~\ref{fig:Labeled Bottom Optics} shows a labeled image of the optics used to couple light into the programmable waveguide and measure the output SHG, and Fig.~\ref{fig:Coupling Setup} shows a labeled cartoon of this apparatus.  Fig.~\ref{fig:Coupling Image} shows a zoomed-in picture of the optical and electrical inputs and outputs on a nonlinear programmable waveguide device. We fiber-coupled a laser into our free-space optical setup and used the knobs on the fiber-coupling mount and a subsequent mirror to align the pump beam to the optical axis of the coupling objectives.  We then used an aspheric lens (C660TME-C; Thorlabs) to couple into our programmable waveguide device and a different aspheric lens (C330TME-B; Thorlabs) to collect the output SHG.  We use different lenses to couple in and out of the device because the input lens is coated for the FH while the output lens is coated for the SH. 

We used a single-lens imaging setup with the output aspheric objective to couple the output second-harmonic into a photomultiplier tube (PMT H10722-20; Hamamatsu Photonics) to measure the power of the generated signal. The PMT was operated with a control voltage of $V_\text{control} =0.5 \ \text{V}$ and separately calibrated to have an output sensitivity of 0.0926 V per nW of optical power.  We multiply this estimated power by 2 to account for a nominal 50 \% coupling efficiency into the PMT (when we used a band-pass filter centered at 790 nm, we multiply by 2.5 to account for the lower transmission of the filter).  For CW experiments, we multiply all measured FH and SH powers by 1.33 to compensate for a 75 \% collection efficiency of the output aspheric objective.  To calculate the peak optical power generated by the waveguide, instead of the average power, we multiply our measured SH power by an additional factor of 2 to account for the AC electrical bias on the device.

To improve the signal-to-noise ratio, we used an optical chopper before coupling the input pump into the waveguide.  The chopper provided a lock-in frequency for detection, allowing us to subtract the background signal more easily. When imaging the mode shapes, we used a Basler (acA2440-35um) camera instead of a PMT and removed the optical chopper from the beam path.  

To apply a voltage bias to the device, we made electrical contact on the \(\mathrm{ITO}\) electrode with a high-voltage probe and on the \(\mathrm{Si}\)-substrate using a ground probe. We used a Tektronix AFG1022 to generate the voltage signal and amplified it using a high voltage amplifier (Trek Model 2220). The applied voltage had an amplitude of 800 V for all devices. The applied electrical frequency depends on the underlying device geometry. Geometries with a smaller spatial footprint can be aligned to receive a higher illumination intensity per unit area because there is less area that needs illumination. As Sec.~\ref{sec:Circuit Model} discusses, the ideal operating electrical frequency increases as the illumination power increases.  Consequently, straight waveguides or small spirals operated closer to 10 Hz, while larger spiral devices operated closer to 5 Hz. For each device, we manually scan the applied electric frequency to see which setting maximizes the generated SHG power.

\section{Estimation of the programmable nonlinearity}
\label{sec:chi2 estimation}
In this section, we describe the model and the experimental results used to estimate the programmable $\chi^{(2)}$ nonlinearity of our device.

\subsection{$\chi^{(2)}$ Estimation Model}
First, we derive a model for SHG following Ref.~\cite{jankowski2024ultrafast}. Let $\mathbf{E}(x,y)$ and $\mathbf{H}(x,y)$ denote the electric and magnetic field profiles of the TM$_{00}$ mode of the waveguide. The physical field is given by $\mathrm{Re}\left[\mathbf{E}(x,y)\,e^{i(kz-\omega t)}\right]$, and likewise for $\mathbf{H}$. We normalize the profiles such that the time-averaged power carried by the mode at unit excitation amplitude is
\begin{align}
    \frac{1}{2}\int\mathrm{d}x\,\mathrm{d}y\,\mathrm{Re}\left(\mathbf{E}\times\mathbf{H}^*\right)\cdot\hat{\mathbf{z}}=P,
\end{align}
where $\hat{\mathbf{z}}$ is the unit vector along the propagation direction, and we take $P=\SI{1}{W}$ throughout. It is convenient to introduce dimensionless field profiles $\mathbf{e}$ and $\mathbf{h}$ via
\begin{align}
    \mathbf{E}(x,y)=\sqrt{\frac{2Z_0P}{n_\text{eff}A_\text{mode}}}\,\mathbf{e}(x,y),\qquad
    \mathbf{H}(x,y)=\sqrt{\frac{2n_\text{eff}P}{Z_0A_\text{mode}}}\,\mathbf{h}(x,y),
\end{align}
where $Z_0\approx\SI{377}{\ohm}$ is the vacuum impedance and $n_\text{eff}$ is the effective index of the mode. We define the area of the mode as
\begin{align}
A_\text{mode}=\int\mathrm{d}x\,\mathrm{d}y\,\mathrm{Re}\left(\mathbf{e}\times\mathbf{h}^*\right)\cdot\hat{\mathbf{z}}.
\end{align}
We fix the remaining freedom in $\mathbf{e}$ and $\mathbf{h}$ by requiring the peak value of $\mathrm{Re}(\mathbf{e}\times\mathbf{h}^*)\cdot\hat{\mathbf{z}}$ to be unity, in which case $A_\text{mode}$ is the ratio of the total mode power to the peak intensity.

We assume that only the modes at the FH ($\omega$) and the SH ($2\omega$) are excited, so that the total electric field in the waveguide takes the form
\begin{align}
    \mathbf{E}(x,y,z,t)=a(z)\,\mathbf{E}^{(\omega)}(x,y)\,e^{ik_1z-i\omega t}+b(z)\,\mathbf{E}^{(2\omega)}(x,y)\,e^{ik_2z-2i\omega t},
\end{align}
where $k_1$ and $k_2$ are the propagation constants of the FH and SH modes, and $a(z)$ and $b(z)$ are slowly varying amplitudes such that $|a|^2=1$ ($|b|^2=1$) corresponds to $P=\SI{1}{W}$ of FH (SH) power. In the presence of a nonlinear polarization $\mathbf{P}_\text{NL}$, the amplitudes evolve according to
\begin{align}
    \partial_za(z)&=\frac{i\omega}{4P}\,e^{-ik_1z}\int\mathrm{d}x\,\mathrm{d}y\,\mathbf{E}^{(\omega)*}(x,y)\cdot\mathbf{P}_\text{NL}^{(\omega)},\\
    \partial_zb(z)&=\frac{2i\omega}{4P}\,e^{-ik_2z}\int\mathrm{d}x\,\mathrm{d}y\,\mathbf{E}^{(2\omega)*}(x,y)\cdot\mathbf{P}_\text{NL}^{(2\omega)}.
\end{align}
Here, $\mathbf{P}_\text{NL}^{(\omega)}$ and $\mathbf{P}_\text{NL}^{(2\omega)}$ denote the complex amplitudes of the nonlinear polarization oscillating at $\omega$ and $2\omega$, respectively. For SHG, the latter is
\begin{align}
P_{\text{NL},i}^{(2\omega)}=\epsilon_0\,a^2\sum_{j,k}d_{ijk}\,E_j^{(\omega)}E_k^{(\omega)}\,e^{2ik_1z},
\end{align}
where $d_{ijk}=\chi^{(2)}_{ijk}/2$ is the second-order nonlinear tensor, and the indices $i,j,k$ run over $x,y,z$. For simplicity, we assume that the nonlinear coupling is present only along $y$, i.e., only $d_{yyy}$ is nonzero, which yields
\begin{align}
    \partial_zb(z)=\frac{2i\omega}{4P}\,\epsilon_0\,d_{yyy}(z)\,a^2\,e^{-i\Delta kz}\int_\text{core}\mathrm{d}x\,\mathrm{d}y\,E_y^{(2\omega)*}\left(E_y^{(\omega)}\right)^2,
\end{align}
where $\Delta k=k_2-2k_1$ is the phase mismatch, and the integral runs over the core region of the waveguide cross-section. We now introduce the amplitudes $A=\sqrt{P}\,a$ and $B=\sqrt{P}\,b$, whose squared moduli give the FH and SH powers, respectively. In terms of these amplitudes, we obtain
\begin{align}
    \partial_zB(z)=i\,\frac{\sqrt{2Z_0}\,\omega\,d_{yyy}(z)}{c\,n_\text{eff}^{(\omega)}\sqrt{n_\text{eff}^{(2\omega)}A_\text{eff}}}\,e^{-i\Delta kz}A^2(z),
    \label{eq:sh-cme}
\end{align}
where we have used $\epsilon_0Z_0=1/c$, and
\begin{align}
A_\text{eff}=\frac{\left(A_\text{mode}^{(\omega)}\right)^2A_\text{mode}^{(2\omega)}}{\left|\int_\text{core}\mathrm{d}x\,\mathrm{d}y\,e_y^{(2\omega)*}\left(e_y^{(\omega)}\right)^2\right|^2}
\end{align}
is the effective area of the nonlinear interaction.

In our programmable nonlinear waveguide, $d_{yyy}(z)$ is switched on and off along $z$ as a square wave with period $\Lambda=2\pi/\Delta k$. We describe this spatial modulation by
\begin{align}
    \label{eq:expansion-r}
    r(z)=\frac{1}{2}\left[\mathrm{sign}\left(\sin\Delta kz\right)+1\right]
    =\frac{1}{2}+\frac{1}{i\pi}\sum_{m=1}^\infty\frac{1}{2m-1}\left(e^{i(2m-1)\Delta kz}-e^{-i(2m-1)\Delta kz}\right).
\end{align}
That is, $r(z)=1$ in the regions where the programming illumination is on and $r(z)=0$ otherwise. In terms of this function, the total nonlinear coefficient is
\begin{align}
    d_{yyy}(z)=\frac{\chi^{(2)}_{yyy}}{2}\,r(z)+\frac{\chi^{(2)}_{yyy,\text{const}}}{2},
\end{align}
where $\chi^{(2)}_{yyy}$ is the programmable nonlinearity and $\chi^{(2)}_{yyy,\text{const}}$ is a static background. When this expression is substituted into Eq.~\eqref{eq:sh-cme}, only the $m=1$ term of Eq.~\eqref{eq:expansion-r} proportional to $e^{i\Delta kz}$ cancels the phase factor $e^{-i\Delta kz}$. The constant terms and the higher harmonics ($m\geq2$) remain phase-mismatched and average out. Thus, keeping only the phase-matched contribution, the SH field evolves as
\begin{align}
    \partial_zB=\kappa_\text{eff}A^2,
\end{align}
with
\begin{align}
\kappa_\text{eff}=\frac{\sqrt{2Z_0}\,\omega\,\chi^{(2)}_{yyy}}{2\pi\,c\,n_\text{eff}^{(\omega)}\sqrt{n_\text{eff}^{(2\omega)}A_\text{eff}}}.
\end{align}
Under the undepleted-pump approximation, the SH power after propagation over a distance $z$ is
\begin{align}
    P_\text{SH}(z)=|\kappa_\text{eff}|^2 z^2P_\text{FH}^2(z=0).
\end{align}
Thus, the normalized slope conversion efficiency of the waveguide, $\eta_0\equiv P_\text{SH}/\left(P_\text{FH}^2z^2\right)$, is
\begin{align}
    \eta_0=|\kappa_\text{eff}|^2.
\end{align}
Conversely, an experimental measurement of $\eta_0$ allows us to estimate the programmable nonlinearity $\chi^{(2)}_{yyy}$ of the device.

\subsection{$\chi^{(2)}$ Experimental Estimation Procedure}
To estimate the $\chi^{(2)}$ nonlinearity, we used a programmable straight waveguide with \(\mathrm{SiN}\) deposited by SVM. The width and height of the waveguide core are \SI{5}{\micro m} and \SI{2}{\micro m}, respectively, giving $A_\text{eff}=\SI{6.1}{\micro m^2}$ for the TM$_{00}$ modes at an FH wavelength of \SI{1580}{nm}. The slope conversion efficiency is estimated to be \SI{6.8e-2}{\%W^{-1}cm^{-2}} for $\chi^{(2)}_{yyy}=\SI{1}{pm/V}$.

In the experiment, we first measured the FH power at the output of the waveguide, $P_\text{FH,measured}$. To estimate the pump power at the location where SHG takes place, we applied the correction $P_\text{FH}=P_\text{FH,measured}/(r_\text{FH,collection}\,r_\text{FH,loss})$. Here, $r_\text{FH,collection}=0.75$ is the estimated efficiency of collecting light from the waveguide onto a free-space detector placed after an aspheric lens, and $r_\text{FH,loss}=\exp(-\alpha_\text{FH}L_\text{prop})$ accounts for the propagation loss in the waveguide. We estimated the distance between the SHG location and the output facet of the waveguide to be $L_\text{prop}=\SI{0.8}{cm}$, and the attenuation coefficient is $\alpha_\text{FH}=\SI{0.32}{dB/cm}$. Overall, we obtained $P_\text{FH}=\SI{15.3}{mW}$.

We then projected a QPM grating over a length $L_\text{QPM}=\SI{0.182}{cm}$. This is a small fraction of the total length of the programmable region, and the position of the section was chosen to maximize the SH output power. To the measured SH power, we applied corrections similar to those for the FH, but with a different attenuation coefficient, $\alpha_\text{SH}=\SI{1.35}{dB/cm}$. Overall, the peak on-chip SH power was estimated to be $P_\text{SH}=\SI{0.46}{nW}$, leading to a slope conversion efficiency of $\eta_0=\SI{5.8e-3}{\%W^{-1}cm^{-2}}$. Consequently, the programmable $\chi^{(2)}$ nonlinearity of the device is estimated to be $\chi^{(2)}_{yyy}=\SI{0.29}{pm/V}$.

\section{Continuous-Wave Frequency Mixing}
\label{sec:appendix CW-NLO}
\subsection{Sum Frequency Generation}
\label{subsec:vanilla SFG}

\begin{figure}
    \centering
    \includegraphics[width=1\linewidth]{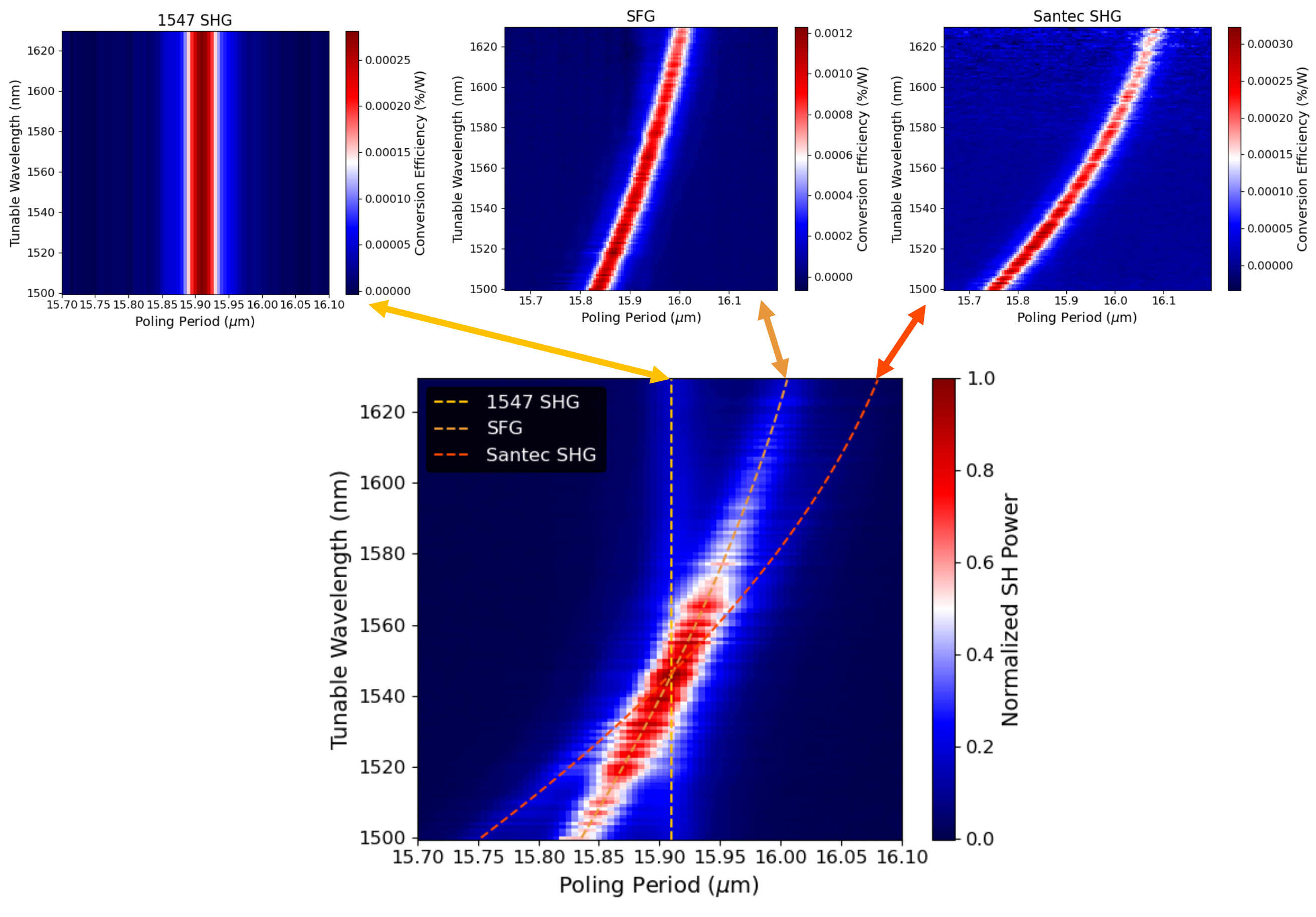}
    \caption{Normalized measured power when scanning the poling period, $\Lambda$, and the Santec tunable wavelength, $\lambda_{\text{santec}}$, when sending both $\lambda_{\text{santec}}$ and $\lambda_{1547}$ into the device.  Each curve is labeled and separately plotted above with the normalized conversion efficiency.}
    \label{fig:SFG Breakdown}
\end{figure}

This section presents experiments that demonstrate programmable, broadband CW second harmonic generation (SHG) and sum frequency generation (SFG) on our platform, allowing us to gain greater control over which frequencies mix in our device.  To perform these experiments, we insert a 50/50 fiber beam-splitter into the fiber path connecting the FH laser to our characterization setup.  One branch coming into the fiber beam splitter is our Santec tunable laser.  The other is a non-tunable source.  In Fig.~\ref{fig:CW-NLO}, the non-tunable laser is at \SI{1442.5}{nm}, while the results in Figure \ref{fig:SFG Breakdown} use a \SI{1547}{nm} non-tunable laser.  Both beams are polarized using fiber-polarization paddles to ensure TM-polarized fields are coupled to the waveguide. The result in Fig.~\ref{fig:CW-NLO} uses a shorter non-tunable wavelength to make dispersion characterization easier (as discussed in Sec.~\ref{sec:GVD_fitting}).  The results here use a \SI{1547}{nm} non-tunable pump to show how the SFG signal can be separated from the SHG signals.

To demonstrate SFG, we run three consecutive experiments.  First, we only couple our tunable laser into the waveguide. We then scan the wavelength of our tunable laser, $\lambda_{\text{santec}}$, and the poling period $\Lambda$ we project onto the device.  Next, we only couple the \SI{1547}{nm} laser, $\lambda_{\text{1547}}$, into the device and sweep $\Lambda$.  These scans allow us to characterize the SHG curves of the individual lasers as $P_{\text{SH}}(\Lambda, \lambda_{\text{santec}})$ and $P_{\text{SH}}(\Lambda, \lambda_{1547})$.  Lastly, we couple both the tunable and \SI{1547}{nm} lasers into our device and sweep $\Lambda$ and $\lambda_{\text{santec}}$, allowing us to characterize $P_{\text{SH}}(\Lambda, \lambda_{\text{santec}}, \lambda_{1547})$.  Unlike $P_{\text{SH}}(\Lambda, \lambda_{\text{santec}})$ and $P_{\text{SH}}(\Lambda, \lambda_{1547})$, $P_{\text{SH}}(\Lambda, \lambda_{\text{santec}}, \lambda_{1547})$ shows three peaks of SH power: two for the SHG of each individual laser source and one for SFG between the two laser sources.  We can confirm that the SFG peak is distinct by analyzing $P_{\text{SH}}(\Lambda, \lambda_{\text{santec}}, \lambda_{1547}) - P_{\text{SH}}(\Lambda, \lambda_{\text{santec}}) - P_{\text{SH}}(\Lambda, \lambda_{1547})$.  This lets us isolate the SFG peak by subtracting away the SHG peaks and confirm that it must come from the interaction of both lasers.  Fig.~\ref{fig:SFG Breakdown} shows the measured conversion efficiency for each SHG and SFG process using \(\mathrm{SiN}\) waveguides fabricated in-house. 

The SFG process in Fig.~\ref{fig:SFG Breakdown} shows $\sim$4 times greater conversion efficiency than the SHG processes of either laser.  This is physically explained by examining the second-order expansion term for the material polarization.  For an SFG process, the electric field is 
\begin{equation}
    E = E_1e^{-i\omega_1t} + E_2e^{-i \omega_2 t}
\end{equation}
When expanded in the second-order term of the polarization, we get
\begin{gather}
    P^{(2)} = \epsilon_0 \chi^{(2)}E^2 \\
    P^{(2)} = \epsilon_0 \chi^{(2)}(E_1^2e^{-2i\omega_1 t} + E_2^2e^{-2i\omega_2 t} + 2E_1E_2e^{-i(\omega_1 + \omega_2)t})
\end{gather}
if $E_1 = E_2 = E$, then
\begin{equation}
    P_{\omega_1 + \omega_2}^{(2)} = 2P_{2\omega_2}^{(2)}=2P_{2\omega_1}^{(2)}
\end{equation}
Because conversion efficiency is calculated by using the power of the SH, the conversion efficiency of the SFG process should be $\approx 4$ times larger for input electric fields with roughly balanced optical power.  The normalized conversion efficiency in Fig.~\ref{fig:SFG Breakdown} for each curve is calculated as 
\begin{gather}
    \eta_{\text{1547}} = \frac{{P}_{\text{SH,1547}}}{{P}_\text{FH,1547}^2} \times100 \% \\  
     \eta_{\text{santec}} = \frac{{P}_{\text{SH,santec}}}{{P}_{\text{FH,santec}}^2} \times100 \% \\   
      \eta_{\text{SFG}} = \frac{{P}_{\text{SH,SFG}}}{{P}_\text{FH,1547} \times {P}_{\text{FH,santec}}} \times100 \%
\end{gather}
where the powers of each FH frequency mode and each SH frequency mixing process are separately characterized.

Lastly, the normalized power plot in Fig.~\ref{fig:SFG Breakdown} shows the generated power decreasing as $\lambda_{\text{santec}}$ is tuned away from 1547 nm.  This is because the free-space optics were optimized for maximizing the amount of 1547 nm light coupled into the waveguide for each laser.  Because of dispersion in the coupling objectives, as the wavelength is tuned away from this value, the focal spot drifts and coupling efficiency goes down.

\subsection{Dispersion Characterization Via Sum Frequency Generation}\label{sec:GVD_fitting}

To characterize the local dispersion environment of different \(\mathrm{SiN}\) waveguide modes, we utilized the real-time reconfigurability of the programmable $\chi^{(2)}$ pattern as an \textit{in situ} momentum spectrometer. By sweeping both the optical pump frequencies and the programmable QPM grating period $\Lambda$, we directly mapped the wavevector mismatch landscape. This allowed us to extract two dispersion parameters that characterize pulse propagation in waveguides: group velocity mismatch (GVM) and group velocity dispersion (GVD).  GVM describes the walk-off between pulses with different central frequencies. GVD describes how fast the modal group velocity changes with frequency, which leads to different rates of pulse broadening.  Mathematically, they are defined as
\begin{gather}
    \text{GVM} = \left.\frac{dk_{\text{SH}}}{d\omega}\right|_{2\omega_0} - \left.\frac{dk_{\text{FH}}}{d\omega}\right|_{\omega_0} = \frac{1}{v_{g,\text{SH}}} - \frac{1}{v_{g,\text{FH}}},
    \label{eq:GVM def} \\
    \text{GVD} = \frac{d^2k}{d\omega^2},
    \label{eq:GVD def}
\end{gather}
where $k(\omega)$ is the propagation constant of the relevant mode and $v_g$ its group velocity. With these definitions the GVM is expressed in fs/mm and the GVD in fs$^2$/mm.

Similar to the experiment described in Sec.~\ref{subsec:vanilla SFG}, the waveguide was pumped with two CW lasers: a fixed-wavelength laser at $\lambda_f = \SI{1442.5}{nm}$ (angular frequency $\omega_f$) and a tunable laser swept from \SIrange{1500}{1630}{nm} in \SI{1}{nm} steps (angular frequency $\omega_t$). At each tunable wavelength, the programmed QPM period was scanned, and the generated signal was recorded with a PMT using lock-in detection. This was done for 2 ranges of poling periods: (1) the range of poling periods that phase matched the $\text{TM}_{00} + \text{TM}_{00} \rightarrow \text{TM}_{00}$ SFG and SHG processes and (2) the range of poling periods that phase matched the $\text{TM}_{00} + \text{TM}_{00} \rightarrow \text{TM}_{20}$ SFG and SHG processes. 

Now, because the PMT records the total generated power, each map (corresponding to one of the processes mentioned above) contains three processes simultaneously: SHG of the fixed pump ($\omega_f+\omega_f$), SFG between the two pumps ($\omega_t+\omega_f$), and SHG of the tunable pump ($\omega_t+\omega_t$). Each appears as a distinct phase-matching ridge in the $(\Lambda,\lambda_t)$ plane (Fig.~\ref{fig:CW-NLO}d): the fixed-pump SHG is a vertical line independent of $\lambda_t$, while the SFG and tunable-pump SHG form curved ridges whose slopes differ by approximately a factor of two. As opposed to the experiment done in Fig.~\ref{fig:SFG Breakdown}, placing the fixed pump outside the tunable band keeps the three ridges separated in period across the whole scan. 

Each ridge is formed where the grating momentum $K = 2\pi/\Lambda$ (first-order QPM) compensates the phase mismatch of the corresponding process,
\begin{equation}
    K(\omega_1,\omega_2) = k_{\text{SH}}(\omega_1+\omega_2) - k_{\text{FH}}(\omega_1) - k_{\text{FH}}(\omega_2).
    \label{eq:K general}
\end{equation}
Now, because the FH and SH bands are quite distant in terms of frequency, a single Taylor expansion cannot describe both. As a result, we expanded $k_{\text{FH}}$ about a reference frequency $\omega_0 = 2\pi c/\lambda_0$ with $\lambda_0 = \SI{1550}{nm}$ and $k_{\text{SH}} = 2\omega_0$ (\SI{775}{nm}). Writing $\Omega_i = \omega_i - \omega_0$ and retaining terms to second order gives us
\begin{equation}
    K(\Omega_1,\Omega_2) \approx \Delta k_0 + \text{GVM}\,(\Omega_1+\Omega_2) + \tfrac{1}{2}\,\text{GVD}_{\text{SH}}\,(\Omega_1+\Omega_2)^2 - \tfrac{1}{2}\,\text{GVD}_{\text{FH}}\,(\Omega_1^2+\Omega_2^2),
    \label{eq:K taylor}
\end{equation}
where $\Delta k_0 = k_{\text{SH}}(2\omega_0) - 2k_{\text{FH}}(\omega_0)$ is the phase mismatch of degenerate SHG at $\lambda_0$, $\text{GVD}_{\text{FH}}$ is evaluated at \SI{1550}{nm} and $\text{GVD}_{\text{SH}}$ at \SI{775}{nm}. For the tunable-pump SHG ($\Omega_1=\Omega_2=\Omega_t$) process, we see that
\begin{equation}
    K_{\text{SHG}}(\Omega_t) = \Delta k_0 + 2\,\text{GVM}\,\Omega_t + \tfrac{1}{2}A\,\Omega_t^2
    \label{eq:K SHG}
\end{equation}
where $A = 4\,\text{GVD}_{\text{SH}} - 2\,\text{GVD}_{\text{FH}}$. Similarly, for the SFG ($\Omega_1=\Omega_t$, $\Omega_2=\Omega_f$) process, we see that
\begin{equation}
    K_{\text{SFG}}(\Omega_t) = K_{\text{SFG},0} + \left(\text{GVM} + \text{GVD}_{\text{SH}}\,\Omega_f\right)\Omega_t + \tfrac{1}{2}B\,\Omega_t^2
    \label{eq:K SFG}
\end{equation}
where $B = \text{GVD}_{\text{SH}} - \text{GVD}_{\text{FH}}$ and $K_{\text{SFG},0} = \Delta k_0 + \text{GVM}\,\Omega_f + \tfrac{1}{2}B\,\Omega_f^2$. The fixed-pump SHG line sits at $K_{\text{SHG}}(\Omega_f)$. Note that the linear coefficient of the SFG ridge is the GVM only when the fixed pump coincides with the reference frequency; otherwise it acquires the correction $\text{GVD}_{\text{SH}}\,\Omega_f$.

The curvatures $A$ and $B$ are linearly independent combinations of the two GVDs, which is what allows them to be separated. This is why the SFG ridge is essential. On the degenerate SHG branch, the FH and SH dispersion enter at every order only through a single combination (the coefficient $A$ at second order), so an SHG-only measurement cannot distinguish $\text{GVD}_{\text{FH}}$ from $\text{GVD}_{\text{SH}}$: any change in one can be absorbed by a compensating change in the other. The fixed-pump SHG line does not lift this degeneracy, since it is the same function evaluated at $\Omega_f$. Its role is instead to anchor the model at a frequency outside the tunable band, constraining $\Delta k_0$ and the GVM. It should be noted that only differences between propagation constants are accessible in this way; the group index of each mode individually, and the absolute effective indices, are not.

The peak ridge positions were extracted by fitting the QPM response in each row of the map. For a grating with longitudinal profile $g(z)$, the response along the period axis is $\propto \left|\int g(z)\,e^{i(K_{\text{PM}} - K)z}\,dz\right|^2$, which for a uniform grating of length $L$ reduces to $\mathrm{sinc}^2[(K_{\text{PM}}-K)L/2]$. Because this is a function of $K$ rather than $\Lambda$, all fitting was performed in $K$. The measured lines were observed to be broader than the ideal $\mathrm{sinc}^2$ of the nominal projected grating length ($\approx$ \SI{7.24}{mm}) and had suppressed side lobes, indicating a non-uniform envelope along $z$ set by the programming-illumination profile and the bias-field distribution. Each ridge was therefore fitted with the transfer function of a super-Gaussian envelope, whose shape was calibrated from that ridge's own composite lineshape (this was obtained by aligning all rows on their approximate centers and averaging). The per-row fit was restricted to the main lobe of the response, with the amplitude, center and a constant background as free parameters. This yielded median center uncertainties of \SIrange{0.1}{0.2}{nm} of period for the TM$_{00}$ + TM$_{00}$ $\rightarrow$ TM$_{00}$ process and \SIrange{0.6}{0.8}{nm} for the TM$_{00}$ + TM$_{00}$ $\rightarrow$ TM$_{20}$ process, well below the respective scan steps of \SI{2}{nm} and \SI{5}{nm}. Wavelength rows between \SI{1556}{nm} and \SI{1582}{nm}, which showed an elevated baseline present at the same wavelengths in every dataset and attributable to modal phasematching, were excluded, as were individual ridge centers deviating by more than $3\sigma$ from the global fit, which arise from crossings with other phase-matching features.

Rather than fitting each ridge with an independent parabola and algebraically inverting Eqs.~\eqref{eq:K SHG} and \eqref{eq:K SFG}, all ridge centers from all three branches were fitted simultaneously to Eq.~\eqref{eq:K taylor} by weighted least squares, with $\{\Delta k_0, \text{GVM}, \text{GVD}_{\text{FH}}, \text{GVD}_{\text{SH}}\}$ as shared parameters. Parameter uncertainties were obtained from the fit covariance and inflated by $\sqrt{\chi^2_\nu}$ (the reduced chi-squared error of the fit) to account for imperfections in the lineshape model. Including third-order dispersion shifts $\text{GVD}_{\text{FH}}$ by approximately \SI{6}{fs^2/mm}, but the two processes disagreed on the third-order coefficient, which indicated residual systematic effects at that order rather than a physical difference. We therefore report results to second order.

As mentioned earlier, we applied this procedure to two mode combinations of the multimode channel waveguide, selected by the programmed grating period: the fundamental process $\text{TM}_{00} + \text{TM}_{00} \rightarrow \text{TM}_{00}$ (periods \SIrange{15.40}{16.10}{\micro m}, \SI{2}{nm} steps) and the higher-order process $\text{TM}_{00} + \text{TM}_{00} \rightarrow \text{TM}_{20}$ (periods \SIrange{18.70}{20.30}{\micro m}, \SI{5}{nm} steps). For the TM$_{20}$ process, the full map additionally contained phase-matching features from other mode combinations. The SFG ridge, which is the brightest feature, and the fixed-pump SHG line at \SI{18.837}{\micro m} were extracted from this map, while the tunable-pump SHG ridge was taken from a separate scan with only the tunable pump present (\SIrange{19.35}{20.395}{\micro m}). The tunable-pump SHG ridge appears in both TM$_{20}$ scans, and its positions agree with no systematic offset (mean difference \SI{2.5}{nm} of period), confirming that the two scans share a common period axis. Both fits were performed jointly, with separate harmonic-band parameters for each process.

The extracted parameters are summarized in Tab.~\ref{tab:dispersion} and Tab.~\ref{tab:dispersion_main}. For the $\text{TM}_{00}$ SH mode, we find $\Delta k_0 = \SI{394.5}{rad/mm}$, corresponding to phase-matched degenerate SHG at $\lambda_0$ for $\Lambda = \SI{15.925}{\micro m}$, a GVM of \SI{44.43 \pm 0.05}{fs/mm} and $\text{GVD}_{\text{SH}} = \SI{135.3 \pm 0.4}{fs^2/mm}$. For the $\text{TM}_{20}$ SH mode, $\Delta k_0 = \SI{316.4}{rad/mm}$ ($\Lambda = \SI{19.860}{\micro m}$), the GVM is \SI{80.40 \pm 0.16}{fs/mm} and $\text{GVD}_{\text{SH}} = \SI{103.5 \pm 3.5}{fs^2/mm}$. The fundamental $\text{TM}_{00}$ mode has anomalous dispersion at \SI{1550}{nm}, $\text{GVD}_{\text{FH}} = \SI{-104.1 \pm 0.9}{fs^2/mm}$.

Because both processes share the same $\text{TM}_{00}$ FH mode, the value of $\text{GVD}_{\text{FH}}$ obtained from each dataset is an independent determination of the same physical quantity. Fitting the two datasets with independent FH dispersion yields $\text{GVD}_{\text{FH}} = \SI{-104.1 \pm 0.9}{fs^2/mm}$ from the $\text{TM}_{00}$ process and \SI{-104.2 \pm 5.1}{fs^2/mm} from the $\text{TM}_{20}$ process, a difference of \SI{-0.1 \pm 5.1}{fs^2/mm}. An F-test comparing the shared and independent fits found no evidence for distinct values ($p = 0.95$). From these results, we argue that the programmable QPM can be utilized as an intrinsic, on-chip platform for mode-resolved dispersion metrology, including in the SH band near \SI{775}{nm}, which is not accessible to interferometric measurements in the telecom band.

We'd like to point out the following:
\begin{itemize}
    \item All quoted uncertainties are statistical. The absolute values are additionally subject to the calibration of the projected grating period, set by the SLM pixel pitch and the demagnification of the programming optics: a fractional error in the period changes $\Delta k_0$, the GVM and both GVDs by the same fractional amount.
    \item The fixed-pump wavelength was manually input as opposed to being a fitted quantity; an error of \SI{1}{nm} in $\lambda_f$ shifts $\text{GVD}_{\text{FH}}$ by approximately \SI{12}{fs^2/mm} and $\text{GVD}_{\text{SH}}$ of the $\text{TM}_{00}$ mode by approximately \SI{2.4}{fs^2/mm}, while changing the GVM by less than \SI{0.1}{fs/mm}.
\end{itemize}  
The data and analysis code are available at \url{https://doi.org/10.5281/zenodo.23072388}.

\begin{table}[h]
\centering
\small
\caption{Dispersion parameters of the straight programmable nonlinear waveguide extracted by programmable SFG and SHG. FH and SH denote the fundamental harmonic at \SI{1550}{nm} and the second harmonic at \SI{775}{nm}. Uncertainties are statistical (one standard deviation) and exclude the period-axis calibration.}
\label{tab:dispersion}
\begin{tabular}{lcc}
\hline\hline
 & $\text{TM}_{00}$(FH) $\leftrightarrow$ $\text{TM}_{00}$(SH) & $\text{TM}_{00}$(FH) $\leftrightarrow$ $\text{TM}_{20}$(SH) \\
\hline
Phase mismatch $\Delta k_0$ (rad/mm) & 394.5 & 316.4 \\
Degenerate-SHG period at \SI{1550}{nm} (\si{\micro m}) & 15.925 & 19.860 \\
GVM (fs/mm) & $44.43 \pm 0.05$ & $80.40 \pm 0.16$ \\
GVD of the SH mode (fs$^2$/mm) & $135.3 \pm 0.4$ & $103.5 \pm 3.5$ \\
GVD of $\text{TM}_{00}$(FH) (fs$^2$/mm), per dataset & $-104.1 \pm 0.9$ & $-104.2 \pm 5.1$ \\
GVD of $\text{TM}_{00}$(FH) (fs$^2$/mm), joint fit & \multicolumn{2}{c}{$-104.1 \pm 0.9$} \\
\hline\hline
\end{tabular}
\end{table}

\subsection{Difference-Frequency Generation}
\label{subsec:DFG}
To characterize difference frequency generation (DFG) in the programmable nonlinear waveguide, we employed a dynamically tuned detection setup to isolate a difference frequency beam. The process was driven by a fixed-wavelength CW pump laser at $\sim$ 802 nm (Thorlabs FPL808P) and a tunable CW seed laser swept from 1615 nm to 1625 nm. Through the DFG process ($\omega_{\text{DFG}} = \omega_{\text{pump}} - \omega_{\text{seed}}$), this configuration generates an optical idler in the $\sim$ 1588–1598 nm band. Prior to measurement, optical filtering was used to isolate the DFG signal from the driving fields. The output light was passed through a dichroic mirror to reject the 802 nm pump, followed by a bandpass filter centered at 1590 nm ($\pm$ 12 nm) to reject the tunable seed laser and background noise.  This bandpass filter constrained the range of DFG wavelengths we could experimentally measure, though there is no fundamental limit to the range of DFG wavelengths that can be produced on our devices. The isolated DFG signal was then coupled into a multimode fiber (MMF).

\begin{figure}
    \centering
    \includegraphics[width=1\linewidth]{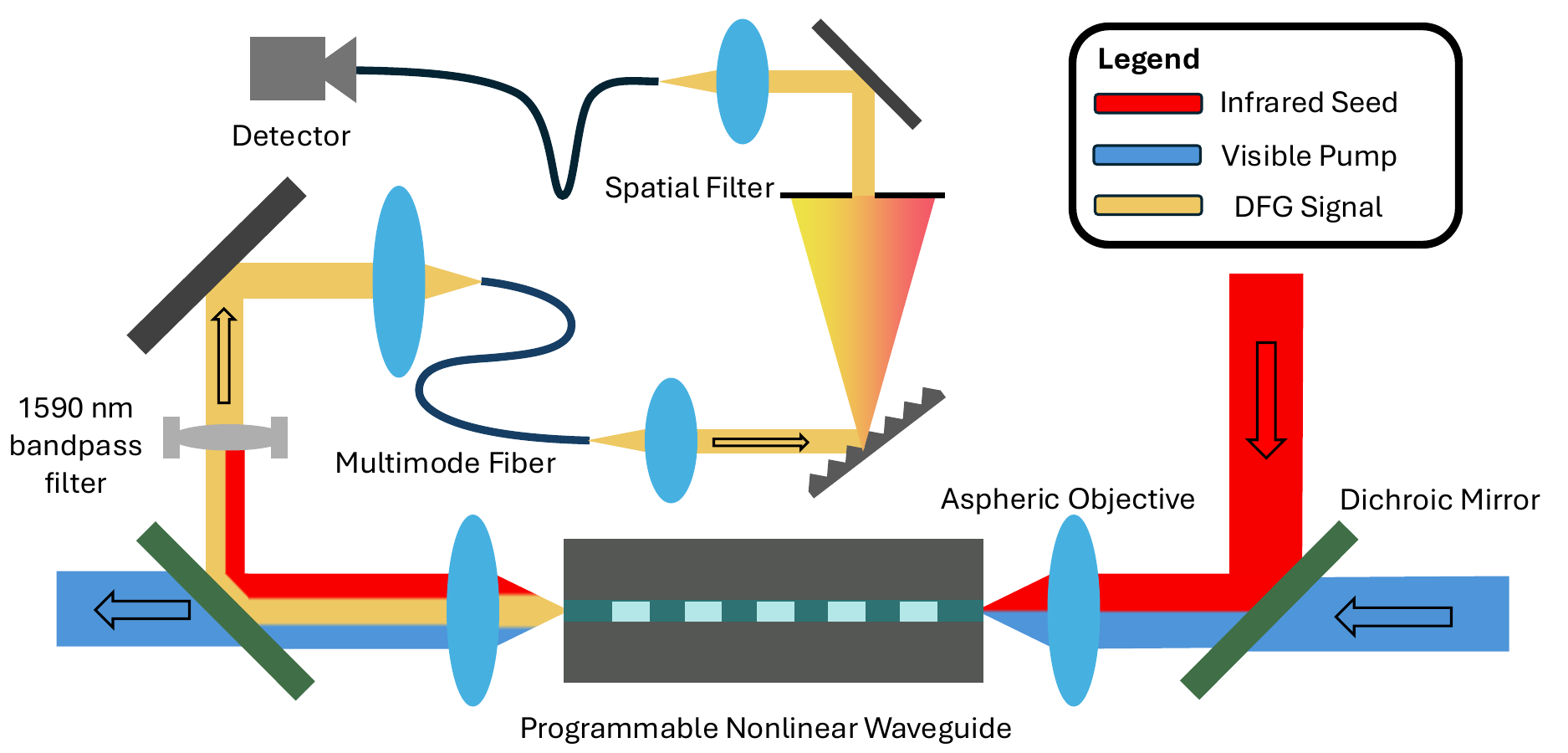}
    \caption{Labeled cartoon diagram of the optical apparatus used to measure the DFG signal generated in the programmable nonlinear waveguide device.}
    \label{fig:DFG Experiment}
\end{figure}

To map the phase-matching landscape (see Fig.~\ref{fig:CW-NLO}), we performed a two-dimensional scan, simultaneously sweeping the tunable seed wavelength and the spatial period of the programmable QPM grating. Because the 1590 nm bandpass filter was insufficient to completely eliminate the tunable seed laser, further spectral isolation was required. Therefore, the MMF output was directed onto a diffraction grating mounted on a motorized rotation stage to spatially separate the remaining seed light from the generated DFG signal. Using a pre-calibrated linear map correlating the physical grating angle to the optical wavelength, the stage was dynamically rotated to steer the diffracted idler beam into a near-infrared single-photon avalanche diode (SPAD; ID Qube Series NIR Free-Running). This active tracking maintained continuous detector coupling throughout the 2D sweep while simultaneously providing an accurate measurement of the generated optical idler wavelength. Fig.~\ref{fig:DFG Experiment} shows a labeled cartoon of the optical setup used to measure DFG on the programmable nonlinear waveguide device.

To extract the conversion efficiency from the 2D SPAD data, we first subtracted the measured detector dark counts. To translate the photon count rate into generated optical power, we used an experimentally measured SPAD quantum efficiency and apparatus collection efficiency. The DFG conversion efficiency was calculated by normalizing the generated power ($P_{\text{DFG}}$) by the product of the incident pump ($P_{\text{pump}}$) and seed ($P_{\text{seed}}$) powers as 
\begin{equation}
    \eta_{\text{DFG}} = \frac{P_{\text{DFG}}}{P_{\text{pump}} P_{\text{seed}}} \times 100 \%
\end{equation}

The 2D DFG data presented in Fig.~\ref{fig:CW-NLO} shows minimal variation in the optimal poling period as the seed wavelength changes.  While the slope of the SFG and SHG plots as a function of the tunable laser depends on the first-order expansion of $k(\delta\omega)$, DFG relies on the second-order expansion of $k(\delta\omega)$.  Because of energy conservation, as the FH seed's frequency is changed by $\delta \omega$, the DFG signal changes by $-\delta\omega$.  This means the momentum mismatch, $\Delta k(\delta \omega)$, changes as
\begin{equation}
    \label{eq:delta_k first exp}
    \Delta k(\delta \omega) = k(\omega_{\text{SH}}) -k(\omega_{\text{FH}} - \delta\omega) - k(\omega_{\text{FH}}+\delta\omega)
\end{equation}
The perturbed wavevector $k(\omega_{\text{FH}}+\delta\omega)$ can be Taylor expanded as 
\begin{equation}
\label{eq:k taylor}
    k(\omega_{\text{FH}}+\delta\omega) = k(\omega_{\text{FH}}) + \beta_1\delta\omega+ \frac{\beta_2}{2}(\delta\omega)^2 \dots
\end{equation}
Plugging Eq.~\eqref{eq:k taylor} into Eq.~\eqref{eq:delta_k first exp} gives
\begin{equation}
    \Delta k(\delta \omega) = k(\omega_{\text{SH}}) - 2k(\omega_{\text{FH}}) - \beta_1\delta\omega +\beta_1\delta\omega-\beta_2(\delta\omega)^2
\end{equation}
when subtracting away $\Delta k_0 = k(\omega_{\text{SH}}) - 2k(\omega_{\text{FH}})$, which has no dependence on $\delta\omega$, we see
\begin{equation}
\label{eq:second-order deltak final}
    \Delta k(\delta \omega) - \Delta k_0 = -\beta_2(\delta\omega)^2
\end{equation}
when neglecting higher order expansion terms in the Taylor series.  From Eq.~\ref{eq:second-order deltak final}, the lowest order dispersion parameter that shifts the ideal poling period for a DFG process is related to the second-order term.  As Sec.~\ref{sec:GVD_fitting} discusses, the second-order dispersion terms for our waveguides are small, so we expect to see minimal variation in the ideal poling period as we scan the pump wavelength over a limited bandwidth.

\subsection{Polarization Experiments}
\label{sec:Polarization Experiments}
This section details experiments that change which field polarizations mix to produce SHG. Mathematically, the nonlinear material polarization is described as 
\begin{equation}
P^{(2)}_i = \epsilon_0 \sum_{j,k} \chi^{(2)}_{ijk} E_j E_k
\end{equation}
where $i$ denotes the basis vector for the polarization, and $j$ and $k$ denote the basis vectors for the input electric fields.  Because the electric field bias in our device is in the $y$ direction, the inversion symmetry of the \(\mathrm{SiN}\) waveguide is broken along that axis.  With this broken inversion symmetry, there are three possible polarization mixing processes: type-0 phase-matching interacts \(\mathrm{TM}\) + \(\mathrm{TM}\) $\xrightarrow{}$ \(\mathrm{TM}\), type-I phase-matching interacts \(\mathrm{TE}\) + \(\mathrm{TE}\) $\xrightarrow{}$ \(\mathrm{TM}\), and type-II phase-matching interacts \(\mathrm{TM}\) + \(\mathrm{TE}\) $\xrightarrow{}$ \(\mathrm{TE}\).  To characterize the relative strength of the $\chi^{(2)}_{ijk}$ components, we use a half-wave plate in front of the input optical beam to select which polarization modes we excite.

Each process has a distinct poling period as the dispersion of the \(\mathrm{TM}\) and \(\mathrm{TE}\) modes varies.  We first set the input polarization to the setting for type-0, type-I, and type-II, respectively, and scanned the poling period to find the point that provided the maximum conversion efficiency. This provided us with three distinct poling periods, $\Lambda_{0}$, $\Lambda_{1}$, and $\Lambda_{2}$ corresponding to each phase-matching process.  We then rotated the half-wave plate angle, $\theta$, and measured the second-harmonic power, which is presented in Fig.~\ref{fig:CW-NLO}.  As a function of the polarization angle $2\theta$, the SH power from the type-0 process varies as $P_{\text{SH}} \propto \operatorname{cos} ^4(2\theta)$, the type-I process varies as $P_{\text{SH}} \propto \operatorname{sin} ^4(2\theta)$, and the type-II process varies as $P_{\text{SH}} \propto \operatorname{cos} ^2(2\theta) \operatorname{sin} ^2(2\theta)$.  These relations can be understood in terms of the expression for $P^{(2)}$ and the polarization of the electric field.  If the optical field is propagating along the z-axis, then we define 
\begin{gather}
    E_{\text{y}} = |E| \operatorname{cos}(2\theta) \\
    E_{\text{x}} = |E| \operatorname{sin}(2\theta)
\end{gather}
For a type-0 process, the two interacting FH fields are $E_{\text{y}}$ and $E_{\text{y}}$, so $P^{(2)} \propto \operatorname{cos}(2\theta)^2$ and $P_{\text{SH}} \propto P^{(2)^2} \propto \operatorname{cos}^4(2\theta)$.  For a type-I process, the two interacting FH fields are $E_{\text{x}}$ and $E_{\text{x}}$, so $P^{(2)} \propto \operatorname{sin}(2\theta)^2$ and $P_{\text{SH}} \propto \operatorname{sin}^4(2\theta)$. Lastly, for a type-II process, the two interacting FH fields are $E_{\text{x}}$ and $E_{\text{y}}$, so $P^{(2)} \propto \operatorname{cos}(2\theta) \operatorname{sin}(2\theta)$ and $P_{\text{SH}} \propto \operatorname{cos}^2(2\theta)\operatorname{sin}^2(2\theta)$.  

Additionally, because $P_{\text{SH}} \propto \chi^{(2)^2}$, we can analyze the relative values of the different tensor components of $\chi^{(2)}$ by analyzing the output SH power in Fig.~\ref{fig:CW-NLO}.  From these measurements, we conclude that 
\begin{equation}
    \lvert \chi^{(2)}_{yxx} \rvert /\lvert \chi^{(2)}_{yyy} \rvert = 0.36
\end{equation}
and 
\begin{equation}
    \lvert \chi^{(2)}_{xyx} \rvert /\lvert \chi^{(2)}_{yyy} \rvert = 0.32
\end{equation}
Under Kleinman symmetry, permutations of the $\chi^{(2)}$ coordinate axes should not change the magnitude of $\chi^{(2)}$.  Our measurements conform with that principle within the error of our experimental methods. This analysis ignores any differences in spatial overlap between the \(\mathrm{TM_{00}}\) and the \(\mathrm{TE_{00}}\) modes, as these are assumed to be negligible compared to the differences in the $\chi^{(2)}$ tensor components.

\section{Poling Wavy Waveguides}
\label{sec:Wavy Waveguides}
\begin{figure}
    \centering
    \includegraphics[width=1\linewidth]{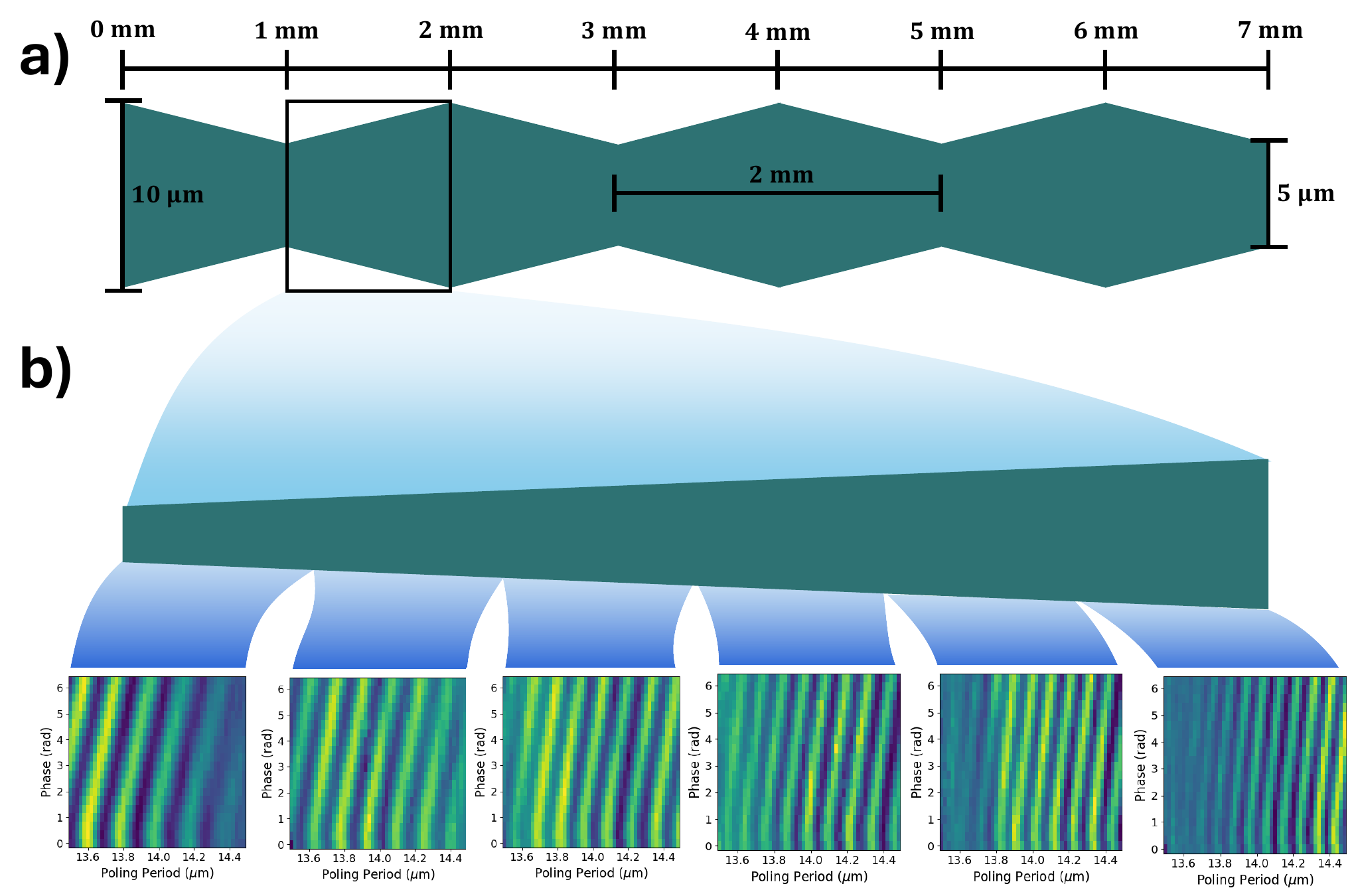}
    \caption{a) Cartoon diagram of the waveguide used in the main experiment, where the maximum and minimum widths of the waveguide are 10 \(\mathrm{\mu m}\) and 5 \(\mathrm{\mu m}\), respectively, and the device linearly changes between these two widths at a geometric period of \SI{2}{mm}.  The total interaction length on this waveguide is \SI{7}{mm}.  b) Poling period and phase sweeps along a section of the waveguide with an adiabatically increasing waveguide width.  The position of each partition along the waveguide is labeled, and the colorbar is associated with the normalized output SHG power from the indicated partition.}
    \label{fig:Adiabatic Interference}
\end{figure}

This section discusses the experimental results and procedures for phase-matching waveguides whose width changes periodically along the axis of propagation (called wavy waveguides in this text).  Previous demonstrations using straight waveguides should, in principle, have the same poling period along the axis of propagation because the dispersion does not change.  While spiral geometries have a spatially varying poling period as the radius of curvature changes, this effect is hard to discern because of misalignment in the programming illumination (as discussed in Sec.~\ref{subsec:spiral phase-mathcing}).  Using waveguides with large width variation allows us to clearly demonstrate the adaptability of our $\chi^{(2)}$ structure to complex underlying waveguide geometries. For this study, we examined waveguides whose widths linearly increased and decreased between 5 \si{\micro \meter} and 10 \si{\micro \meter} over a geometric period of \SI{1}{mm} and \SI{2}{mm}.  To optimize the $\chi^{(2)}$ structure, we split the waveguide into many partitions and sweep the poling period and phase on each partition to maximize the coherent constructive interference between the SHG produced in each partition (see Alg.~\ref{alg:insitu_opt_wavy}).  Fig.~\ref{fig:Adiabatic Interference} shows the normalized SHG interference as we swept the phase and poling period along a selection of partitions of a waveguide with a geometric period of \SI{2}{mm}.

\begin{algorithm}[htbp]
\caption{\textit{In situ} QPM optimization for the wavy waveguides}
\label{alg:insitu_opt_wavy}
\begin{algorithmic}[1]
\Require 
    \Statex $N$: Number of partitioned waveguide sections (e.g., $32$)
    \Statex $\mathbf{P}$: Set of test periods (e.g., $[13.50, 13.52, \dots, 14.50]\,\mu\mathrm{m}$)
    \Statex $\mathbf{\Theta}$: Set of test phases (e.g., $20$ steps from $0$ to $2\pi$)
    \Statex $\lambda_{\text{pump}}$: Target CW pump wavelength (e.g., $1550\,\mathrm{nm}$)
    \Statex $\vec{\Lambda}_{\text{default}}$: Initial array of uniform QPM periods
    \Statex $\textproc{Project\_QPM\_Grating}(\vec{\Lambda}, \vec{\Phi})$: Function to project the grating onto the waveguide
\Ensure 
    \Statex $\vec{\Lambda}_{\text{opt}}$, $\vec{\Phi}_{\text{opt}}$: Arrays of optimized local periods and phases
\vspace{0.2cm}
\State Initialize $\vec{\Lambda}_{\text{opt}} \gets \vec{\Lambda}_{\text{default}}$ \Comment{Default uniform period array}
\State Initialize $\vec{\Phi}_{\text{opt}} \gets \vec{0}$
\State Set CW laser input wavelength to $\lambda_{\text{pump}}$

\For{$j = 1$ \textbf{to} $N$} \Comment{Iterate through sections sequentially}
    \State $P_{\text{max}} \gets 0$
    \State $\Lambda_{\text{best}} \gets \vec{\Lambda}_{\text{opt}}[j]$
    \State $\Phi_{\text{best}} \gets 0$
    
    \For{\textbf{each} $\Lambda \in \mathbf{P}$} \Comment{Sweep local period}
        \For{\textbf{each} $\Phi \in \mathbf{\Theta}$} \Comment{Sweep local phase}
            
            \State $\vec{\Lambda}_{\text{opt}}[j] \gets \Lambda$
            \State $\vec{\Phi}_{\text{opt}}[j] \gets \Phi$
            
            \State $\textproc{Project\_QPM\_Grating}(\vec{\Lambda}_{\text{opt}}, \vec{\Phi}_{\text{opt}})$
            \State $P_{\text{SHG}} \gets \textproc{Measure\_Total\_SHG\_Power}()$
            
            \If{$P_{\text{SHG}} > P_{\text{max}}$}
                \State $P_{\text{max}} \gets P_{\text{SHG}}$
                \State $\Lambda_{\text{best}} \gets \Lambda$
                \State $\Phi_{\text{best}} \gets \Phi$
            \EndIf
            
        \EndFor
    \EndFor
    
    \State $\vec{\Lambda}_{\text{opt}}[j] \gets \Lambda_{\text{best}}$ \Comment{Lock in optimal parameters}
    \State $\vec{\Phi}_{\text{opt}}[j] \gets \Phi_{\text{best}}$ \Comment{for the current section}
    
\EndFor

\State \Return $\vec{\Lambda}_{\text{opt}}, \vec{\Phi}_{\text{opt}}$
\end{algorithmic}
\end{algorithm}

\begin{figure}
    \centering
    \includegraphics[width=1\linewidth]{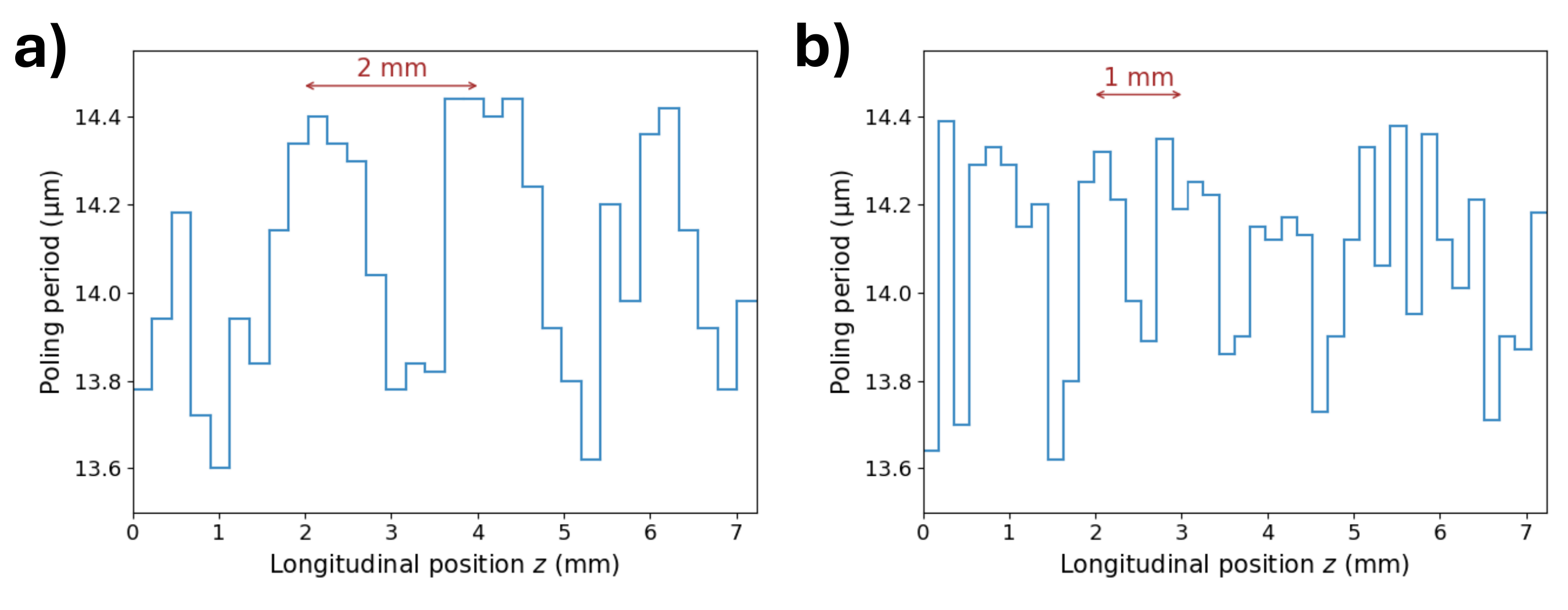}
    \caption{Optimal poling period versus longitudinal position on a wavy waveguide.  Panel a) shows the results for a waveguide with a geometric period of \SI{2}{mm} and panel b) shows the results for a waveguide with a geometric period of \SI{1}{mm}.}
    \label{fig:Adiabatic Position vs PP}
\end{figure}
While Fig.~\ref{fig:Adiabatic Interference} shows the poling period and phase interference optimization curves for a waveguide geometric period of \SI{2}{mm}, the same procedure was used on waveguides with a geometric period of \SI{1}{mm} as well.  Fig.~\ref{fig:Adiabatic Position vs PP} shows the optimal poling period vs spatial position for waveguides with \SI{2}{mm} and \SI{1}{mm} geometric periods.  As is shown, the spatial period of the optimized poling period tracks the underlying geometric period of the waveguide.  This is because, as Fig.~\ref{fig:Dispersion Properties} in Sec.~\ref{subsec:dispersion properties} shows, the poling period needed to phase-match a nonlinear process depends on the waveguide geometry.  Consequently, a changing underlying geometry means the ideal poling period should change along with it, which is what we observe.

\begin{figure}
    \centering
    \includegraphics[width=1\linewidth]{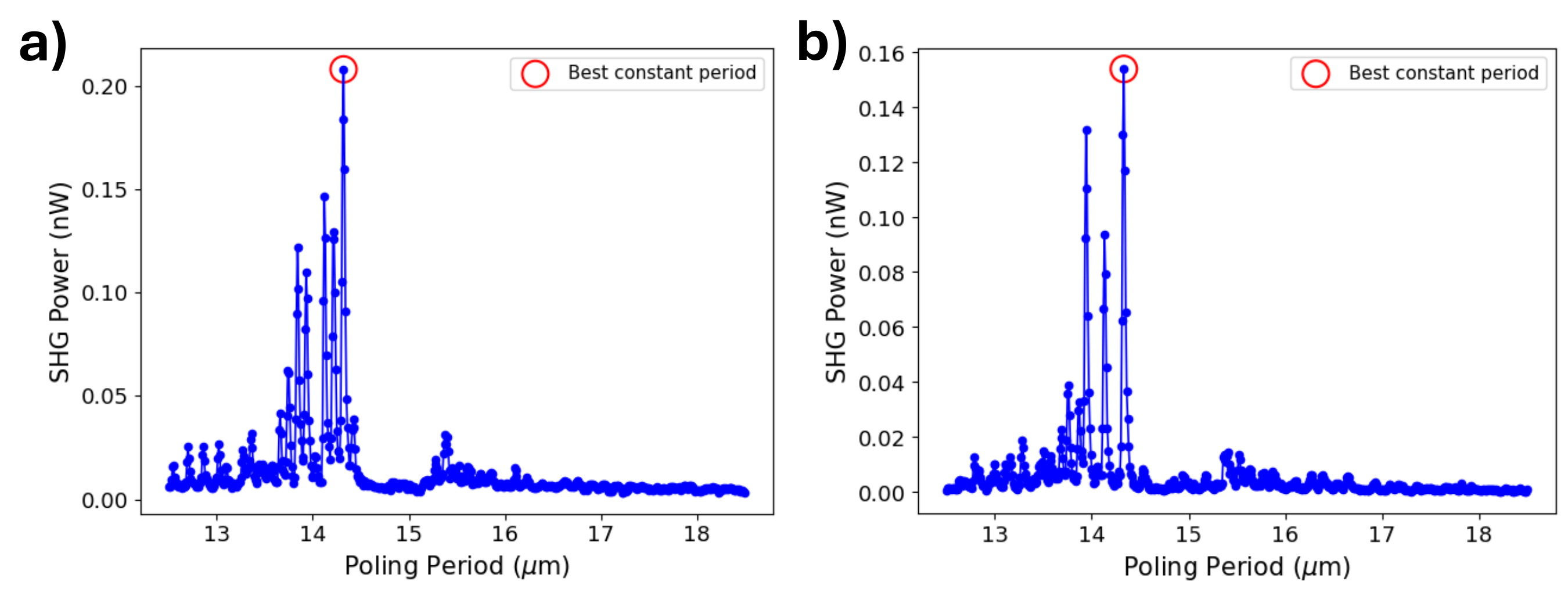}
    \caption{Generated SHG power for a $\chi^{(2)}$ pattern with a constant poling period as the poling period is swept.  a) Shows the result for a waveguide with a geometric period of \SI{2}{mm} and b) shows the result for a waveguide with a geometric period of \SI{1}{mm}.  The constant poling period that maximized SHG power was then used as a baseline, unoptimized response in Fig.~\ref{fig:Adiabatic Transfer Function}.}
    \label{fig:Wavy_waveguide_poling_scan}
\end{figure}
\begin{figure}
    \centering
    \includegraphics[width=1\linewidth]{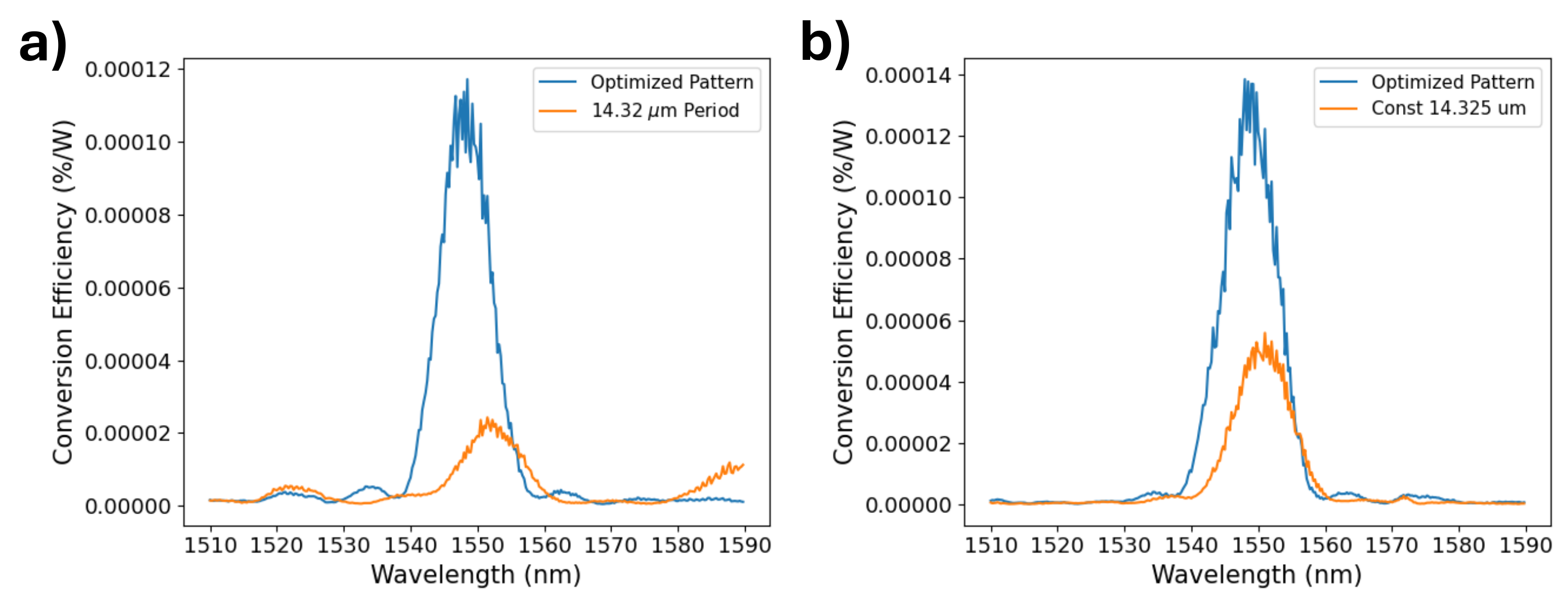}
    \caption{Transfer functions from $\chi^{(2)}$ patterns with a constant poling period and spatially optimized structure.  Panel a) shows the results for wavy waveguides with a geometric period of \SI{2}{mm} and panel b) shows the results for structures with a geometric period of \SI{1}{mm}.  The $\chi^{(2)}$ pattern with a constant period used the poling period that maximized the SHG signal, as shown in Fig.~\ref{fig:Wavy_waveguide_poling_scan} for a waveguide with a \SI{2}{mm} geometric period.}
    \label{fig:Adiabatic Transfer Function}
\end{figure}

To see the enhanced nonlinear performance of the optimized $\chi^{(2)}$ pattern, we measured the transfer function of the wavy waveguide using an optimized $\chi^{(2)}$ pattern and compared it to a $\chi^{(2)}$ pattern with a constant poling period.  The $\chi^{(2)}$ pattern with a constant period uses the $\Lambda$ that maximizes SHG signal (as shown in Fig.~\ref{fig:Wavy_waveguide_poling_scan}).  Fig.~\ref{fig:Adiabatic Transfer Function} shows the transfer function from the optimized $\chi^{(2)}$ structure and the transfer function from the constant $\Lambda$ structure for waveguides with \SI{2}{mm} and \SI{1}{mm} geometric periods.  The optimized transfer function not only shows much higher nonlinear conversion efficiency, it also more effectively takes the shape of a $\operatorname{sinc}^2$ function.

While the transfer functions presented in Fig.~\ref{fig:Adiabatic Transfer Function} show the advantage of in situ optimization of the $\chi^{(2)}$ pattern to enhance nonlinearity on devices with complex geometries, there are still challenges to the approach presented here.  Firstly, as is discussed in Sec.~\ref{subsec:spiral phase-mathcing}, imperfections in the programming illumination likely mean a constant poling period on the SLM does not translate to a constant poling period being projected onto the waveguide.  So some of the active optimization of the $\chi^{(2)}$ pattern accounts for aberrations in the programming illumination instead of dispersion variations caused by the waveguide's changing width.  Furthermore, the waveguide in Fig.~\ref{fig:Adiabatic Interference} shows \SI{7}{mm} of interaction length instead of \SI{5.5}{mm} as labeled in Tab.~\ref{tab:possible devices}.  This discrepancy (which is explained in detail in Sec.~\ref{sec:possible devices}) is caused by non-uniform optical power in the illumination optics.  Consequently, the partitions near the edge of the programming illumination have weaker nonlinear performance, which makes it harder to optimize the poling structure on those partitions.

As a more fundamental limit, as the geometric period becomes smaller, more partitions are needed to properly digitize the changing waveguide width. Smaller partitions lead to reduced SHG power and broader $\operatorname{sinc}^2$ transfer functions for each partition. This makes it more challenging to identify the optimal poling period because the background noise starts to dominate over the generated signal. For geometric periods on the order of the poling period, a different approach to $\chi^{(2)}$ optimization is likely needed. Instead of approximating the ideal $\chi^{(2)}$ pattern by stitching together discrete sections with a constant poling period, allowing the poling period to continuously chirp as a function of position might lead to better nonlinear performance. Still, the results shown in this section show that adaptability to complex underlying waveguide geometries is possible with a programmable $\chi^{(2)}$ and leaves open the possibility for more complex optimization schemes in the future.

\section{Poling Spiral Waveguides}
\label{sec:spirals}
\subsection{Iterative Phase-Matching}
\label{subsec:spiral phase-mathcing}
This section describes our process for optimizing and aligning the programmable $\chi^{(2)}$ structure on spiral waveguide geometries.  Through this process, we demonstrate that a programmable $\chi^{(2)}$ provides the flexibility to ensure constructive interference of SHG over longer waveguide geometries.  This discussion also illuminates non-fundamental experimental challenges from non-ideal conditions in the illumination optics.

While it is possible to map an exact shape generated on the SLM to the lithographically defined spiral on-chip, this presents the challenge of spatial alignment.  For example, if a circular arc has a constant angular periodicity $\omega$, the translational periodicity depends on the radial position as $\Lambda = r \omega$, where $r$ is the radial position. So if $\Lambda_0$ is calibrated for some radial position $r_0$, a small radial offset of $\Delta r$ leads to a translational period offset of $\Delta \Lambda = \Delta r \omega$. Because of the discretized spatial resolution of 3.77 \si{\micro\meter} given by the SLM pixel geometry and imaging demagnification, it is challenging to position the projected $\chi^{(2)}$ structure perfectly on top of the lithographically-defined spiral.  Instead of trying to perfectly align the programming illumination with the physical spiral, our approach was to create an illumination pattern with many parameters that can be actively optimized using experimental feedback.

\begin{figure}
    \centering
    \includegraphics[width=1\linewidth]{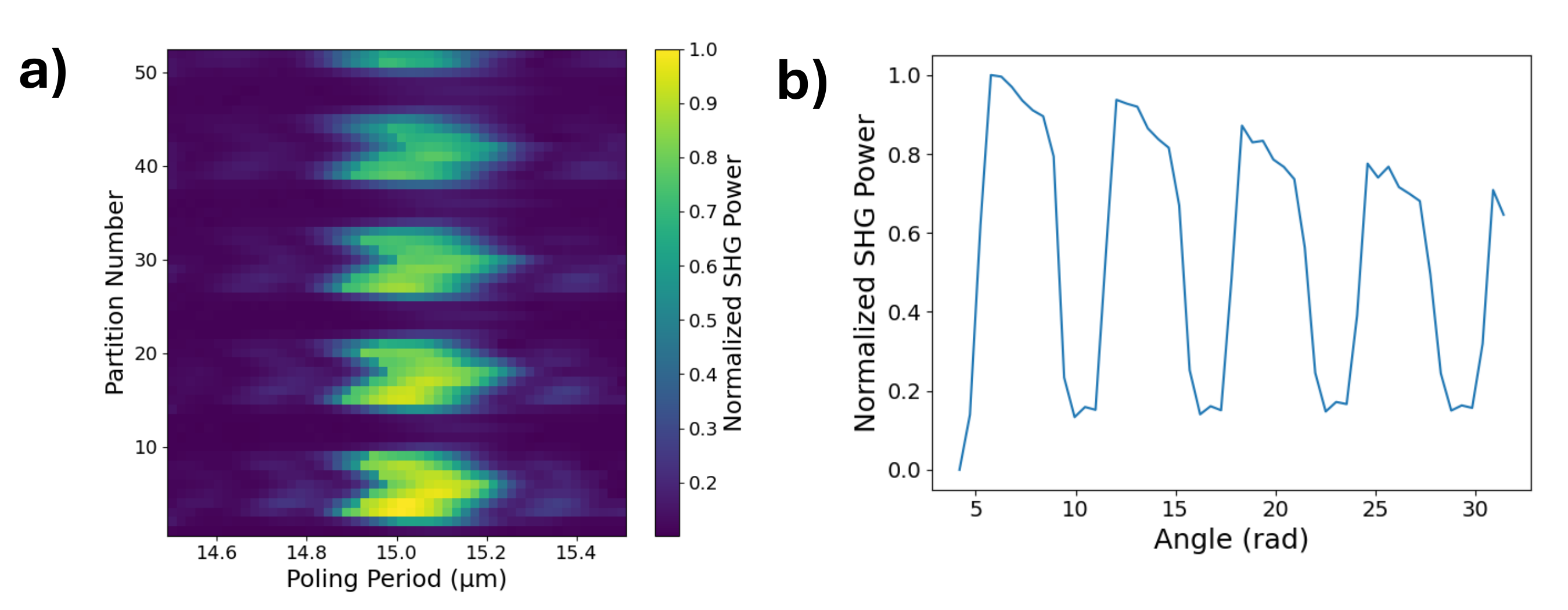}
    \caption{a) Normalized measured SHG power as the poling period is swept on each partition of a 10.1 cm long spiral pumped with a pulsed FH source.  b) Maximum normalized SHG power for each partition, where the angle corresponds to the partition's angle with respect to the entrance at the top of the spiral.  The partitions shown here are on the first half of the spiral, starting from the entrance to the middle.}
    \label{fig:pulsed spiral}
\end{figure}

To do this, our illumination geometry consisted of many circular arcs projected onto the Archimedean part of the physical spiral.  The relative position of each illumination arc towards the other illumination arcs is defined by the spiral generating equation (as described in Sec.~\ref{subsec:Mode Shapes}) and hyperparameters describing the spiral's central position and the illumination optics' demagnification ratio.  The hyperparameters are hand-tuned by looking at the spatial overlap between the projected and physical spirals.  The physical offset between the projected pattern and the physical waveguide creates an alignment ``dispersion'', which can be minimized through iterative adjustments of the projected spiral's hyperparameters.  Fig.~\ref{fig:pulsed spiral}(a) shows the normalized power measured out of a 10.1 cm long spiral as the poling period is swept for each partition.  Additionally, Fig.~\ref{fig:pulsed spiral}(b) shows the normalized maximum power generated for each partition as a function of angle.  Unlike later experimental results, Fig.~\ref{fig:pulsed spiral} uses a pulsed laser as the FH pump source instead of a CW laser.  The higher peak power of the pulsed laser allowed us to individually characterize the response of each partition.

As Fig.~\ref{fig:pulsed spiral} shows, there is a clear dependence on a partition's position for the amount of SHG power produced there. One reason is that the illumination spot is not perfectly uniform when projected onto the spiral.  As Sec.~\ref{sec:Circuit Model} and Fig.~\ref{fig:Illumination Contrast} show, if different sections have different illumination intensities but the same electrical frequency, their nonlinear performance varies dramatically.  Furthermore, focal aberrations resulting from unwanted tilt on the SLM cause some parts of the projected pattern to be more in focus than others. Although we take steps to minimize these effects, they are difficult to completely eliminate using our current optical setup.  These difficulties become more pronounced for devices with larger spatial footprints, as focal aberrations and illumination non-uniformities become harder to minimize.  Additionally, by looking at Fig.~\ref{fig:pulsed spiral}(b), it is clear that the generated SHG power in each partition decreases with increasing angle.  This is because of pump attenuation due to FH loss.

Once the projection optics are aligned and characterized, we switch the pulsed laser to a CW source.  Because a CW source does not suffer from SH walk-off, interference effects must be carefully accounted for.  To optimize the $\chi^{(2)}$ pattern for a CW pump, we use Alg.~\ref{alg:insitu_opt_wavy} from Sec.~\ref{sec:Wavy Waveguides} to iteratively optimize the phase and period of the grating on each spiral partition. This procedure ensures that each spiral partition is phase-matched and that all the SHG interferes constructively.

\begin{figure}
    \centering
    \includegraphics[width=1\linewidth]{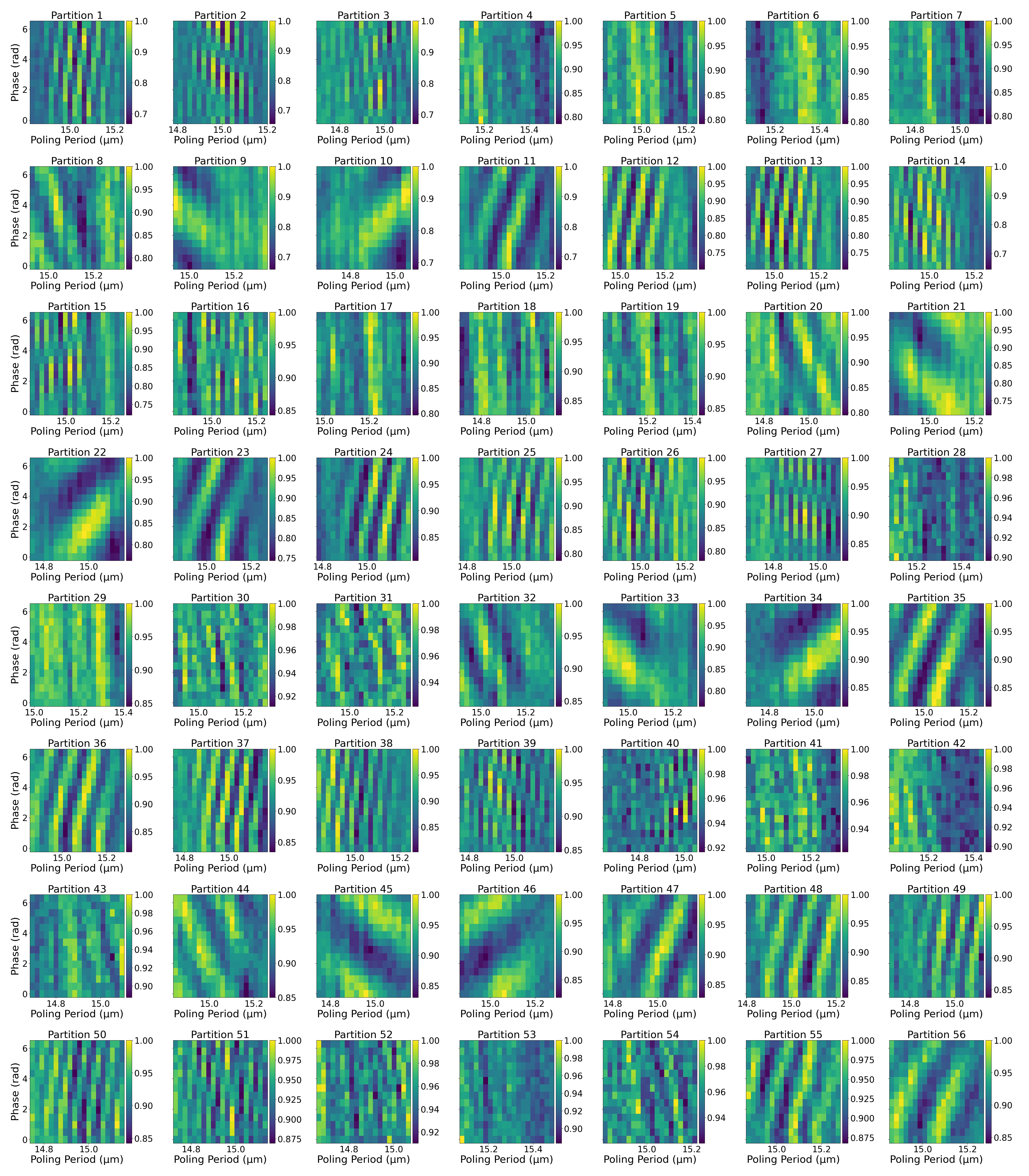}
    \caption{Measured SHG out of a 10.1 cm long spiral waveguide as the phase and poling period of each partition is changed. Partition 1 is the first partition to be optimized, and these plots only show the partitions from the first half of the spiral.}
    \label{fig:phase-pp-scan}
\end{figure}

Fig.~\ref{fig:phase-pp-scan} shows the measured normalized SHG power from a 10.1 cm long spiral with a 1590 nm pump as the phase and poling period of each partition is scanned.  The first 56 partitions of a spiral with 108 total partitions are shown.  The coherent interference effects shown in Fig.~\ref{fig:phase-pp-scan} are repeatable across different spirals and wavelengths.  The best poling period of the projected arc varies depending on the partition position.  This is because the radius of curvature gradually changes and there is imperfect alignment between the projected pattern and the etched waveguide.  

Importantly, this demonstration shows that programmable $\chi^{(2)}$ allows for the phase of each $\chi^{(2)}$ grating to be adjusted in real-time to compensate for phase drifts and dispersion changes in the underlying spiral geometry.  Furthermore, because constructive and destructive interference behavior is visible as the partition phase changes, we can confirm that there is coherent interaction between the SH produced in different partitions. 

\subsection{Super-Linear Scaling Measurements}
\label{subsec:superlinear}

The previous section described how the programmable $\chi^{(2)}$ pattern is partitioned and then optimized using real-time feedback to maximize the constructive interference between the SH produced in each waveguide partition.  In this section, we describe the relationship between increased poling distance and output SHG power using a CW FH pump.  To perform these measurements, we started with no programming illumination and iteratively activated the next partition along the propagation route using optimized values for the grating phase and period. We measured the output SHG power for each phase-matching length. 

\begin{figure}
    \centering
    \includegraphics[width=1\linewidth]{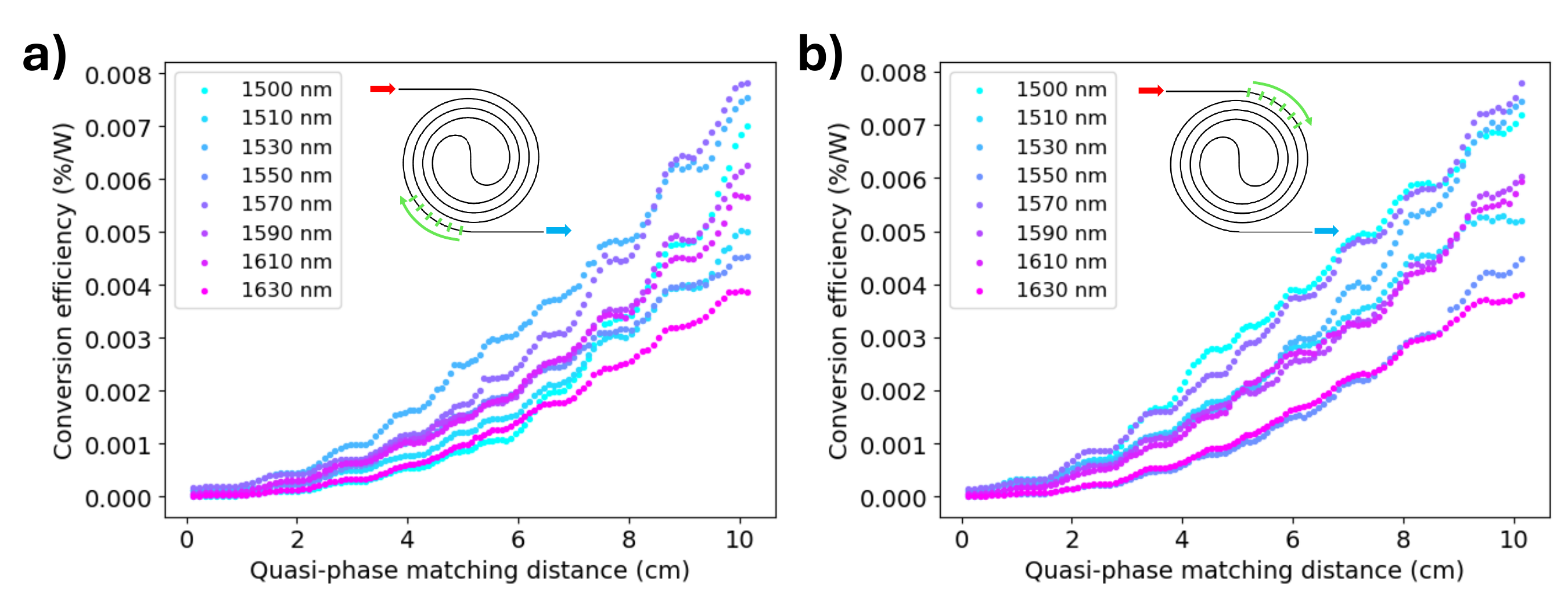}
    \caption{a) Backward scaling of the measured SHG conversion efficiency with increased poling distance on \(\mathrm{SiN}\) fabricated in-house. b) Forward scaling of the measured SHG conversion efficiency with increased poling distance using the same devices and final poling structures as in panel a).}
    \label{fig:SRN3 Scaling}
\end{figure}
\begin{figure}
    \centering
    \includegraphics[width=1\linewidth]{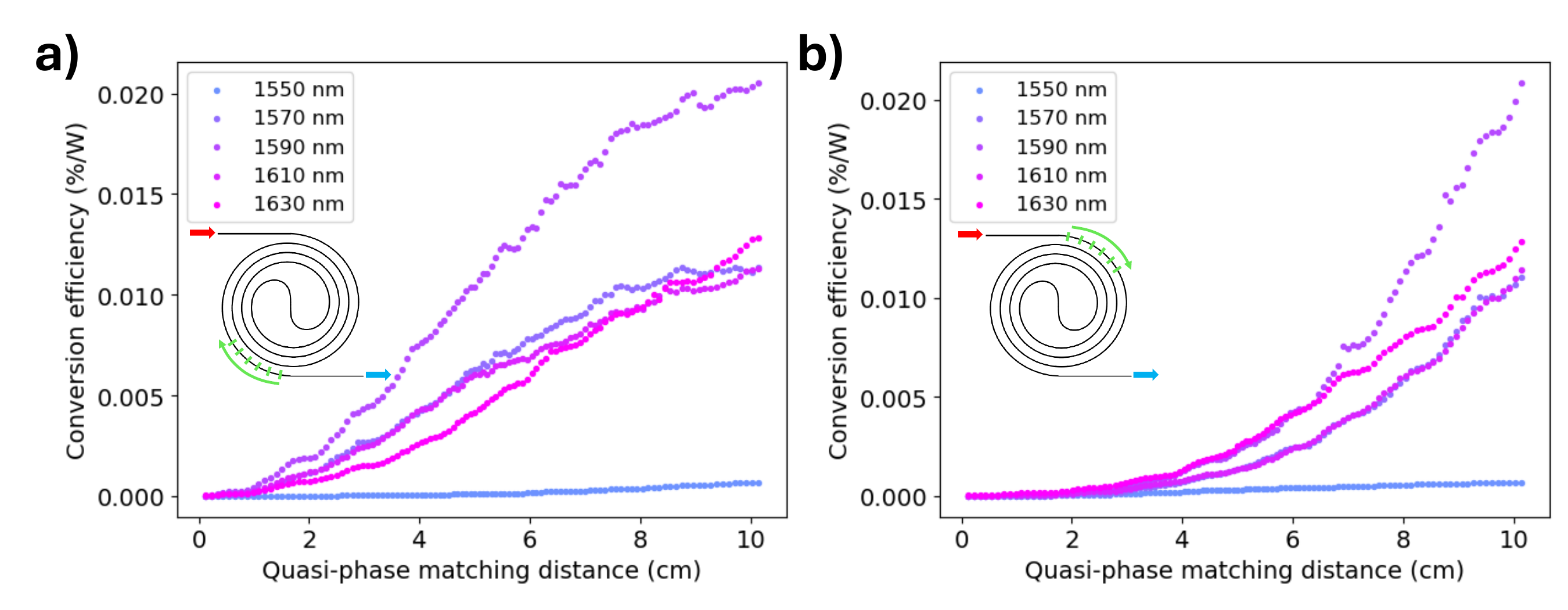}
    \caption{a) Backward scaling of measured SHG conversion efficiency with increased poling distance on \(\mathrm{SiN}\) fabricated by SVM.  b) Forward scaling of measured SHG conversion efficiency with increased poling distance using the same devices and final poling structures as in panel a).}
    \label{fig:SVM Scaling}
\end{figure}

Fig.~\ref{fig:SRN3 Scaling} shows the measured output SHG conversion efficiency using \(\mathrm{SiN}\) fabricated in-house as we phase-match from the beginning to the end of the spiral (forward scaling) and from the end to the beginning of the spiral (backward scaling).  Fig.~\ref{fig:SVM Scaling} shows the same thing for \(\mathrm{SiN}\) deposited by SVM. \(\mathrm{SiN}\) waveguides deposited by SVM show $\sim 2 \times$ greater nonlinear conversion efficiency above 1600 nm than \(\mathrm{SiN}\) waveguides deposited in-house.  This difference is due to the higher silicon content in these waveguides, which provides a higher nonlinearity.  The downside of this material is the narrower bandwidth, as the loss peak near 1520 nm (as shown in Fig.~\ref{fig:Loss Spectra}) becomes severely detrimental for performance over longer interaction lengths.

Even though both waveguides presented in Fig.~\ref{fig:SRN3 Scaling} and Fig.~\ref{fig:SVM Scaling} had their poling structure optimized using the same technique, the dependence on increased poling distance varies dramatically.  In addition to constructive interference between different sections, the loss of the FH and SH light plays an important role in establishing the scaling conditions.  In the case of FH loss, the pump is weaker at the end of the waveguide than at the beginning.  This means the rate of SH power generation decreases with increased propagation length.  For SH loss, the generated signal suffers propagation loss from the point where it is produced to the waveguide output.  

To more rigorously examine the effect of loss on the scaling of SHG, we designed numerical simulations that account for waveguide loss when estimating nonlinear behavior over longer interaction lengths.  The evolution of the SH field (assuming $\Delta k = 0$) is
\begin{equation}
\label{eq:field spiral evolution}
    \frac{dA_{\text{SH}}}{dz} = \sqrt{\eta}P_{\text{FH}} - \frac{\alpha_{\text{SH}}}{2}A_{\text{SH}}
\end{equation}
where $A_{\text{SH}}$ is the field amplitude of the SH, $\eta$ is the slope conversion efficiency (in units of $\%$/(Wcm$^2$)), and $\alpha_{\text{SH}}$ is the SH loss (in units 1/cm).  Because SH loss is measured for power, the loss term is divided by a factor of 2 for field amplitude loss in Eq.~\eqref{eq:field spiral evolution}.  $P_{\text{FH}}$ evolves as 
\begin{equation}
    P_{\text{FH}} = P_0e^{-\alpha_{FH}z}
\end{equation}
so Eq.~\eqref{eq:field spiral evolution} can be solved as 
\begin{equation}
    \label{eq:SH field complete}
    A_{\text{SH}} = \sqrt{\eta}P_0\frac{e^{-\alpha_{\text{FH}}L} - e^{\frac{-\alpha_{\text{SH}}L}{2}}}{\frac{\alpha_{\text{SH}}}{2} - \alpha_{\text{FH}}}
\end{equation}
and the power in the SH mode can be solved as $P_{\text{SH}} =A_{\text{SH}}^2$.  Using the expression in Eq.~\eqref{eq:SH field complete}, we solve for the evolution of the SH field inside of a spiral of length $L$ with finite FH and SH losses. This is noticeably different from the lossless case, where the scaling is predicted to be 
\begin{equation}
\label{eq:lossless SHG}
    P_{\text{SH}} = \eta P_0^2L^2
\end{equation}

\begin{figure}
    \centering
    \includegraphics[width=1\linewidth]{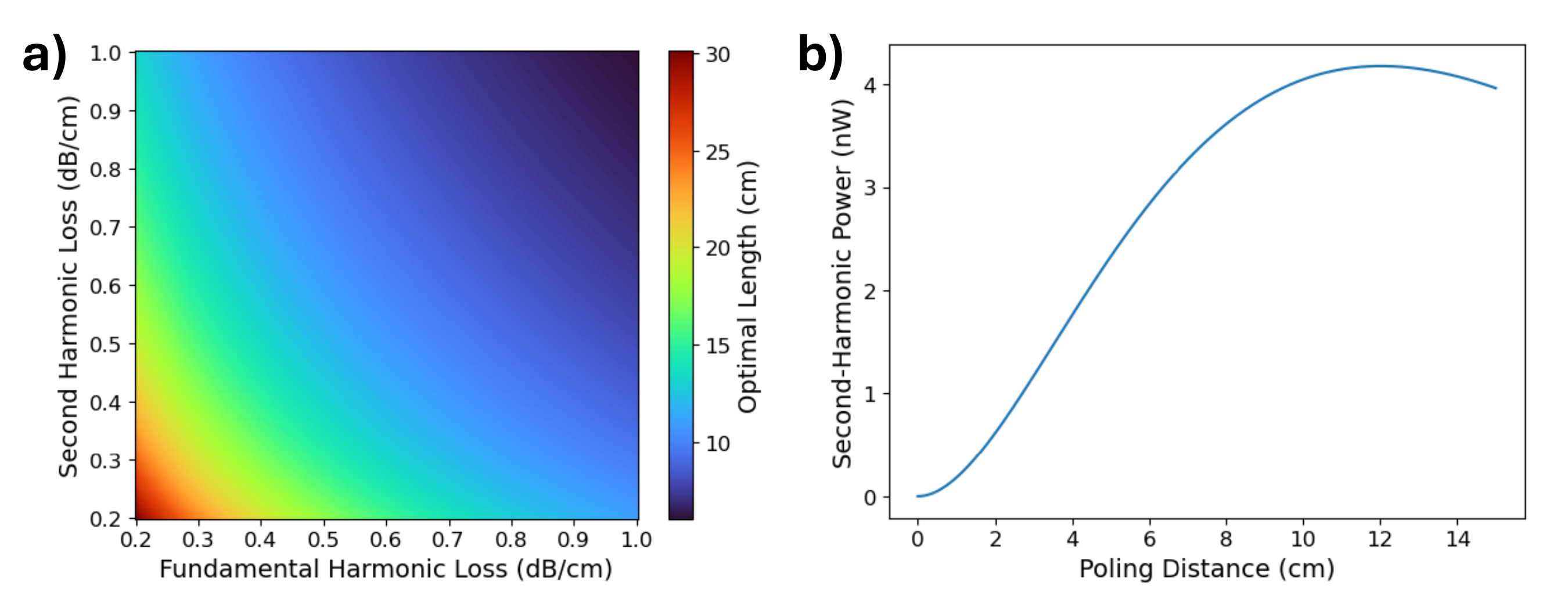}
    \caption{a) The poling distance that provides the maximum SHG inside a long nonlinear programmable waveguide device. b) Simulation of the generated second-harmonic power for 6 mW of FH pumping at the waveguide input, 0.4 dB/cm of FH loss, 0.65 dB/cm of SH loss, and $5.9 \times 10^{-4}$ \(\mathrm{\%/(Wcm^2)}\) slope conversion efficiency.  These are similar to the measured parameters for \(\mathrm{SiN}\) deposited in-house.}
    \label{fig:internal_spiral}
\end{figure}
\begin{figure}
    \centering
    \includegraphics[width=1\linewidth]{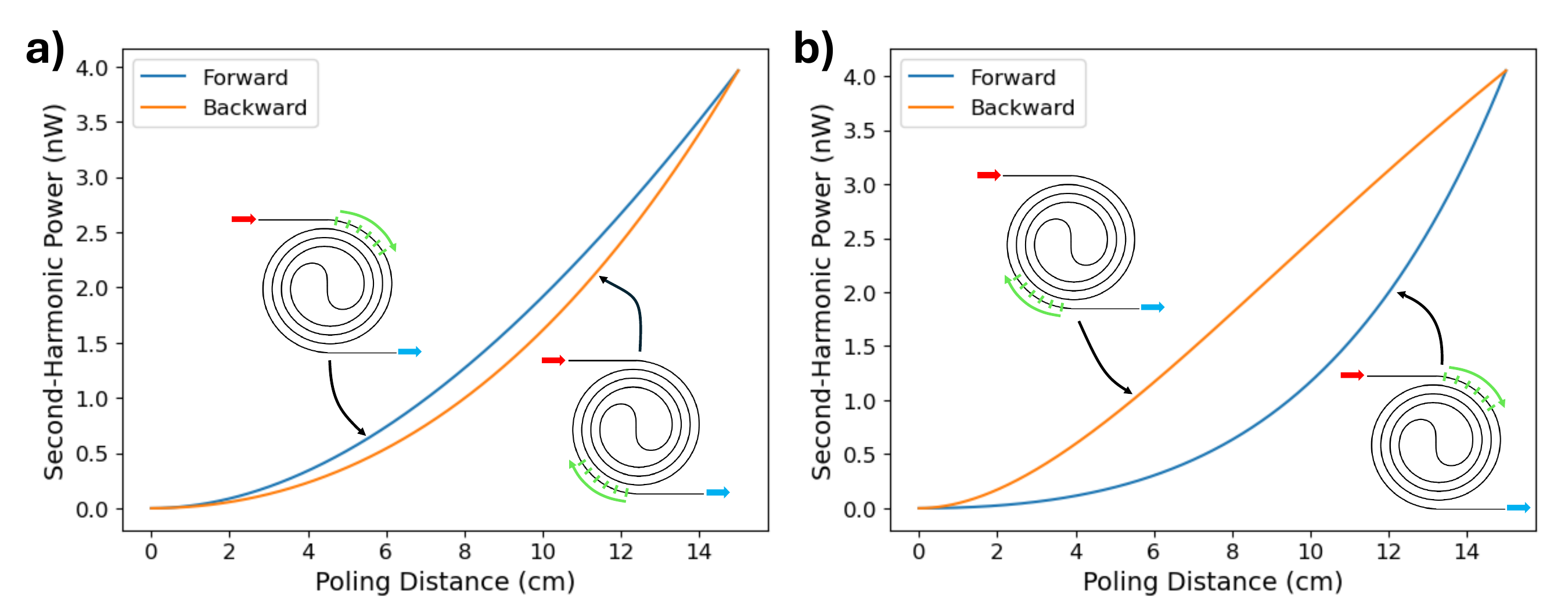}
    \caption{Simulation of the expected measured forward and backward scaling of a) \(\mathrm{SiN}\) deposited in-house and b) \(\mathrm{SiN}\) deposited by SVM.  The values used in this simulation are taken from Tab.~\ref{tab:possible devices} and Sec.~\ref{sec:loss characterization} for the straight versions of each \(\mathrm{SiN}\) material platform.  The input power in these simulations is 6 mW.  Additionally, there is a cartoon spiral next to each line depicting the physical meaning of forward and backward scaling.}
    \label{fig:for-back sim}
\end{figure}

Fig.~\ref{fig:internal_spiral}(a) shows a simulation of the ideal poling distances for spirals with different FH and SH losses.  Fig.~\ref{fig:internal_spiral}(b) shows the expected internal scaling of SHG power in a \(\mathrm{SiN}\) waveguide deposited in-house using experimentally characterized parameters.  While Eq.~\eqref{eq:lossless SHG} would predict quadratic scaling of SHG power with increased interaction distance, the saturation behavior shown in Fig.~\ref{fig:internal_spiral} is a byproduct of waveguide loss expressed more fully in Eq.~\eqref{eq:SH field complete}.

Despite the saturation effect simulated in Fig.~\ref{fig:internal_spiral}, the measured results in Fig.~\ref{fig:SRN3 Scaling} and Fig.~\ref{fig:SVM Scaling} show a consistent upward super-linear scaling of the SHG conversion efficiency.  This is because Fig.~\ref{fig:internal_spiral} simulates the effects of longer propagation and poling distance, while the results in Fig.~\ref{fig:SRN3 Scaling} and Fig.~\ref{fig:SVM Scaling} only show the effect of longer poling distance for a fixed propagation distance.  As the poling distance increases for a fixed propagation distance, the SH signal still must propagate through the remainder of the spiral to reach the detector.  This means the SH loss attenuates the measured signal, reducing the measured power.  

Interestingly, the forward and backward scaling of the waveguide differ because of the underlying FH and SH losses.  As shown in the previous experimental measurements of forward and backward scaling, these relationships can approximately be fit by a power law described as 
\begin{equation}
    P_{\text{SH}} \propto L^a
\end{equation}
where $L$ is the nonlinear interaction distance and $a$ is a power-law scaling factor.  In the case where SH loss is zero and FH loss is high, the beginning of the spiral provides the most power at the output facet, so $a \approx 1$ for forward scaling while $a > 1$ for backward scaling.  Conversely, for a waveguide where SH loss is high and FH loss is zero, the end of the waveguide provides most of the power at the output facet.  This means $a>1$ for the forward scaling while $a \approx 1$ for the backward scaling.  Fig.~\ref{fig:for-back sim} shows the simulated forward and backward scaling relationships that would be expected during an experimental measurement for \(\mathrm{SiN}\) deposited in-house and by SVM.  The different scaling relations shown in Fig.~\ref{fig:for-back sim} agree with the trends shown in Fig.~\ref{fig:SRN3 Scaling} and Fig.~\ref{fig:SVM Scaling}.  The difference in forward and backward scaling is especially notable for \(\mathrm{SiN}\) deposited by SVM because $\text{Loss}_{\text{SH}} \gg 2\times\text{Loss}_{\text{FH}}$.  

When examining the experimentally measured forward and backward scalings in Fig.~\ref{fig:SRN3 Scaling} and Fig.~\ref{fig:SVM Scaling}, there are two additional discrepancies with the simulations in Fig.~\ref{fig:for-back sim} and Fig.~\ref{fig:internal_spiral}: lower conversion efficiency than expected and non-uniform scaling of the nonlinear conversion efficiency.  This is because of the non-uniform illumination intensity and focal aberrations in the programming illumination, as discussed in Sec.~\ref{subsec:spiral phase-mathcing}.  Consequently, the slope conversion efficiency has a strong spatial dependence, which reduces the overall conversion efficiency and causes non-uniform scaling with increased interaction length.  

It should also be noted that different wavelengths have different measured SHG conversion efficiencies after $\chi^{(2)}$ optimization.  While some of this is a result of changes in field confinement and $\chi^{(2)}$ dispersion as a function of wavelength, most of this change can be attributed to experimental error.  More specifically, the coupling conditions of the FH pump and SH signal were realigned for each wavelength, the illumination optics were realigned, and the $\chi^{(2)}$ pattern was re-optimized.  While each free-parameter was tuned to maximize the measured SH power, it is possible that some parameters were set to sub-optimal conditions.

\subsection{Phase-Matching Longer Spirals}
\label{subsec:transfer function}
The previous section discussed the advantages of longer phase-matching distances for a fixed total device propagation length. This section discusses the advantage that our programmable $\chi^{(2)}$ technique shows by phase-matching longer distances on spirals of differing device propagation length. By comparing the relative width and height of the transfer functions, it becomes clear that our programmable $\chi^{(2)}$ approach improves nonlinear performance by adaptively poling longer propagation lengths. The transfer function of an SHG process is described as 
\begin{equation}
    P_{\text{SHG}} = \eta_{\text{norm}}P_{FH}^2L^2\operatorname{sinc}(\frac{\Delta k L}{2})^2
\end{equation}
Unlike earlier demonstrations, where $\Delta k$ was changed by tuning the poling period $\Lambda$, $\Delta k$ is now changed by scanning the wavelength of the FH, $\lambda_{\text{FH}}$.  

While the rate at which $\Delta k$ changes with $\lambda_{\text{FH}}$ depends on the dispersion of the waveguide, because of the factor of $L$ within the $\operatorname{sinc}^2$ function, the width of the transfer function scales as $\frac{1}{L}$.  Because of the $L^2$ term in the expression for $P_{\text{SHG}}$, the height of the transfer function should scale quadratically with $L$ in the lossless limit.

\begin{figure}
    \centering
    \includegraphics[width=1\linewidth]{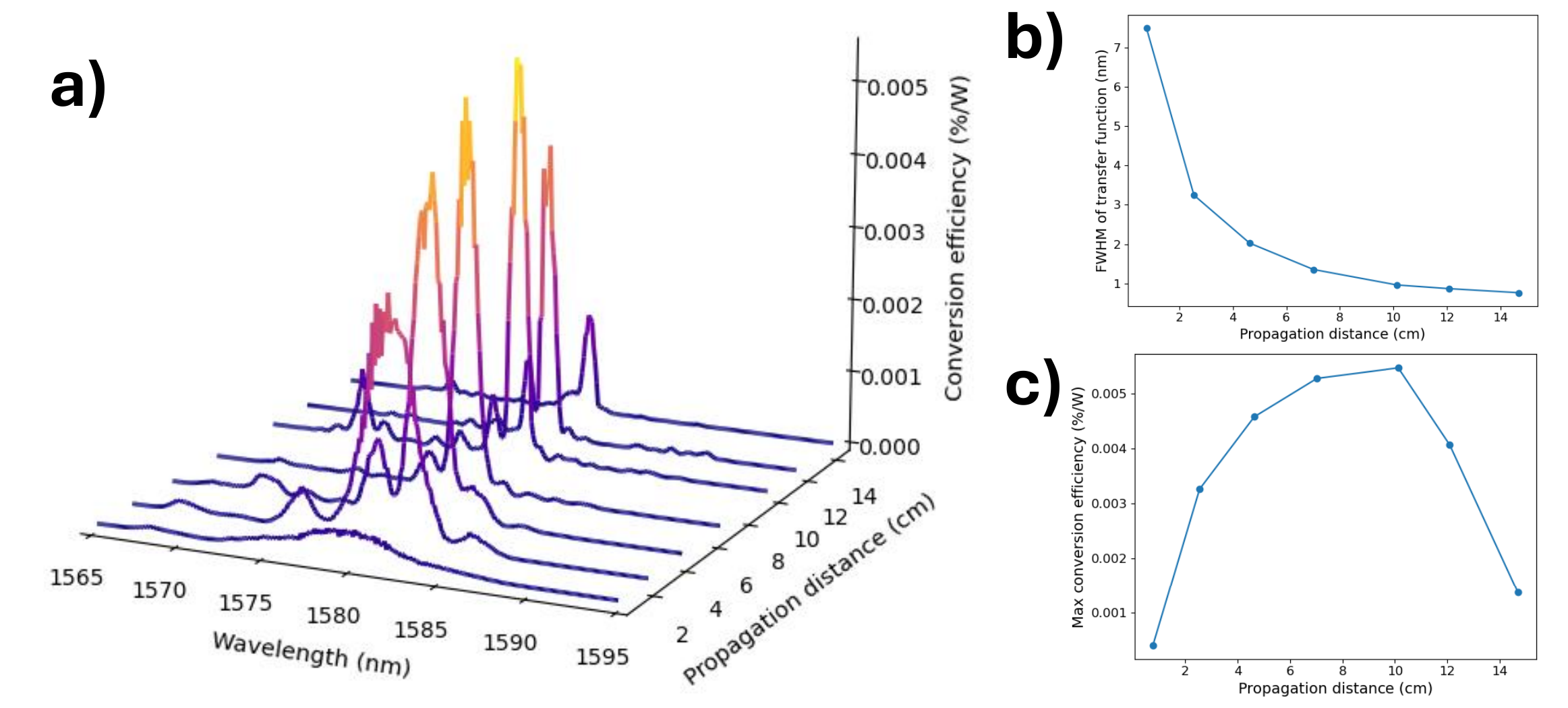}
    \caption{a) Measured transfer functions for spiral geometries of different lengths.  Each spiral has its poling structure optimized to maximize nonlinear conversion efficiency.  The color of each point corresponds to the magnitude of the conversion efficiency.  b) Fitted FWHM of each transfer function as the spiral length increases. c) Maximum measured SHG conversion efficiency as the spiral length increases.}
    \label{fig:Comparitive Transfer Function}
\end{figure}

Fig.~\ref{fig:Comparitive Transfer Function}(a) shows the measured transfer functions as the length of the underlying spiral waveguide is increased.  The $\chi^{(2)}$ pattern is optimized for each spiral using the procedure in Sec.~\ref{subsec:spiral phase-mathcing}.  Fig.~\ref{fig:Comparitive Transfer Function}(b) shows how the full width at half maximum (FWHM) of the transfer function decreases as the spiral length increases, and Fig.~\ref{fig:Comparitive Transfer Function}(c) shows the maximum measured SHG conversion efficiency as the spiral length increases.  The FWHM is fitted by approximating the center of the $\operatorname{sinc}^2$ function as a Gaussian.  The width of the transfer function scales approximately as $\frac{1}{L}$, but the height of the transfer function scales far more slowly than $L^2$ and shows saturation behavior around 10 cm of length.  This behavior is caused by internal SH and FH loss within the spiral and non-ideal behavior in the illumination optics, as discussed in Sec.~\ref{subsec:superlinear}.

When taken together, Sec.~\ref{sec:spirals} shows that our adaptive method for patterning $\chi^{(2)}$ is able to phase-match increasingly longer interaction lengths to achieve better nonlinear performance.  Coherent addition of the optical fields is shown in Sec.~\ref{subsec:spiral phase-mathcing}.  The advantage of longer phase-matching distances for a device of fixed propagation distance is shown in Sec.~\ref{subsec:superlinear} and the advantage of phase-matching longer propagation distances is shown in Sec.~\ref{subsec:transfer function}.  While there are limitations in our approach due to imperfections in the illumination optics, FH loss, and SH loss, these are not fundamental and can be alleviated with better devices and experimental methods.  Importantly, these results show that longer phase-matching distances are possible using adaptive $\chi^{(2)}$ and that these longer phase-matching distances provide a path to scaling the nonlinear performance of this approach to programmable nonlinear photonics.

\section{Spatial Mode Shaping} 
\label{sec:spatial modes}
\subsection{Higher-Order Spatial Mode Generation}
\label{subsec:many modes}
Because of the large cross-sectional geometry of the devices used in these experiments, it is highly likely that light butt-coupled into the exposed waveguide facet excites multiple spatial modes simultaneously. While previous sections focused primarily on interactions between the $\mathrm{TM_{00}}$ FH mode and the $\mathrm{TM_{00}}$ SH mode, it is possible to couple a variety of the excited FH spatial modes to different SH spatial modes simply by altering the programmable $\chi^{(2)}$ grating. For the sake of simplicity, we assume $\chi^{(2)}$ only has a finite element in the $yyy$ component and that the electric fields implicitly denote their $y$-components. When accounting for interactions between spatial modes, the nonlinear mode coupling is proportional to the spatial overlap integral of the interacting fields. However, because we are dealing with guided waveguide modes rather than free-space optics, the overlap integral must be appropriately normalized by the respective modal areas. The normalized overlap ($\Gamma$) is given by:
\begin{equation}
\label{eq:gamma eq}
    \Gamma=\frac{\iint \chi^{(2)} E_1(x, y)E_2(x, y)E_3^*(x, y)dxdy}{\sqrt{\iint \mathrm{Re}(\mathbf{E}_1\times \mathbf{H}_1^*)\cdot \hat{\mathbf{z}}\, dxdy} \sqrt{\iint \mathrm{Re}(\mathbf{E}_2\times \mathbf{H}_2^*)\cdot \hat{\mathbf{z}}\, dxdy} \sqrt{\iint \mathrm{Re}(\mathbf{E}_3\times \mathbf{H}_3^*)\cdot \hat{\mathbf{z}}\, dxdy}}
\end{equation}
where $E_1$ and $E_2$ are the interacting FH electric fields, and $E_3^*$ is the complex conjugate of the generated SH electric field. In the denominator, $\mathbf{E}_{1,2,3}$ and $\mathbf{H}_{1,2,3}^*$ represent the full vectorial electric and complex-conjugate magnetic fields for each respective mode, and $\hat{\mathbf{z}}$ is the unit vector along the axis of propagation. Because the fields are integrated over the entire transverse $x$ and $y$ plane, the parity of the interacting modes strictly dictates whether a nonlinear process is physically allowed. For example, while the process $\mathrm{TM_{00}}$ + $\mathrm{TM_{01}}$ $\xrightarrow{}$ $\mathrm{TM_{01}}$ is allowed by parity, the process $\mathrm{TM_{01}}$ + $\mathrm{TM_{01}}$ $\xrightarrow{}$ $\mathrm{TM_{01}}$ evaluates to zero and is forbidden. More generally, there cannot be an odd number of odd-parity modes in either spatial dimension for a process to proceed. This assumes that the distribution of the induced $\chi^{(2)}$ has inversion symmetry for $x \mapsto -x$ and $y \mapsto -y$, which is a valid approximation for the device architectures demonstrated in this work.

Tab.~\ref{tab:spatial combinations} details the required poling period and normalized field overlap for a selection of possible nonlinear processes using the first five TM modes presented in Fig.~\ref{fig:mode shape}. Here, normalized overlap is defined as $(\frac{\Gamma}{\Gamma_0})^2$, where $\Gamma_0$ represents the baseline overlap of the purely fundamental process ($\mathrm{TM_{00}}$ + $\mathrm{TM_{00}}$ $\xrightarrow{}$ $\mathrm{TM_{00}}$) with $\Gamma$ defined in Eq.~\eqref{eq:gamma eq}. This metric directly compares the theoretical efficiency of a selected higher-order process against the fundamental process when accounting solely for the spatial overlap of the interacting fields.

\begin{table}[h!]
\centering
\begin{tabular}{ccccc} 
\toprule
\(\mathrm{E_1}\) (FH) & \(\mathrm{E_2}\) (FH) & \(\mathrm{E_3}\) (SH) & \(\mathrm{\Lambda \ (\mu m)}\) & \(\mathrm{(\frac{\Gamma}{\Gamma_0}})^2\) \\
\midrule
\(\mathrm{TM_{00}}\) & \(\mathrm{TM_{00}}\) & \(\mathrm{TM_{00}}\) & 12.82 & 1 \\
\(\mathrm{TM_{00}}\) & \(\mathrm{TM_{00}}\) & \(\mathrm{TM_{20}}\) & 15.72 & 0.0306 \\
\(\mathrm{TM_{00}}\) & \(\mathrm{TM_{10}}\)& \(\mathrm{TM_{10}}\)& 12.03 & 0.648 \\
\(\mathrm{TM_{00}}\) & \(\mathrm{TM_{20}}\) & \(\mathrm{TM_{20}}\) & 10.9 & 0.6 \\
\(\mathrm{TM_{00}}\) & \(\mathrm{TM_{01}}\)& \(\mathrm{TM_{01}}\)& 9.36 & 0.621 \\
\(\mathrm{TM_{10}}\)& \(\mathrm{TM_{10}}\)& \(\mathrm{TM_{00}}\) & 10.7 & 0.624 \\
\(\mathrm{TM_{10}}\)& \(\mathrm{TM_{10}}\)& \(\mathrm{TM_{20}}\) & 11.82 & 0.361 \\
\(\mathrm{TM_{20}}\) & \(\mathrm{TM_{20}}\) & \(\mathrm{TM_{00}}\) & 7.45 & 0.57 \\
\(\mathrm{TM_{30}}\) & \(\mathrm{TM_{30}}\) & \(\mathrm{TM_{00}}\) & 5.38 & 0.151 \\
\bottomrule
\end{tabular}
\caption{Poling periods (\(\mathrm{\Lambda}\)) and normalized field overlap ($\frac{\Gamma}{\Gamma_0}$) for different FH spatial modes at 1550 nm and SH spatial modes at 775 nm.}
\label{tab:spatial combinations}
\end{table}

While Tab.~\ref{tab:spatial combinations} only highlights 9 combinations, utilizing just the five FH modes shown yields 24 distinct nonlinear processes between three interacting modes that possess non-zero spatial overlaps. If the induced $\chi^{(2)}$ grating possessed full two-dimensional spatial programmability in both the transverse $x$ and $y$ dimensions, the parity restrictions could be lifted, enabling 75 distinct three-mode interaction processes. In general, with $N$ distinct FH and $M$ distinct SH modes interacting, there are
\begin{equation}
\frac{MN(N+1)}{2}
\end{equation}
possible mode combinations assuming $\chi^{(2)}$ can be fully programmed in space.

To experimentally identify the nonlinear processes enabled on our device, we swept a wide range of programmed poling periods and continuously monitored the output for SHG. Fig.~\ref{fig:modal}(b) presents the results of this broad sweep, with a selection of the peaks labeled by the output SH mode shape they generate. As demonstrated, we can selectively generate $\mathrm{TM_{00}}$, $\mathrm{TM_{10}}$, $\mathrm{TM_{20}}$, and $\mathrm{TM_{01}}$ modes within the exact same device simply by reconfiguring the software-defined poling period.

When comparing the results in Fig.~\ref{fig:modal} and the predicted behavior in Tab.~\ref{tab:spatial combinations}, several experimental discrepancies become apparent.  For example, the predicted poling period for the $\mathrm{TM_{00}}+\mathrm{TM_{00}}\xrightarrow{}\mathrm{TM_{00}}$ process is 13.18 \si{\micro\meter}, while the experimentally measured value is 14.55 \si{\micro\meter}.  While the experiment was conducted with annealed \(\mathrm{SiN}\) deposited in-house, the model uses the dispersion relation for unannealed \(\mathrm{SiN}\) deposited by SVM.  As Sec.~\ref{subsec:dispersion properties} discusses in greater detail, the broadband dispersion properties of annealed \(\mathrm{SiN}\) deposited in-house are inaccessible using ellipsometry due to pre-annealing waveguide patterning.  Consequently, there is no large, unpatterned slab region that we can characterize.  Additionally, the experiment in Fig.~\ref{fig:modal} is conducted at 1590 nm while the results in Tab.~\ref{tab:spatial combinations} were calculated at 1550 nm.

Although there are slight experiment-theory gaps in the specific phase-matching periods between Fig.~\ref{fig:modal} and Tab.~\ref{tab:spatial combinations}, several physical processes can still be clearly identified. For example, the strongest peak occurs at a poling period of 14.55 \si{\micro\meter}, and another peak producing the exact same $\mathrm{TM_{00}}$ SH mode shape occurs at a poling period of 43.65 \si{\micro\meter}. An identical relationship is observed for the $\mathrm{TM_{10}}$ SH mode, with peaks occurring at 12.65 \si{\micro\meter} and 37.83 \si{\micro\meter}. This exact $3 \times$ scaling between the resonant poling periods is a direct consequence of higher-order phase matching. Because the optical illumination pattern closely approximates a square wave, the induced electric field distribution—and therefore the $\chi^{(2)}$ grating—contains higher-order Fourier components. Even though we primarily utilize the fundamental spatial frequency ($n=1$, where $k_1 = \Delta k$) for efficient phase-matching, it is also possible to satisfy the momentum conservation condition using the odd harmonics of the grating ($n=3, 5, 7\ldots$). Based on these relations, we identify the process at $\Lambda = 14.55$ \si{\micro\meter} and $\Lambda = 43.55$ \si{\micro\meter} as the first- and third-order phase-matching of $\mathrm{TM_{00}}$ + $\mathrm{TM_{00}}$ $\xrightarrow{}$ $\mathrm{TM_{00}}$ and the process at $\Lambda = 12.65$ \si{\micro\meter} and $\Lambda = 37.83$ \si{\micro\meter} as the first- and third-order matching of $\mathrm{TM_{00}}$ + $\mathrm{TM_{10}}$ $\xrightarrow{}$ $\mathrm{TM_{10}}$.

\subsection{Modal Phase-Matching}
\label{subsec:Modal phase-matching}
\begin{figure}
    \centering
    \includegraphics[width=0.7\linewidth]{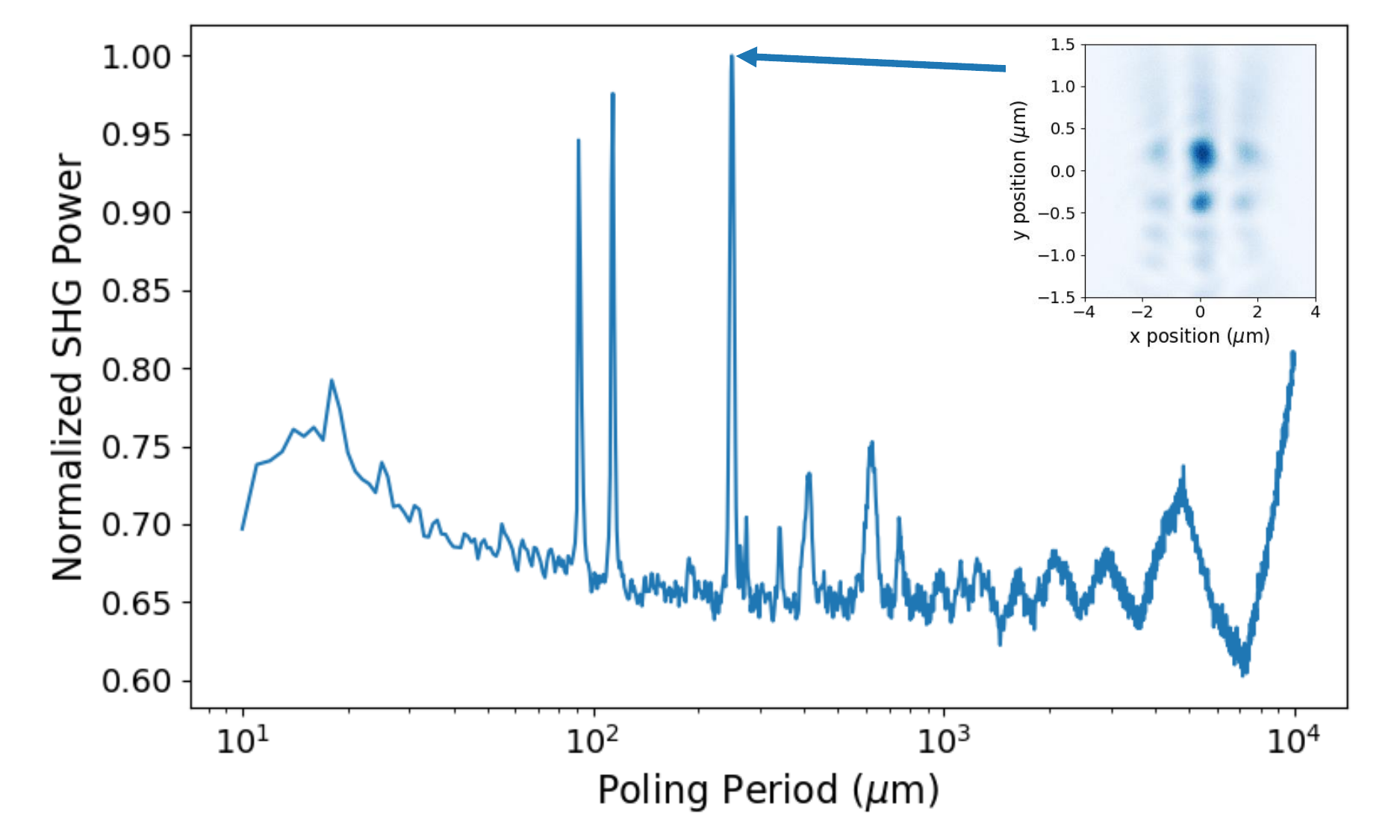}
    \caption{Measured SH power, normalized to the maximum power observed, as the poling period was changed.  Data in this figure are from a \(\mathrm{SiN}\) waveguide deposited by SVM pumped with a CW 1601.5 nm laser.  The inset image shows the measured SH mode-shape for the poling period that generated the most SHG power.}
    \label{fig:Modal Phase-matching}
\end{figure}

Occasionally, when scanning the pump wavelength, we observed an elevated SHG baseline power with no applied illumination pattern. This effect is caused by modal phase-matching, where the waveguide dispersion causes the wavevector mismatch ($\Delta k$) to be near zero for certain spatial mode mixing processes without any external patterning of the $\chi^{(2)}$. Fig.~\ref{fig:Modal Phase-matching} shows the measured SH power from an \(\mathrm{SiN}\) waveguide pumped with a 1601.5 nm CW source as the programmable poling period was scanned.  Fig.~\ref{fig:Modal Phase-matching} is unique because the background SH power compared to the maximum SH power shows a baseline ratio no lower than 60 \%.  This is dissimilar to other SHG data shown in Fig.~\ref{fig:CW-NLO}, where the background SHG power at non-phase-matched poling periods is vanishingly small compared to the maximum SHG power.  Additionally, Fig.~\ref{fig:Modal Phase-matching} is unique because the poling period that shows the maximum conversion efficiency is at 249 \si{\micro\meter}.  While the exact FH modes that interact in this process are unknown, we verified that the SH generates into the \(\mathrm{TM_{\text{21}}}\) mode.

Although demonstrating modal phase-matching was not the aim of this work, the highly multimode nature of our waveguides means many nonlinear processes are possible.  Modal phase-matching could present a way to dramatically reduce the waveguide dimensions while maintaining long poling periods.  This could be done by engineering the waveguide dispersion such that the desired process is nearly phase-matched, which means the poling period will increase into the regime allowed by the limitations of field fringing and imaging resolution, as discussed in Sec.~\ref{subsec:field fringing} and Sec.~\ref{sec:optical setup}.

\subsection{Multimode Interference Measurements}
\label{subsec:MM Interference}
Sec.~\ref{subsec:many modes} discusses how the programmable nonlinear device can generate a variety of spatial modes in the SH.  This section describes results that interfere different SH spatial modes to produce new output amplitude and phase profiles.  Instead of patterning the entire length of the straight waveguide with the same poling period, we partition the waveguide into several sections and independently tune the poling period of each section and the relative phases between the sections.

\begin{figure}
    \centering
    \includegraphics[width=0.7\linewidth]{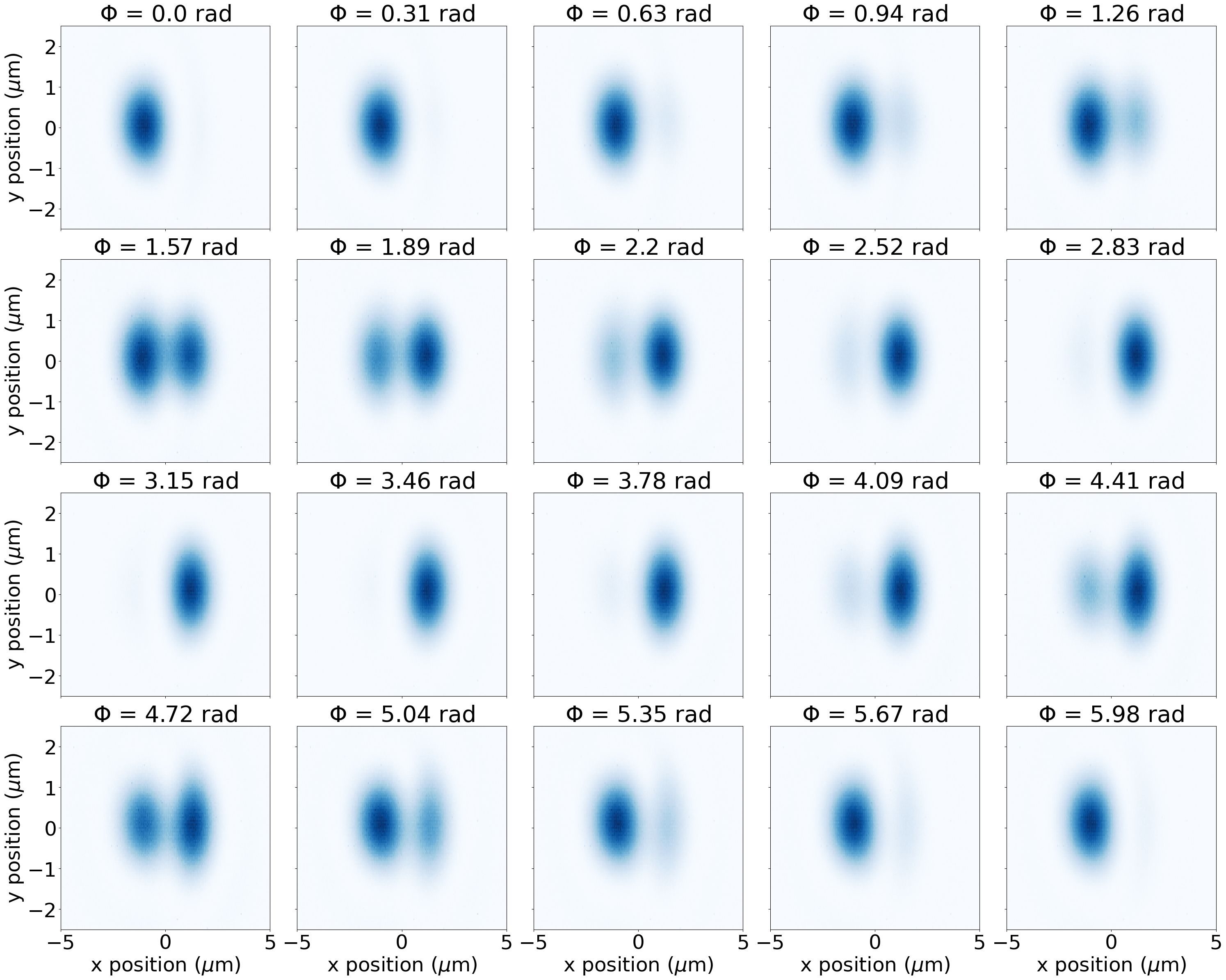}
    \caption{Images of SHG as the relative phase between a \(\mathrm{TM_{00}}\) mode and a \(\mathrm{TM_{10}}\) mode is changed.}
    \label{fig:TM00-TM10}
\end{figure}
\begin{figure}
    \centering
    \includegraphics[width=0.7\linewidth]{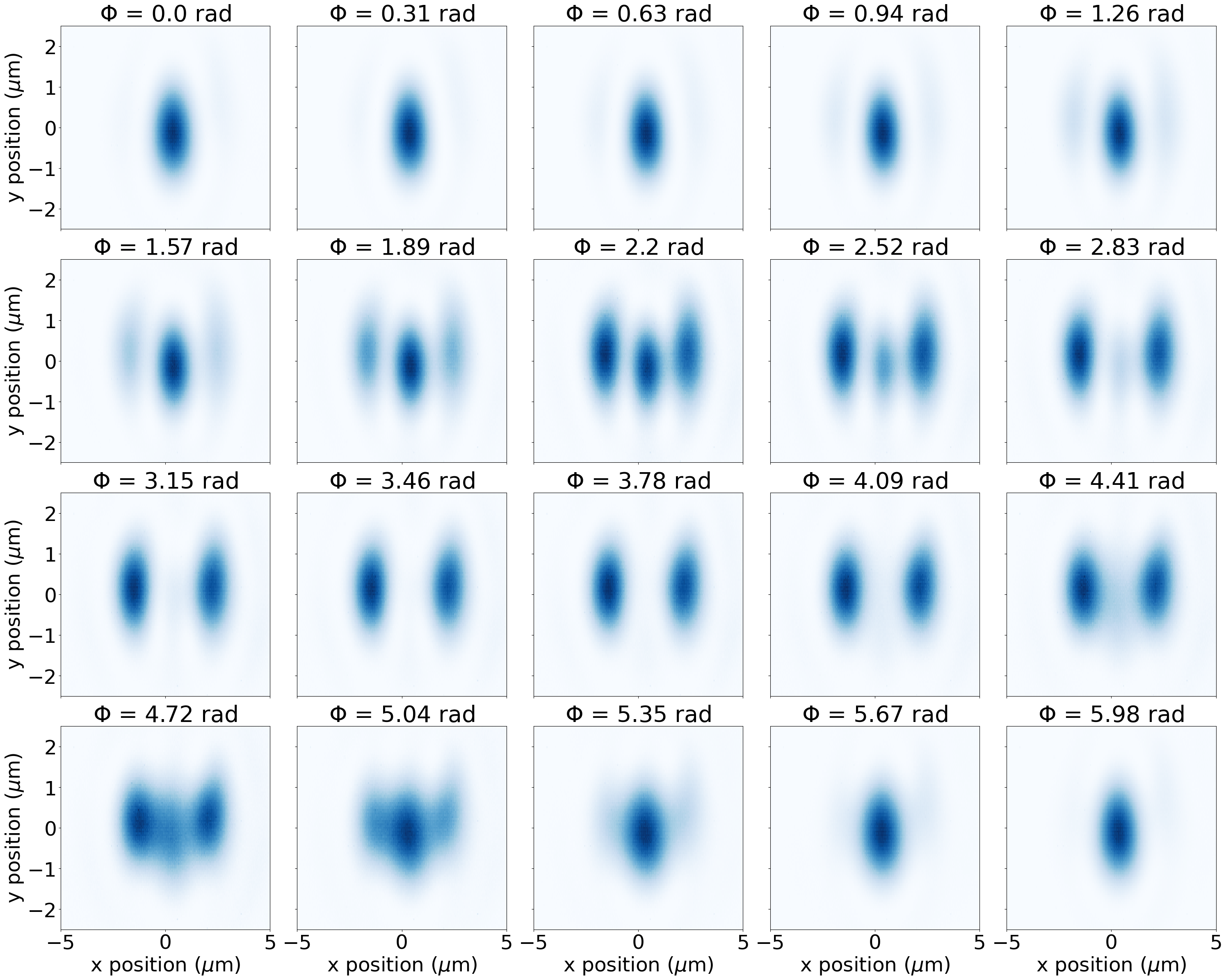}
    \caption{Images of SHG as the relative phase between a \(\mathrm{TM_{00}}\) mode and a \(\mathrm{TM_{20}}\) mode is changed.}
    \label{fig:TM00-TM20}
\end{figure}
\begin{figure}
    \centering
    \includegraphics[width=0.7\linewidth]{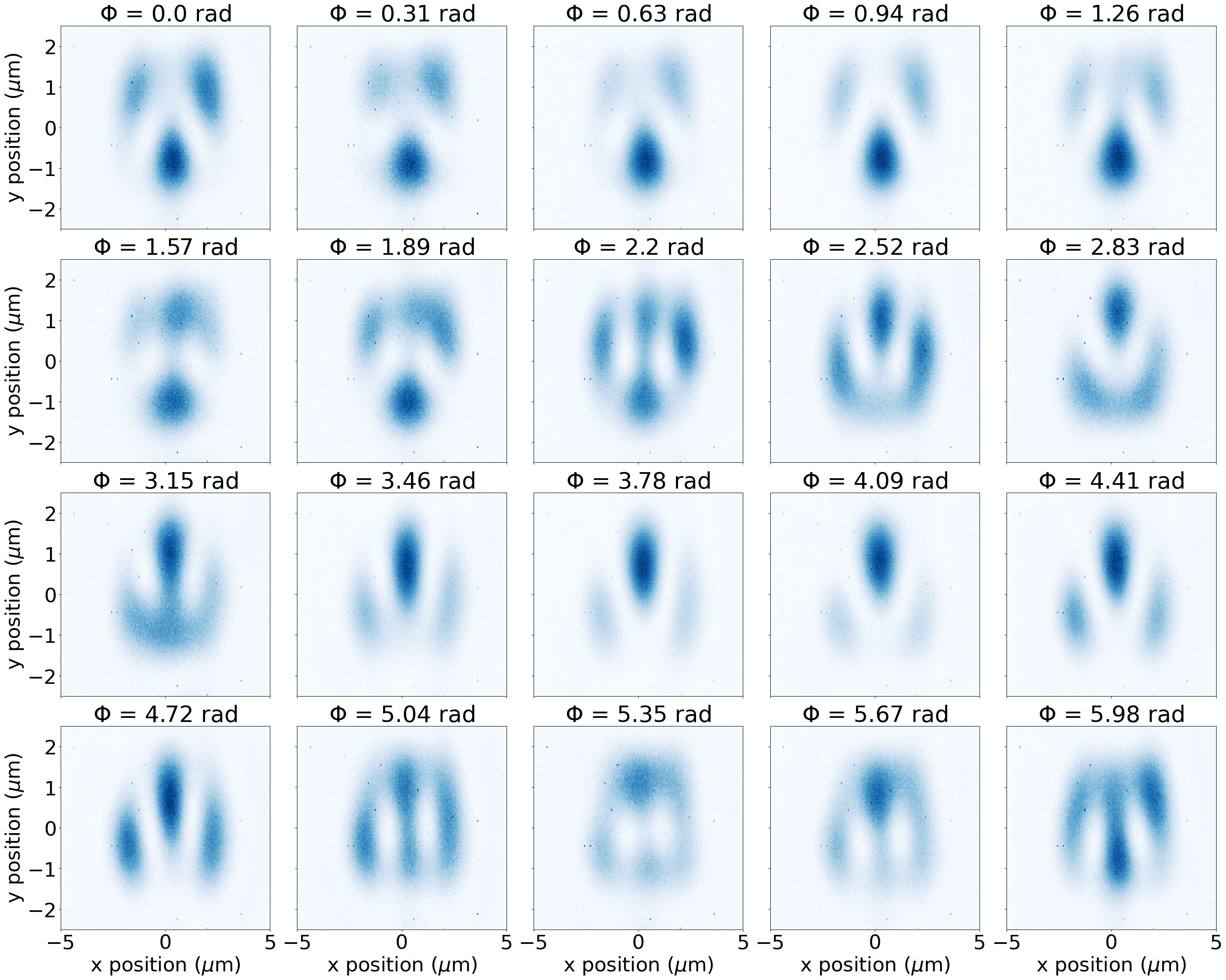}
    \caption{Images of SHG as the relative phase between a \(\mathrm{TM_{01}}\) mode and a \(\mathrm{TM_{20}}\) mode is changed.}
    \label{fig:TM01-TM20}
\end{figure}

Fig.~\ref{fig:TM00-TM10} shows interference between a \(\mathrm{TM_{00}}\) mode and a \(\mathrm{TM_{10}}\) mode, Fig.~\ref{fig:TM00-TM20} shows interference between a \(\mathrm{TM_{00}}\) mode and a \(\mathrm{TM_{20}}\) mode, and Fig.~\ref{fig:TM01-TM20} shows interference between a \(\mathrm{TM_{01}}\) mode and a \(\mathrm{TM_{20}}\) mode.  In these experiments, the length of each grating was initialized such that the power in each mode is roughly identical.  

\subsection{Orbital Angular Momentum Mode Generation and Characterization}
\label{subsec:OAM generation}
Sec.~\ref{subsec:MM Interference} discusses how the nonlinear waveguide device can actively change the interference between different spatial modes generated on-chip.  In this section, we demonstrate the on-chip generation and \textit{in situ} phase tomography of an orbital angular momentum (OAM) mode. OAM modes are characterized by a helical phase front that rotates in the transverse plane orthogonal to the propagation direction \cite{yao2011OAM, yang2026OAMoverview}. While such modes are conventionally generated using free-space optics such as tailored diffraction gratings \cite{yao2011OAM}, our approach leverages the sequential ``stacking'' of distinct programmable nonlinear processes along a single waveguide to synthesize and characterize these states directly on-chip.

\begin{figure}
    \centering
    \includegraphics[width=1\linewidth]{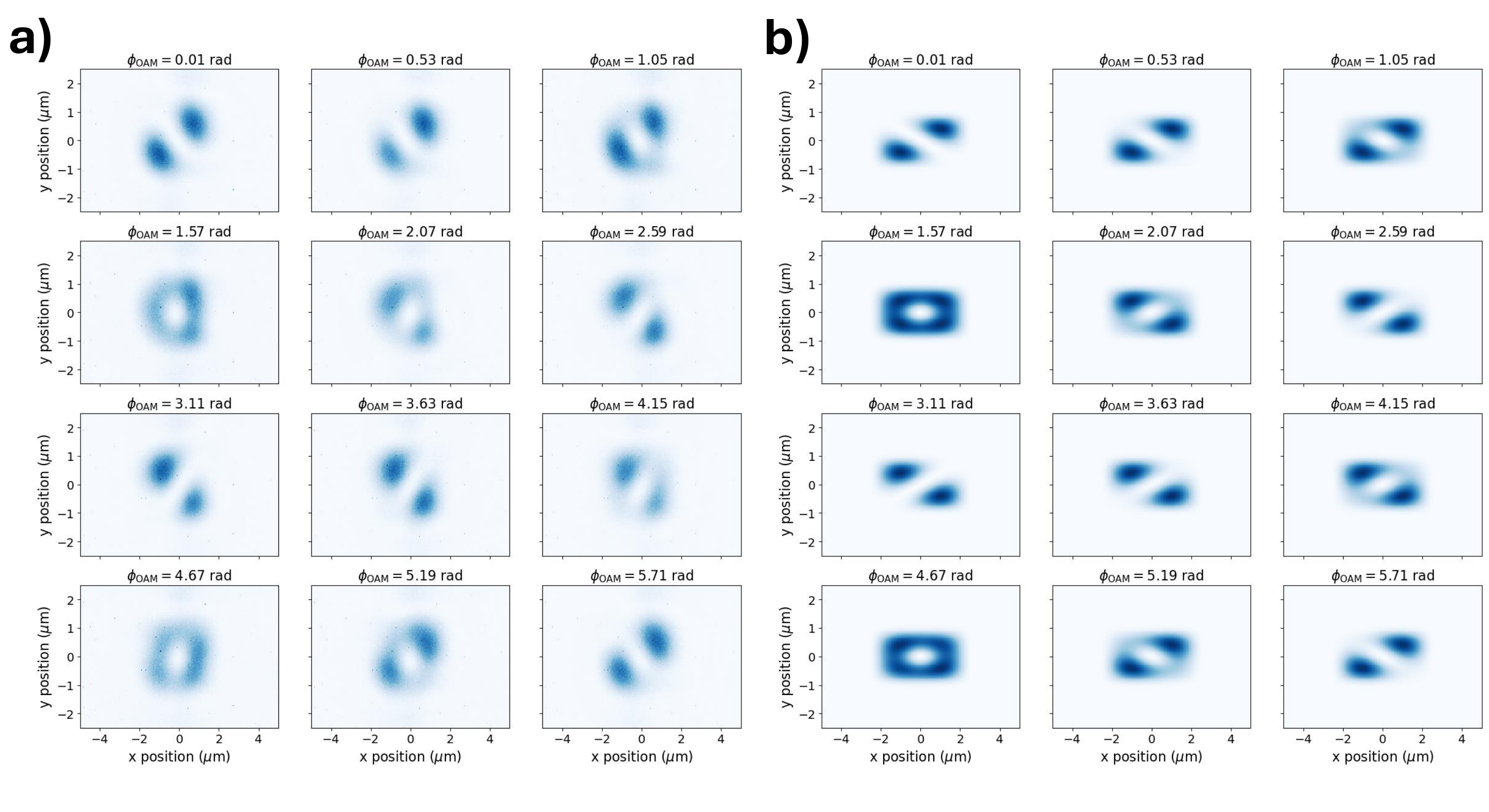}
    \caption{a) Experimentally measured intensity as the relative phase between a \(\mathrm{TM_{10}}\) mode and a \(\mathrm{TM_{01}}\) mode is changed.  b) Theoretically predicted field intensity as the relative phase between a \(\mathrm{TM_{10}}\) mode and a \(\mathrm{TM_{01}}\) mode is changed. }
    \label{fig:OAM intensity}
\end{figure}

To generate the OAM mode, we partition the programmable waveguide into two distinct longitudinal segments. The first segment is programmed with a QPM period $\Lambda_1$ chosen to phase-match the SHG of the $\mathrm{TM}_{10}$ mode, while the second segment is programmed with a period $\Lambda_2$ to generate the $\mathrm{TM}_{01}$ mode. The complex electric field of the resulting superposition can be expressed as:$$E_{\text{OAM}}(x,y) = A_{10} E_{10}(x,y) + A_{01} E_{01}(x,y) e^{i\phi_{\text{OAM}}}$$where $A_{10}$ and $A_{01}$ are the respective real amplitudes of the generated modes, $E_{10}(x,y)$ and $E_{01}(x,y)$ are their normalized transverse spatial profiles, and $\phi_{\text{OAM}}$ is their relative phase. In our experiment, we equalize the mode amplitudes ($A_{10} = A_{01} = A$) by adjusting the lengths of the programmed segments, and we deterministically tune the relative phase $\phi_{\text{OAM}}$ by shifting the longitudinal spatial position of the second QPM grating. Fig.~\ref{fig:OAM intensity} shows the measured and predicted field intensities for OAM modes as the relative phase between the \(\mathrm{TM_{10}}\) and \(\mathrm{TM_{01}}\) modes is changed.

When the relative phase is tuned to $\phi_{\text{OAM}} = \pm \pi/2$, the superposition becomes 
\begin{equation}
    E_{\text{OAM}}(x,y) = A \left[ E_{10}(x,y) \pm i E_{01}(x,y) \right]
\end{equation} 
The resulting optical intensity, $I_{\text{OAM}}(x,y) \propto |E_{\text{OAM}}(x,y)|^2$, has the characteristic ``donut'' shape associated with an OAM state of topological charge $l = \pm 1$. To verify the helical phase structure of the generated mode, we utilize the programmable waveguide to perform on-chip phase tomography. We used a third longitudinal segment programmed with a period $\Lambda_3$ chosen to generate the fundamental $\mathrm{TM}_{00}$ mode. Because the $\mathrm{TM}_{00}$ mode possesses a spatially uniform phase profile across the core, it serves as an ideal intrinsic local oscillator (LO) for interferometry. The total electric field at the output of the waveguide is the coherent sum of the OAM mode and this reference mode:$$E_{\text{tot}}(x,y) = |E_{\text{OAM}}(x,y)| e^{i\Phi_{\text{OAM}}(x,y)} + |E_{\text{ref}}(x,y)| e^{i\phi_{\text{ref}}}$$where $\Phi_{\text{OAM}}(x,y)$ is the spatially varying phase profile of the OAM mode we wish to reconstruct, $|E_{\text{ref}}(x,y)|$ is the amplitude of the $\mathrm{TM}_{00}$ reference, and $\phi_{\text{ref}}$ is the globally tunable phase of the reference mode (controlled by translating the $\Lambda_3$ grating along the propagation axis). The corresponding interference intensity captured by the camera for a given reference phase $\phi_{\text{ref}}$ is:$$I(x,y, \phi_{\text{ref}}) = I_{\text{OAM}}(x,y) + I_{\text{ref}}(x,y) + 2 \sqrt{I_{\text{OAM}}(x,y) I_{\text{ref}}(x,y)} \cos(\Phi_{\text{OAM}}(x,y) - \phi_{\text{ref}})$$

\begin{figure}
    \centering
    \includegraphics[width=1\linewidth]{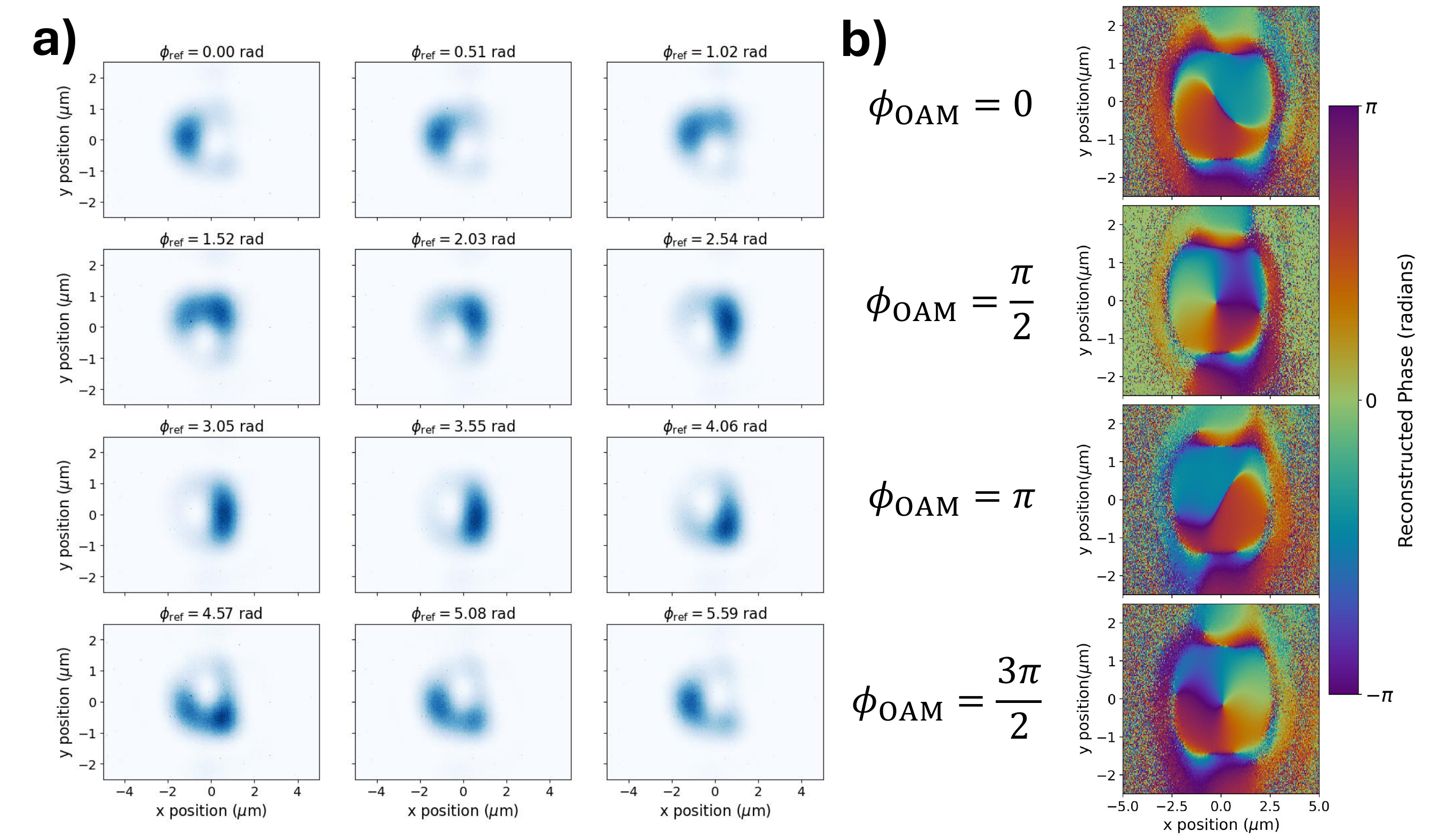}
    \caption{a) Measured SH mode profile for \(\mathrm{\phi_{OAM}} = \frac{3\pi}{2}\) as \(\mathrm{\phi_{ref}}\) of the \(\mathrm{TM_{00}}\) mode was swept.  b) Reconstructed phase profiles of \(\mathrm{TM_{10}}\) and \(\mathrm{TM_{01}}\) interferences with different relative angles between the modes.}
    \label{fig:OAM phase tomography}
\end{figure}

To robustly retrieve the spatially resolved phase profile $\Phi_{\text{OAM}}(x,y)$ against experimental noise, we employ an overdetermined $N$-step phase-shifting interferometry algorithm. We sequentially sweep the reference phase through $N \ge 100$ discrete steps, capturing an interference pattern $I_k(x,y)$ at each phase shift $\phi_k$. By analyzing the harmonic modulation of the intensity at each transverse coordinate $(x,y)$ across the comprehensive phase sweep, the target phase profile is uniquely and accurately extracted via:$$\Phi_{\text{OAM}}(x,y) = \arctan \left( \frac{\sum_{k=1}^N I_k(x,y) \sin(\phi_k)}{\sum_{k=1}^N I_k(x,y) \cos(\phi_k)} \right)$$ Fig.~\ref{fig:OAM phase tomography}(a) shows a few images of the interference between the local oscillator ($\mathrm{TM}_{00}$) and different superpositions of the $\mathrm{TM}_{01}$ and $\mathrm{TM}_{10}$ modes for different relative phases of the local oscillator. Fig.~\ref{fig:OAM phase tomography}(b) shows the retrieved spatial phase profile of different superpositions of the $\mathrm{TM}_{01}$ and $\mathrm{TM}_{10}$ modes.

Finally, we note that the devices utilized in these experiments feature a rectangular waveguide cross-section. Consequently, the $\mathrm{TM}_{10}$ and $\mathrm{TM}_{01}$ modes are not perfectly symmetric and exhibit slight geometric asymmetries in their spatial extent, which manifest as sidelobes rather than a perfectly isotropic donut profile. Future implementations utilizing a perfectly square waveguide geometry would enforce strict spatial symmetry between the constituent modes, thereby suppressing these sidelobes and yielding a pristine, symmetric OAM intensity profile.

\section{Spectral Shaping of Second Harmonic Pulse}
\label{sec:spectral shaping}
\subsection{Optimization Routine}
\label{subsec:spectral-cross-section}
Previous sections focused on experiments utilizing a CW light source for SHG.  While pulsed FH sources suffer from walk-off effects due to finite waveguide GVM, they enable broadband SH output pulses. In this section, we describe how we engineer the output spectrum of a SHG pulse using programmable $\chi^{(2)}$.  

Fig.~\ref{fig:Spectral Broadening} shows how the output SH spectrum changes as the programmable poling period ($\Lambda$) and the FH average pulse power ($P_{\text{FH}}$) increase.  The implications of changing $P_{\text{FH}}$ will be discussed further in Sec.~\ref{subsec:Spec Broadening}, but these experiments determine the effective bandwidth that an SH pulse can be shaped over and which poling periods are required to shape in this bandwidth.  Using this range of poling periods and the full tunability of our programmable $\chi^{(2)}$ platform, we separate a waveguide structure into several partitions that are each described by a poling period ($\Lambda$), phase ($\phi$), and grating intensity ($I$).  In this experiment, $I$ is defined as the on-state pixel value between 0 (totally off) and 1 (totally on) in a poling partition.  Unlike CW experiments, where the intensity of a poling partition was always 1, spectral pulse-shaping takes advantage of this intensity degree-of-freedom to gain additional control over the output spectrum.  For a waveguide with $N$ partitions, this means the poling structure is characterized by $3N$ parameters.  While CW experiments could be efficiently optimized with brute-force scanning of the poling period and phase, spectral-pulse shaping has greater interplay between the different parameters characterizing the poling structure.  To discover the optimal poling structure in a more time-efficient manner, we use a random perturbative optimization routine to refine our parameters.  

\begin{figure}
    \centering
    \includegraphics[width=1.0\linewidth]{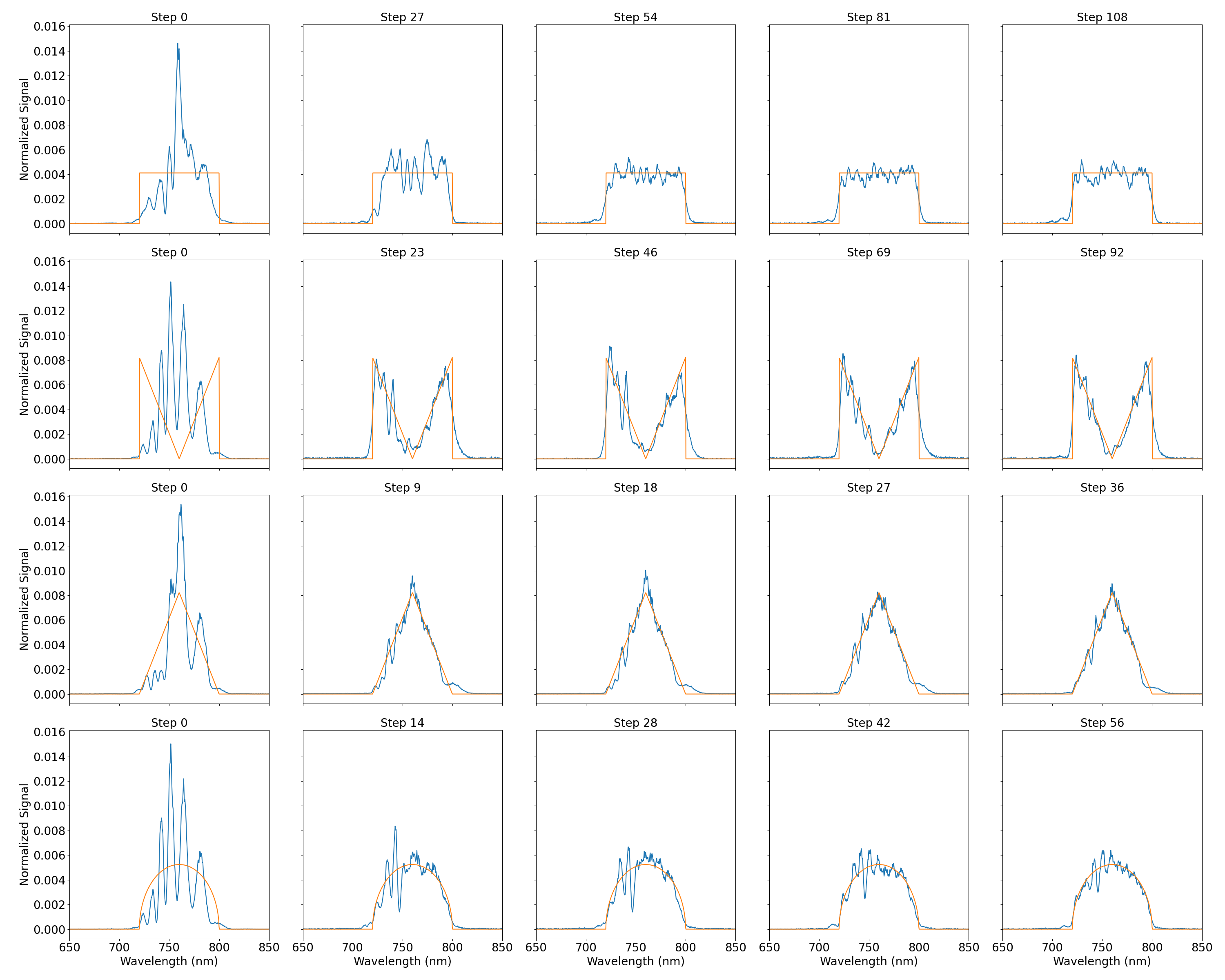}
    \caption{Measured output SH spectrum as the square, inverse-triangle, triangle, and circular spectral shapes were trained.  The step count during optimization is listed above each panel, and all optimizations were performed over the same 80 nm bandwidth.  The target spectral shape is in orange and the experimentally measured shape is in blue.  Plots are normalized such that the sum under each curve is 1.}
    \label{fig:Optimization Path Geometry}
\end{figure}
\begin{figure}
    \centering
    \includegraphics[width=1\linewidth]{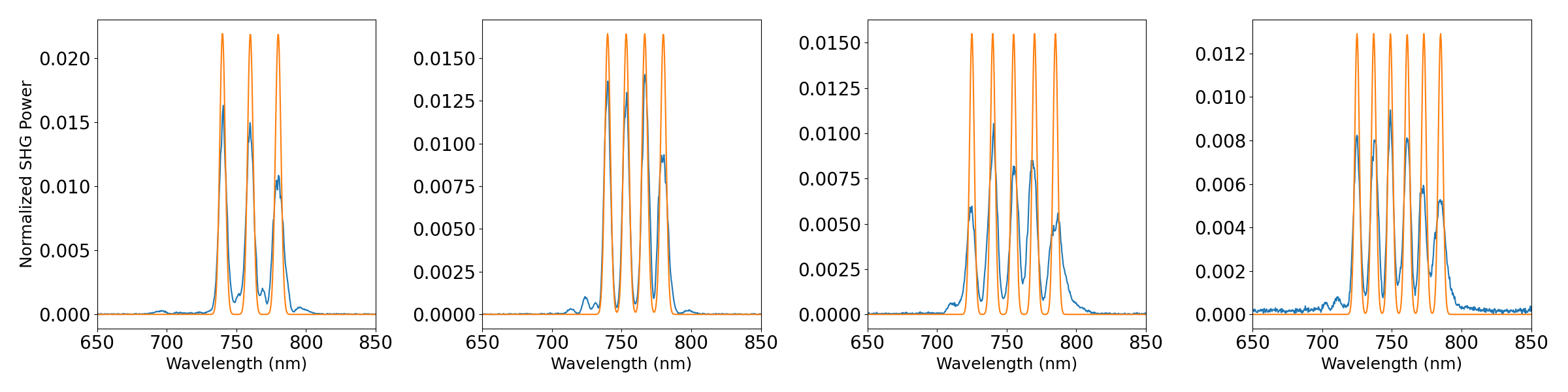}
    \caption{Optimized output spectrum with three, four, five, and six Gaussian peaks. The target spectral shape is in orange and the experimentally measured shape is in blue. Plots are normalized such that the area is 1.}
    \label{fig:Multiple Peaks}
\end{figure}

During optimization, we randomly perturb each set of parameters.  First, we perturb all the poling periods within a range of $\Delta\Lambda$, all the phases within a range of $\Delta\phi$, and all the intensities within a range of $\Delta I$.  For a perturbation to be accepted, it must minimize the mean squared error between a normalized experimental spectrum and normalized target spectrum.  This process is repeated $M$ times per parameter set, and we loop through the three parameter sets $K$ times.  This process is repeated another $G$ times with the perturbation scale, $\Delta \Lambda$, $\Delta \phi$, and $\Delta I$, reduced by a factor of $\tau$.  Lastly, to avoid getting stuck in a local minimum, we rerun these optimizations $W$ times.  While these parameters were adjusted for each optimization trial, common parameters implemented in this work are $\Delta\Lambda = 0.005$ \si{\micro\meter}, $\Delta\theta = 0.15$ radians, $\Delta I = 0.05$, $M=10$, $K=30$, $G = 3$, $\tau = 4$, and $W=9$. 

Fig.~\ref{fig:Optimization Path Geometry} shows how different geometric shapes presented in the main text are steadily fit to their respective target spectra with successive optimization steps. Fig.~\ref{fig:Multiple Peaks} shows optimized output spectra where the targets contain different numbers of Gaussian peaks in the SH. These results show the flexibility of our optimization routine to achieve many possible output spectra using the same pump laser and waveguide device.

To collect these and other spectral-shaping results, we use the same optical setup presented in Sec.~\ref{sec:optical setup}.  Instead of using a PMT to measure the total optical power or a camera to image the mode shape, we use a spectrometer (OCEAN-FX-VIS-NIR-ES; Ocean Optics) to measure the generated signal.  To prevent detector saturation, we insert absorptive filters into the optical path between the output objective and spectrometer.  We use an ELMO pulsed laser (ELMO-HP; Menlo Systems) to pump our waveguides and place a filter between it and the waveguide to control the FH power, as will be discussed in Sec.~\ref{subsec:Spec Broadening}. The results in Sec.~\ref{subsec:spectral-cross-section} and Sec.~\ref{subsec:Spec Broadening} use straight waveguides with \(\mathrm{SiN}\) deposited in-house.

\subsection{Spectral Broadening with Self-phase Modulation}
\label{subsec:Spec Broadening}
In addition to the programmable $\chi^{(2)}$ nonlinearity, the silicon nitride that constitutes the waveguides has an intrinsic $\chi^{(3)}$ nonlinearity. Because the third-order nonlinear polarization scales as the cube of the electric field, whereas the second-order polarization scales as its square, $\chi^{(3)}$ contributions tend to be more pronounced in devices with tighter field confinement. Compared to the slab waveguide used for spectral engineering in Ref.~\cite{Ryo-paper}, the etched waveguides used in this work offer tighter transverse field confinement, allowing us to demonstrate a unique interplay between $\chi^{(2)}$ and $\chi^{(3)}$ nonlinearities.

\begin{figure}
    \centering
    \includegraphics[width=0.6\linewidth]{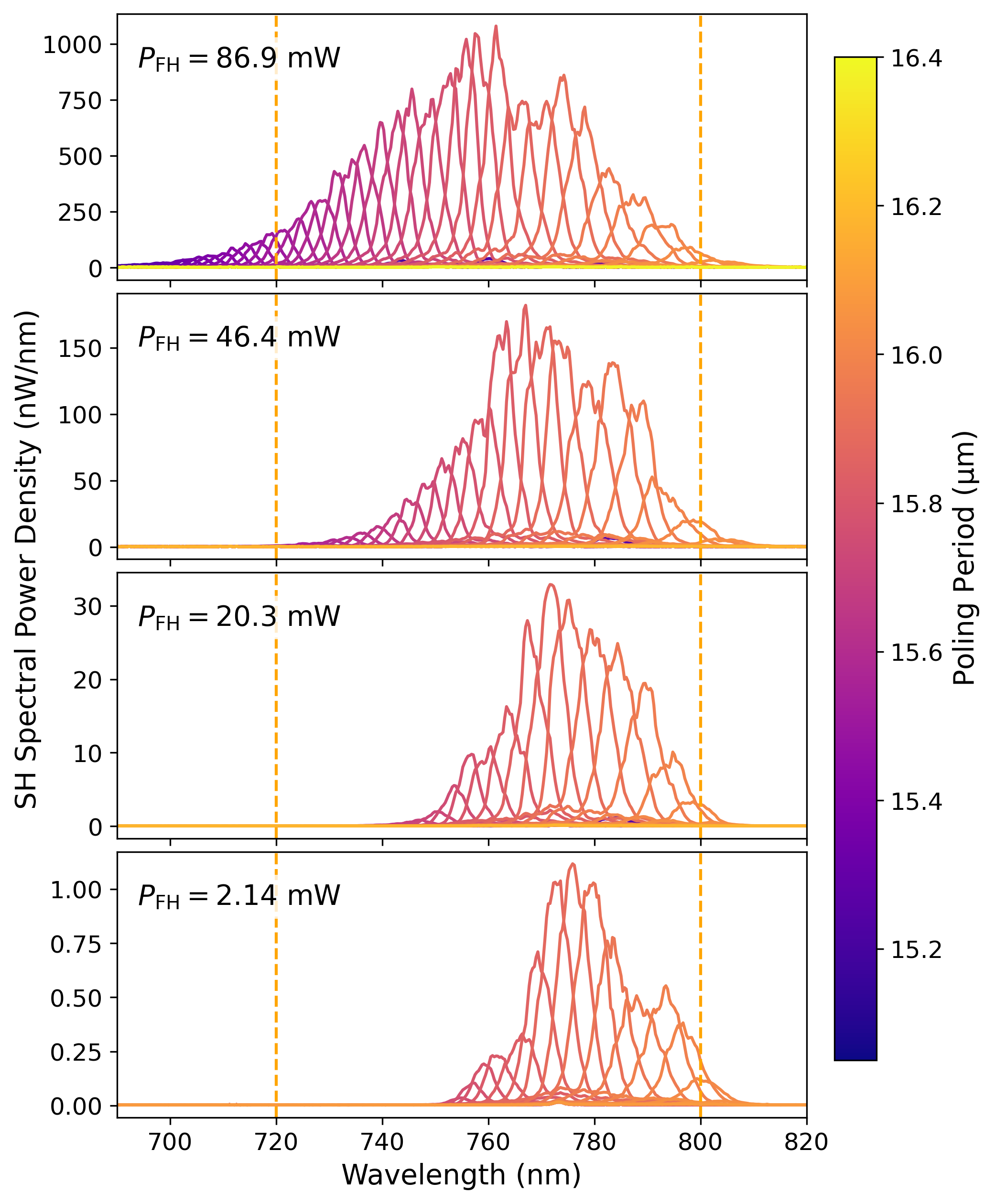}
    \caption{Measured SH spectra as the poling period and average pump power of an FH pulsed laser are swept.  The dashed orange lines show the bandwidth of spectral shaping tasks performed using \(\mathrm{SiN}\) deposited in-house.}
    \label{fig:Spectral Broadening}
\end{figure}

When we pumped the waveguide with ultrashort pulses of varying power, we observed that the pump spectrum broadens due to $\chi^{(3)}$ self-phase modulation (SPM). A consequence of this spectral broadening is shown in Fig.~\ref{fig:Spectral Broadening}, where we plot the output SHG spectrum for various QPM periods. The SHG spectrum broadened as the pump power was increased, reflecting the broadening of the pump spectrum. The average pump power was tuned by attenuating the laser before it was coupled into the programmable waveguide device. At $P_{\text{FH}} = \SI{86.9}{mW}$, we achieved SH spectral shaping over a bandwidth of \SI{80}{nm}, twice that demonstrated previously with a programmable nonlinear slab waveguide~\cite{Ryo-paper}.

In the experiment, we did not use the maximum available pulse power, because at high pump power dispersive waves emerged at wavelengths close to that of the SHG, producing a strong background that impaired our ability to generate arbitrary spectra. This effect could be avoided by engineering the waveguide dispersion to shift the dispersive waves away from the SH band.

\subsection{Spectral Shaping over Time}

\begin{figure}
    \centering
    \includegraphics[width=1\linewidth]{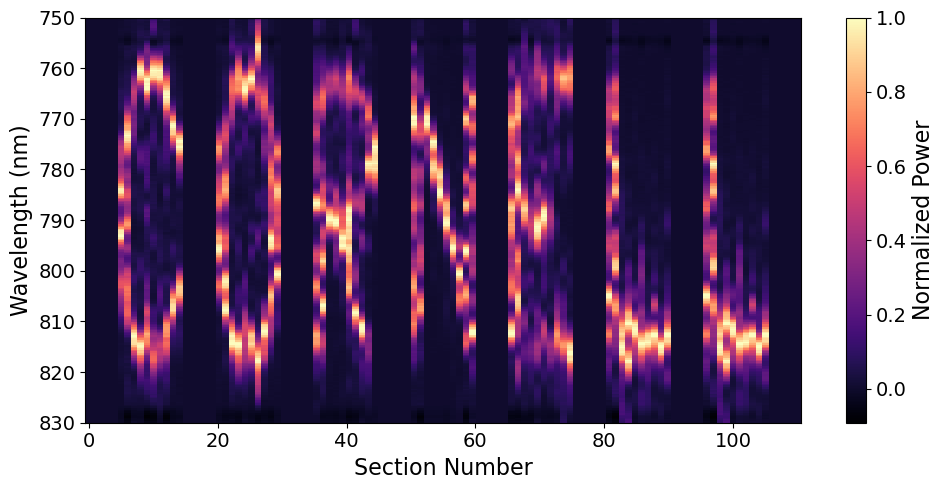}
    \caption{Optimized output spectra spelling the word ``CORNELL'' when placed in the appropriate order.}
    \label{fig:Temporal Shaping}
\end{figure}

In addition to shaping an individual spectrum, we can shape many spectra in a reprogrammable order.  When the spectrum is plotted as a function of time, this allows us to shape different patterns in a two-dimensional space.  Fig.~\ref{fig:Temporal Shaping} shows how, using the same optimization procedure in Sec.~\ref{subsec:spectral-cross-section}, we can shape many spectra and change the poling structure such that the spectra, as a function of time, spell ``CORNELL''.  Unlike previous demonstrations in Sec.~\ref{sec:spectral shaping} which used \(\mathrm{SiN}\) deposited in-house, this result used \(\mathrm{SiN}\) deposited by SVM.  While the SH spectral shaping bandwidth remained 80 nm, the range of wavelengths was shifted due to the different material dispersion.  The time axis was left in units of partition number because the temporal spacing between sections can be tuned anywhere above the 1 second response time of the SLM.

While previous work \cite{Ryo-paper} demonstrated a similar result as presented in Fig.~\ref{fig:Temporal Shaping}, the result in this work shows that channel waveguides can show the same flexibility to generate many arbitrary output spectra using one optimization routine.  Furthermore, it emphasizes the advantage of using channel waveguides to utilize the native $\chi^{(3)}$ to broaden the bandwidth over which $\chi^{(2)}$ spectral shaping takes place.

\section{Device Improvements}
\label{sec:Device Improvements}
\subsection{SiN Fabrication Improvements}

As discussed in Sec.~\ref{sec:loss characterization}, the FH losses of our waveguides are on the order of 0.3 dB/cm.  This is much higher than the state-of-the-art for \(\mathrm{SiN}\) integrated spiral waveguides, which can demonstrate losses as low as 2.4 dB/m \cite{liu2021highyield, Lipson-complete-SiN}.  While some of our elevated loss can be attributed to radiative losses through the substrate and photoconductor (as described in Sec.~\ref{subsec:leakage}), much of the loss can be attributed to material loss stemming from the use of PECVD \(\mathrm{SiN}\) instead of LPCVD \(\mathrm{SiN}\).  While LPCVD has fewer defects because of the higher deposition temperature, LPCVD films suffer from higher tensile film stress \cite{Lipson-PECVD}, making 2 \si{\micro\meter} film deposition challenging.  This compels us to use PECVD \(\mathrm{SiN}\) despite its comparatively worse optical performance.

One future fabrication route to avoid the stress issues of LPCVD \(\mathrm{SiN}\) is using a photonic damascene process \cite{Pfeiffer2016damascene, liu2021highyield}. Instead of depositing a film and etching away most of the material to pattern waveguides, the photonic damascene process relies upon patterning a thick \(\mathrm{SiO_2}\) layer and then using LPCVD to deposit \(\mathrm{SiN}\) into the etched trenches.  The trench walls allow for stress to be relieved locally instead of accumulating along the wafer-scale film.  Consequently, a thicker film can be deposited while taking advantage of the superior optical quality of LPCVD \(\mathrm{SiN}\).  An additional advantage of this fabrication process is that narrow gaps can be easily patterned at waveguide couplers, as there is no need to fill high aspect-ratio features \cite{dirani2018damage}.  In the context of the devices presented in this paper, the photonic damascene process not only offers a path to reduce propagation losses, but it also offers a compelling route to demonstrating programmable QPM on resonant structures.  As discussed in Ref.~\cite{Ryo-paper}, resonators allow the nonlinear conversion efficiency to be dramatically scaled because the resonant structure enables higher field confinement.  While resonant structures limit which spectral regions can be phase-matched (as resonant conditions must be met), these structures offer a compact method to enhance device performance.

\subsection{Hybrid Phase Matching with an Asymmetric Film Stack}
\label{sec:hybrid-phase-matching}
In our approach to programmable nonlinearity, it is challenging to realize a QPM grating period below the few-micron scale because of electric field fringing and projection illumination resolution, which limits the applicability of the photoconductive-programming technique to situations where the inherent phase mismatch of the interacting light waves is large. A possible resolution to this challenge is to engineer the device structures so that the phase mismatch of the interacting modes is largely canceled before applying the programmable $\chi^{(2)}$ hologram, and we use programmable QPM on top to compensate for the residual phase mismatch. Specifically, here we show that modal phase matching (which is experimentally discussed in Sec.~\ref{subsec:Modal phase-matching}) can be a powerful means to not only realize approximate phase matching but also group-velocity matching, potentially enabling highly efficient and broadband frequency conversions.

\begin{figure}[h]
    \centering
    \includegraphics[width=\linewidth]{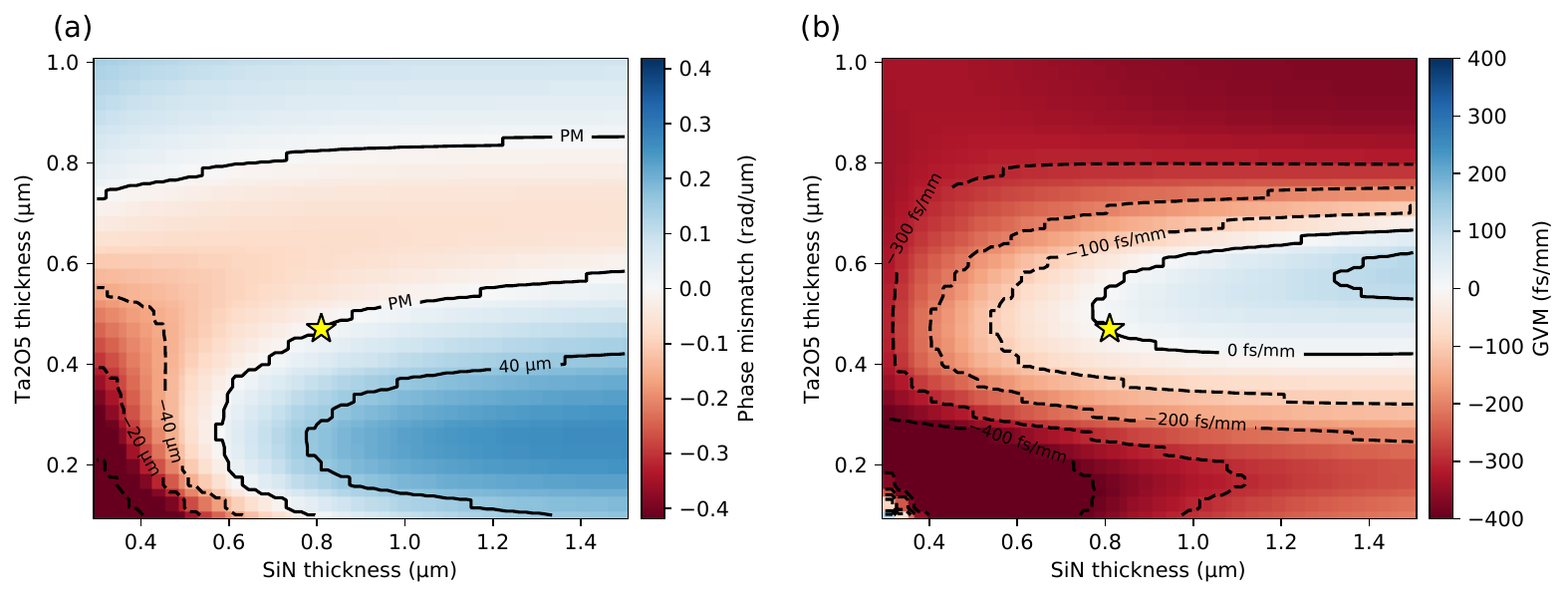}
    \caption{(a) The phase mismatch $\Delta k$ and (b) GVM between the FH \(\mathrm{TM_0}\) mode and the SH \(\mathrm{TM_1}\) modes for various thicknesses of $d_\mathrm{SiN}$ and $d_\mathrm{Ta_2O_5}$~\cite{black2021tantala}. The yellow stars represent a device geometry where both the phase mismatch and the GVM are close to zero.}
    \label{fig:slab-modal-phase-matching}
\end{figure}

For concreteness, let us consider a slab waveguide consisting of \(\mathrm{SiO_2}\) claddings and a bilayer core comprised of \(\mathrm{SiN}\) and \(\mathrm{Ta_2O_5}\) with thicknesses of $d_\mathrm{SiN}$ and $d_\mathrm{Ta_2O_5}$, respectively. For the frequency conversion process, we consider the \(\mathrm{TM_0}\) mode of the FH at \SI{1560}{nm} and the \(\mathrm{TM_1}\) mode of the SH at \SI{780}{nm}. Typically, higher-order modes have lower indices than the fundamental mode. Thus, the \(\mathrm{TM_1}\) mode of the SH can have a similar index to the \(\mathrm{TM_0}\) mode of the FH, which relaxes the phase-matching condition. In Fig.~\ref{fig:slab-modal-phase-matching}(a), we show the values of the phase mismatch $\Delta k$ for various $d_\mathrm{SiN}$ and $d_\mathrm{Ta_2O_5}$. We can see a contour over which $\Delta k$ is essentially zero, and even away from there, a large area has a corresponding QPM period greater than \SI{40}{\micro m}. Furthermore, in Fig.~\ref{fig:slab-modal-phase-matching}(b), we show the GVM between the FH and the SH modes. Notably, we can find a combination of the film thicknesses so that both the phase mismatch and GVM are close to zero. This is a notable point since zero GVM eliminates temporal walkoff, allowing us to realize highly efficient broadband frequency conversions with ultrashort pulses~\cite{jankowski2021dispersion}.

We note that the asymmetric film stack is advantageous from the perspective of nonlinear coupling. Because of the asymmetry of the dielectric permittivity of the core materials ($6\sim 7$ for \(\mathrm{SiN}\) and $25$ for \(\mathrm{Ta_2O_5}\)), the bias electric field is expected to predominantly impinge on the \(\mathrm{SiN}\) layer. This means the electric-field-induced $\chi^{(2)}$ nonlinearity would also have an asymmetric distribution in the vertical direction, which ensures that the mode overlap integral between the \(\mathrm{TM_0}\) and \(\mathrm{TM_1}\) modes stays large~\cite{du2023modal}.

\subsection{Potential Improvement on the Photoconductor}
\label{sec:photoconductor-improvement}
The maximum update speed of the programmable nonlinearity is upper-bounded by the photoconductor. More specifically, we can write the bright state and dark state time constants of the device as
\begin{align}
    \tau_j=\frac{\epsilon_0\epsilon_\mathrm{wg} d_\mathrm{PC}}{\sigma_jd_\mathrm{wg}},
\end{align}
where $j\in\{\mathrm{b, d}\}$ denotes the labels ``bright'' and ``dark'', $\epsilon_\mathrm{wg}$ is the relative permittivity of the waveguide, and $d_\mathrm{wg}$ and $d_\mathrm{PC}$ are the thicknesses of the waveguide and photoconductive layers, respectively. To realize considerable programmable nonlinearity, the frequency of the bias voltage has to be between $(2\pi\tau_\mathrm{b})^{-1}$ and $(2\pi\tau_\mathrm{d})^{-1}$, and the maximum update speed of the device is set by the bright-state time constant $\tau_\mathrm{b}$. For the case of \(\mathrm{Si}\)-rich \(\mathrm{SiN}\), we found $\sigma_\text{b}\approx \SI{1e-8}{\ohm^{-1}m^{-1}}$ and $\sigma_\text{d}\approx \SI{1e-10}{\ohm^{-1}m^{-1}}$, corresponding to $\tau_\mathrm{b}\approx \SI{8}{ms}$, or a maximum update speed of $(2\pi\tau_\mathrm{b})^{-1}\approx \SI{20}{Hz}$. Here, we assumed $d_\mathrm{PC}=\SI{9}{\micro m}$, $d_\mathrm{wg}=\SI{4}{\micro m}$, and $\epsilon_\mathrm{wg}=3.9$.

The speed of the device could be significantly improved by employing photoconductive materials that have stronger photoconductivities. For instance, Ref.~\cite{kumata1986a-Si} studied the photoconductivity of amorphous Si (a-Si) and reported the bright and dark state conductivities of $\sigma_\text{b}=\SI{1e-5}{\ohm^{-1}m^{-1}}$ and $\sigma_\text{d}=\SI{1e-10}{\ohm^{-1}m^{-1}}$, respectively, at an illumination intensity of \SI{0.3}{mW/cm^2} using a He-Ne laser. This bright-state conductivity corresponds to a time constant of $\tau_\mathrm{b}=\SI{8}{\micro s}$, or equivalently, a maximum update speed of $(2\pi\tau_\mathrm{b})^{-1}\approx \SI{20}{kHz}$. Assuming that the photoconductivity scales linearly with illumination intensity, at an illumination intensity of $\SI{18}{mW/cm^2}$, which is similar to the intensity we used in this work, we would have $(2\pi\tau_\mathrm{b})^{-1}\approx \SI{1.2}{MHz}$. These numbers suggest that there are realistic routes to reach operating speeds exceeding a megahertz with improvements to the photoconductor.

\subsection{Enabling DC Operation and Enhancing Nonlinearity with Dynamic Modulation of Programming Illumination}
\label{sec:dynamic-illumination}
Presently, our approach to realizing programmable nonlinearity requires the use of an AC bias voltage so that the effective impedance of the waveguide core sits between the bright and dark state impedances of the photoconductor (see Sec.~\ref{sec:Circuit Model}). Consequently, the amplitude of the realized nonlinearity is not constant in time but is periodically modulated, which hinders the application of the device to situations requiring continuous operation. Here, we propose an approach that circumvents this limitation and realizes DC operation of the device with a temporally constant effective $\chi^{(2)}$ nonlinearity via dynamical modulation of the programming illumination. Notably, this scheme could also enhance the programmable nonlinearity by a factor of $2$ without changing the device design.

\begin{figure}[h]
    \centering
    \includegraphics[width=0.5\linewidth]{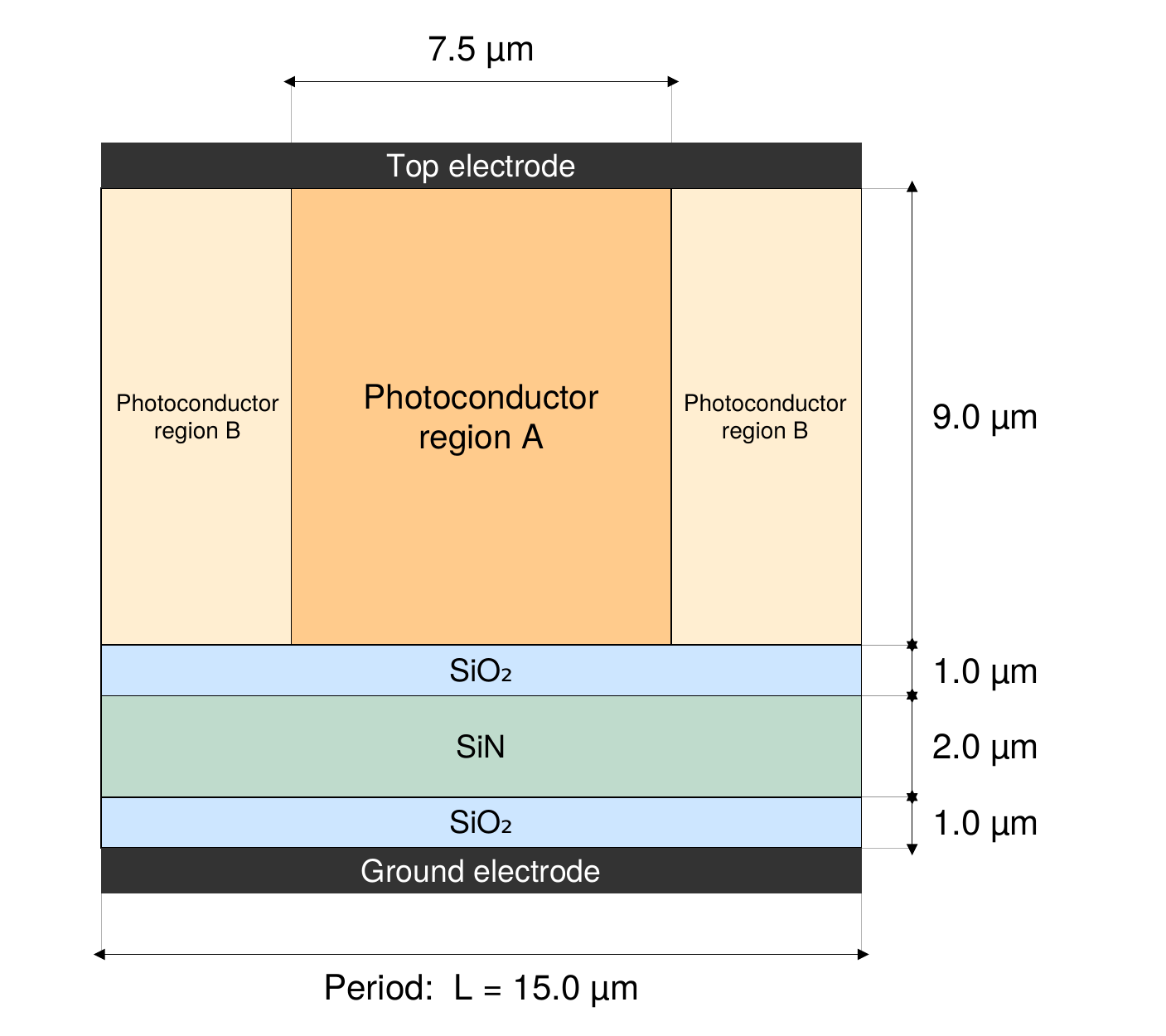}
    \caption{Illustration of the model of a programmable nonlinear waveguide. We assumed relative permittivities of $3.9$, $6$, and $8$ for \(\mathrm{SiO_2}\), SiN, and the photoconductor, respectively. The bright and dark state conductivities are $\sigma_\text{b}=\SI{1e-7}{S/m}$ and $\sigma_\text{d}=\SI{1e-10}{S/m}$, respectively. A periodic boundary condition with period $L=\SI{15}{\micro m}$ is set for the horizontal direction.}
    \label{fig:photoconductor_illustration}
\end{figure}

To introduce the dynamic illumination scheme, we first construct a simplified model of a programmable nonlinear waveguide, shown in Fig.~\ref{fig:photoconductor_illustration}. On top of the SiN waveguide, a photoconductor layer and a top electrode are stacked along the vertical (i.e., $y$) direction. The substrate and top electrode are connected to the ground and the bias voltage source, respectively. We assume a periodic boundary condition along the horizontal (i.e., $x$) direction with a nominal period of $L=\SI{15}{\micro m}$, and we partition the photoconductor into two regions, A and B, of equal width.

\begin{figure}[h]
    \centering
    \includegraphics[width=\linewidth]{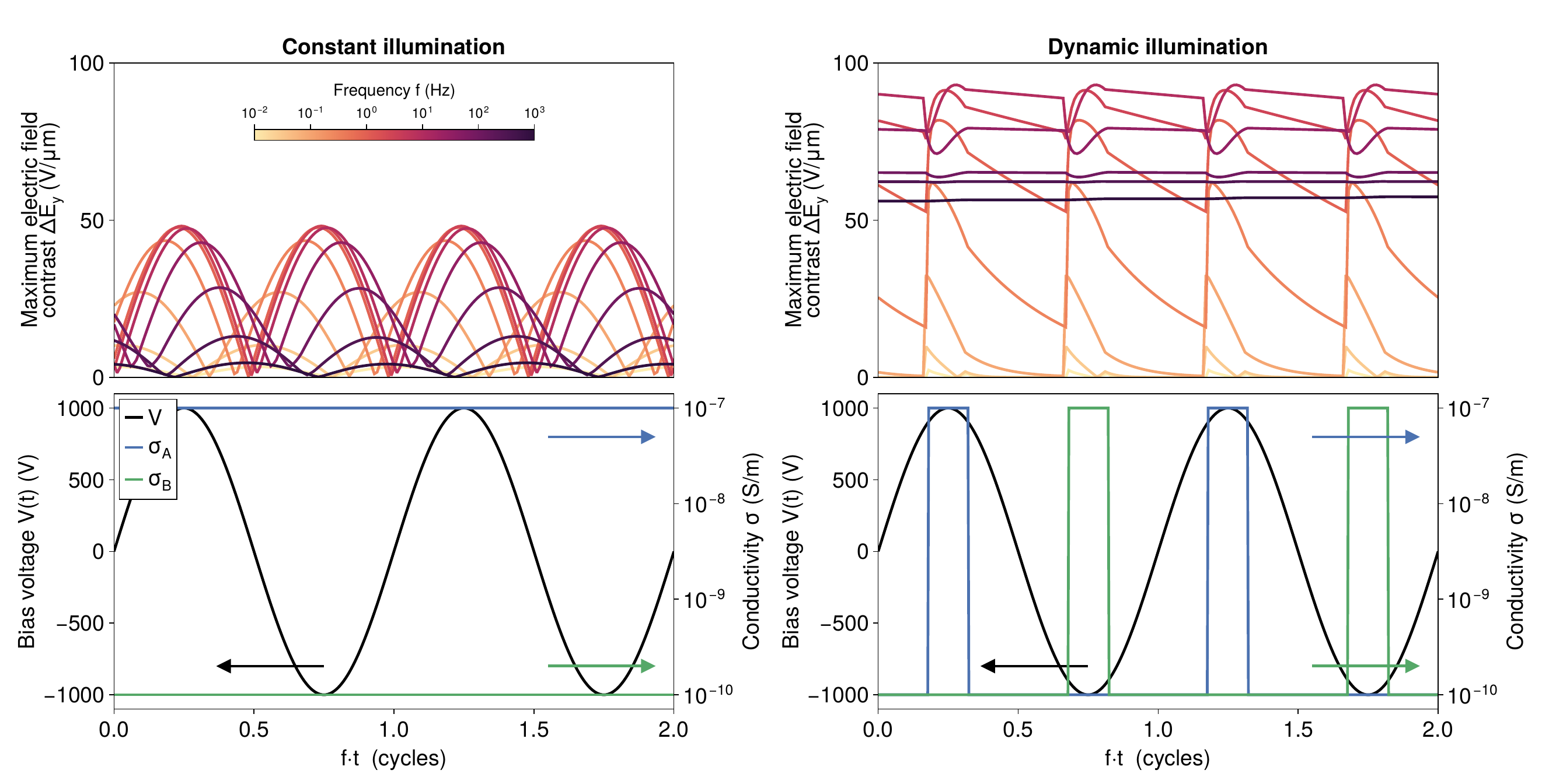}
    \caption{The simulated operation of a programmable nonlinear waveguide under constant programming illumination (left column) and dynamic illumination (right column). We partitioned the system illustrated in Fig.~\ref{fig:photoconductor_illustration} into a $120\times 52$ grid and used a finite-element method~\cite{Badia2020Gridap} to solve for the time evolution of the electric potential. The profiles of the bias voltage and conductivities are shown in the bottom row.}
    \label{fig:photoconductor_dynamic}
\end{figure}

Normal operation of the programmable nonlinear waveguide under constant programming illumination can be modeled as a situation in which the conductivity of region A, $\sigma_\text{A}$, is increased by the illumination relative to that of region B, $\sigma_\text{B}$. Intuitively, this raises the electric field inside the core beneath region A relative to that beneath region B, creating a contrast in the vertical electric field, $\Delta E_y$. This contrast in turn produces a spatial variation in the electric-field-induced $\chi^{(2)}$ nonlinearity, realizing a QPM grating. Because the programming illumination is constant in time, $\sigma_\text{A}$ and $\sigma_\text{B}$ are also time-independent. We assume the top electrode is connected to an AC bias voltage source $V=V_0\sin(2\pi f t)$ with frequency $f$. In Fig.~\ref{fig:photoconductor_dynamic}, we show the dynamics of $\Delta E_y$ simulated for various frequencies $f$ using a finite-element method. The time evolution of $\Delta E_y$ exhibits a clear periodic oscillation, which is the aforementioned limitation that hinders the continuous operation of the device.

In the dynamic illumination scheme, by contrast, we switch the programming illumination pattern in sync with the bias voltage. Specifically, only when the bias voltage exceeds a positive threshold $V_\text{th}$ do we illuminate region A, switching its conductivity to the bright-state value $\sigma_\text{b}$. Likewise, we illuminate region B only when the bias voltage falls below a negative threshold $-V_\text{th}$. Otherwise, the conductivity of the photoconductor remains at the dark state value $\sigma_\text{d}\ll\sigma_\text{b}$. This dynamic modulation rectifies the electric field inside the core, producing nearly time-invariant positive and negative electric fields $E_y$ beneath regions A and B, respectively. The results of the finite-element simulation are shown in Fig.~\ref{fig:photoconductor_dynamic}, where $\Delta E_y$ clearly remains above a certain level and converges to a constant as $f$ increases.

\begin{figure}[h]
    \centering
    \includegraphics[width=\linewidth]{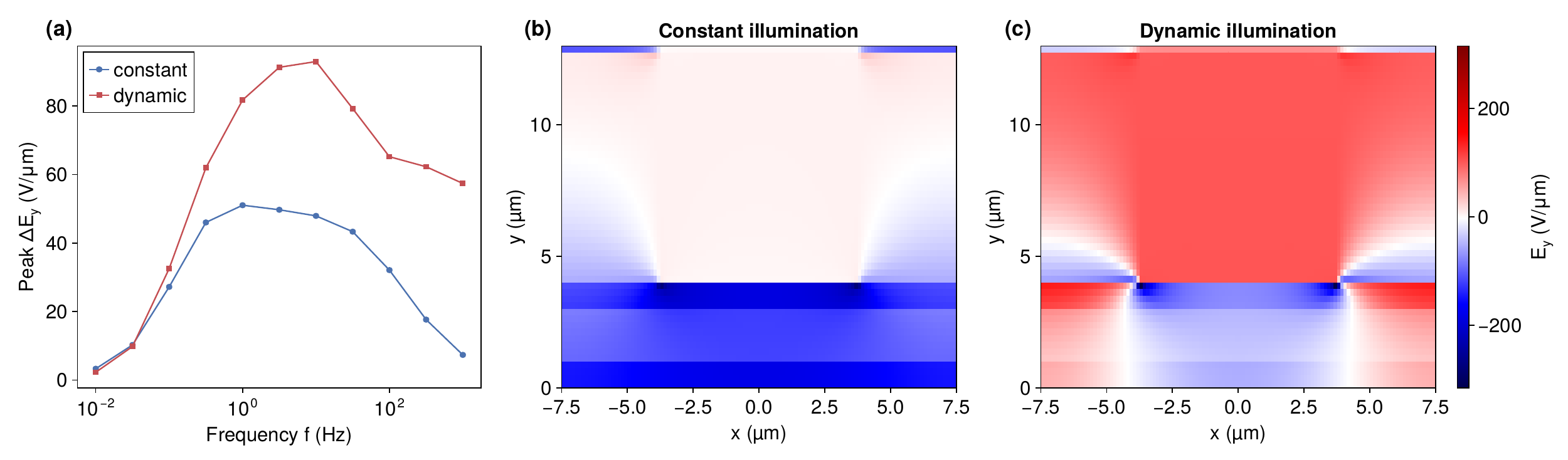}
    \caption{(a) The maximum field contrast $\Delta E_y$ attainable with the constant and dynamic illumination schemes for various frequencies $f$. The distribution of the vertical electric field $E_y$ under (b) constant illumination ($ft=0.4$) and (c) dynamic illumination ($ft=0.62$). The same model and simulations as in Fig.~\ref{fig:photoconductor_dynamic} are used.}
    \label{fig:photoconductor_field}
\end{figure}

It is worth pointing out that the dynamic illumination scheme can realize a $\Delta E_y$ approximately twice as large as that of the constant illumination scheme, indicating that we could nearly double the effective $\chi^{(2)}$ nonlinearity. In Fig.~\ref{fig:photoconductor_field}(a), we show the maximum attainable $\Delta E_y$ as a function of $f$ for both schemes. The $\Delta E_y$ attainable with the dynamic scheme is nearly twice that of the constant scheme. Intuitively, this is because the constant illumination can produce only a fully positive or fully negative distribution of $E_y$ inside the core (see Fig.~\ref{fig:photoconductor_field}(b)), whereas the dynamic illumination can produce both positive and negative $E_y$ simultaneously, which increases the dynamic range (see Fig.~\ref{fig:photoconductor_field}(c)).

\end{document}